2012

# FOUNDATIONS OF SCIENTIFIC RESEARCH

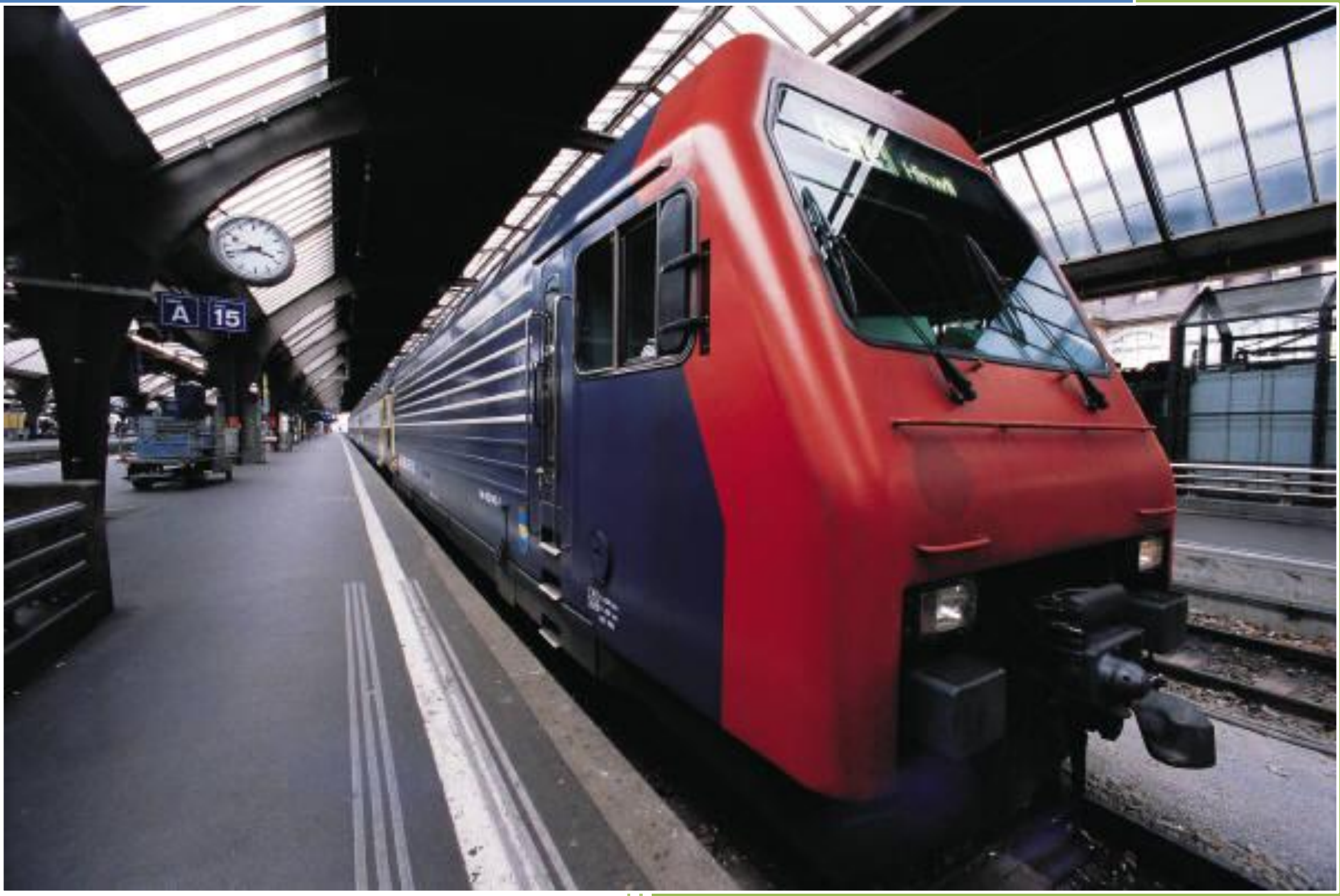

N. M. Glazunov

National Aviation University

25.11.2012

CONTENTS

## Preface

During years 2008 – 2011 author gives several courses on “Foundations of Scientific Research” at Computer Science Faculty of the National Aviation University (Kiev).

This text presents material to lectures of the courses. Some sections of the text are sufficiently complete, but in some cases these are sketchs without references to Foundations of Research Activities (FSR). Really this is the first version of the manual and author plan to edit, modify and extend the version. Some reasons impose the author to post it as e-print. Author compiled material from many sources and hope that it gives various points of view on Foundations of Research Activities.

Ars longa, vita brevis

## INTRODUCTION

Mastering the discipline “Foundations of Scientific Research” (Foundations of Research Activities) is aimed at training students in methodological foundations and organization of scientific research; organization of reference and information retrieval on the topic of research in system of scientific and technical libraries and by local and global computer information networks; analysis and evaluation of information and research and development processes in civil aviation and in another fields of national economy; guidance, principles and facilities of optimization of scientific research; preparation of facts, which documenting results of research scientific work (scientific report, article, talk, theses, etc.)

The main tasks of the discipline are to familiarize students with basic terminology, theoretical and experimental methods of scientific research as well as methods of analysis of observed results, their practical use and documentation facilities. The tasks of mastering the discipline “Foundations of scientific research” are the following:

- to learn professional terminology of scientific research;
- to be able to perform the reference and information retrieval on the topic of research;
- to be able to formulate methodological foundations of scientific research on specialty;
- to understand the organization of scientific research;
- to make scientific report (talk) on professional and socio-political topics defined by this syllabus.

Practical skills in the foundations of scientific research enable students to be aware of world scientific results and new technologies, to understand novel scientific results, papers, computer manuals, software documentation, and additional literature with the aim of professional decisions-making. Prolific knowledge and good practical skills in the foundations of scientific research allow students to study in novel scientific results, make investigations, reports, summaries and comments, develop scientific projects and be engaged in foundations of scientific research.

As a result of mastering the discipline a student shall

KNOW:

- basic professional and technical terminology on the disciplines defined by the academic curriculum;
- categorical apparatus of scientific research;
- main rules of handling scientific and technical literature;
- aim and tasks of scientific research;
- methodology and methods of scientific research;
- classification of methods by the level of investigation, by the state of the organization of scientific research, by the character of cognitive activity;
- types of exposition results of scientific research;
- peculiarities of students research activities.

LEARNING OUTCOMES:

- organize and carry out scientific research by oneself;
- carry out information retrieval of scientific literature;
- competently work with scientific information sources;
- take out optimal research methods by the content and aim of the scientific task.

The ideas in this manual have been derived from many sources [1-19,25]. Here I will try to acknowledge those that are explicitly attributable to other authors. Most of the other ideas are part of Scientific Research folklore. To try to attribute them to anyone would be impossible. Also in the manual we use texts from Wikipedia and some another papers and books. The author thanks his students A. Babaryka, V. Burenkov, K.Vasyanovich, D. Eremenko, A. Kachinskaya, L. Mel’nikova, O. Samusenko,

I. Tatomyr, V. Trush and others for texts of lectures, labs and homeworks on the discipline “Foundations of Scientific Research”. The list of references is indicated in Literature section at the end of the manual.

## 1. GENERAL NOTIONS ABOUT SCIENTIFIC RESEARCH

Science is the process of gathering, comparing, and evaluating proposed models against observables. A model can be a simulation, mathematical or chemical formula, or set of proposed steps. Under science we will understand natural sciences, mathematical sciences and applied sciences with special emphasis on computer sciences. In sone cases we will distinguish mathematics as the language of science. From school and university mathematical cources we know that reseachers (in the case these are schoolgirls, schoolboys, students) can clearly distinguish what is known from what is unknown at each stage of mathematical discovery. Science is like mathematics in that researchers in both disciplines can clearly distinguish what is known from what is unknown at each stage of scientific discovery. Models, in both science and mathematics, need to be internally consistent and also ought to be falsifiable (capable of disproof). In mathematics, a statement need not yet be proven; at such a stage, that statement would be called a conjecture. But when a statement has attained mathematical proof, that statement gains a kind of immortality which is highly prized by mathematicians, and for which some mathematicians devote their lives.

The hypothesis that people understand the world also by building mental models raises fundamental issues for all the fields of cognitive science. For instance in the framework of computer science there are a questions: How can a person's model of the word be reflected in a computer system? What languages and tools are needed to describe such models and relate them to outside systems? Can the models support a computer interface that people would find easy to use ?

Here we will consider basic notions about scientific research, research methods, stages of scientific research, motion of scientific research, scientific search. In some cases biside with the term “scientific research” we will use the term “scientific activety”.

At first we illustrate the ontology based approach to design the course Foundations of Research Activities. This is a course with the problem domains “Computer sciences”, “Software Engeneering“, “Electromagnetism”, “Relativity Theory (Gravitation)” and “Quantum Mechenics” that enables the student to both apply and expand previous content knowledge toward the endeavour of engaging in an open-ended, student-centered investigation in the pursuit of an answer to a question or problem of interest. Some background in concept analtsis, electromagnetism, special and general relativity and quantum theory are presented. The particular feature of the course is studying and applying computer-assisted methods and technologies to justification of conjectures (hypotheses). In our course, justification of conjectures encompasses those tasks that include gathering and analysis of data, go into testing conjectures, taking account of mathematical and computer-assisted methods of mathematical proof of the conjecture. Justification of conjectures is critical to the success of the solution of a problem. Design involves problem-solving and creativity.

Then, following to Wiki and some another sources, recall more traditional information about research and about scientific research.

At first recall definitions of two terms (Concept Map, Conception (Theory)) that will use in our course.

Concept Map: A schematic device for representing the relationships between concepts and ideas. The boxes represent ideas or relevant features of the phenomenon (i.e. concepts) and the lines represent connections between these ideas or relevant features. The lines are labeled to indicate the type of connection.

Conception (Theory): A general term used to describe beliefs, knowledge, preferences, mental images, and other similar aspects of a t lecturer’s mental structure.

Research is scientific or critical investigation aimed at discovering and interpreting facts. Research may use the scientific method, but need not do so.

Scientific research relies on the application of the scientific method, a harnessing of curiosity. This research provides scientific information and theories for the explanation of the nature and the properties of the world around us. It makes practical applications possible. Scientific research is funded by public authorities, by charitable organisations and by private groups, including many companies. Scientific research can be subdivided into different classifications according to their academic and application disciplines.

Recall some classifications:

Basic research. Applied research.
Exploratory research. Constructive research. Empirical research
Primary research. Secondary research.

Generally, research is understood to follow a certain structural process. The goal of the research process is to produce new knowledge, which takes three main forms (although, as previously discussed, the boundaries between them may be fuzzy):

- Exploratory research, which structures and identifies new problems
- Constructive research, which develops solutions to a problem
- Empirical research, which tests the feasibility of a solution using empirical evidence

Research is often conducted using the hourglass model. The hourglass model starts with a broad spectrum for research, focusing in on the required information through the methodology of the project (like the neck of the hourglass), then expands the research in the form of discussion and results.

Though step order may vary depending on the subject matter and researcher, the following steps are usually part of most formal research, both basic and applied:

- Formation of the topic
- Hypothesis
- Conceptual definitions
- Operational definitions
- Gathering of data
- Analysis of data
- Test, revising of hypothesis
- Conclusion, iteration if necessary

A common misunderstanding is that by this method a hypothesis can be proven or tested. Generally a hypothesis is used to make predictions that can be tested by observing the outcome of an experiment. If the outcome is inconsistent with the hypothesis, then the hypothesis is rejected. However, if the outcome is consistent with the hypothesis, the experiment is said to support the hypothesis. This careful language is used because researchers recognize that alternative hypotheses may also be consistent with the observations. In this sense, a hypothesis can never be proven, but rather only supported by surviving rounds of scientific testing and, eventually, becoming widely thought of as true (or better, predictive), but this is not the same as it having been proven. A useful hypothesis allows prediction and within the accuracy of observation of the time, the prediction will be verified. As the accuracy of observation improves with time, the hypothesis may no longer provide an accurate prediction. In this case a new hypothesis will arise to challenge the old, and to the extent that the new hypothesis makes more accurate predictions than the old, the new will supplant it.

## 1.1. Scientific method

Scientific method [1-4,6-8] refers to a body of techniques for investigating phenomena, acquiring new knowledge, or correcting and integrating previous knowledge. To be termed scientific, a method of inquiry must be based on gathering observable, empirical and measurable evidence subject to specific principles of reasoning. A scientific method consists of the collection of data through observation and experimentation, and the formulation and testing of hypotheses.

As have indicated in cited references knowledge is more than a static encoding of facts, it also includes the ability to use those facts in interacting with the world.

There is the operative definition: Knowledge - attach purpose and competence to information potential to generate action

Although procedures vary from one field of inquiry to another, identifiable features distinguish scientific inquiry from other methodologies of knowledge. Scientific researchers propose hypotheses as explanations of phenomena, and design experimental studies to test these hypotheses. These steps must be repeatable in order to dependably predict any future results. Theories that encompass wider domains of inquiry may bind many independently-derived hypotheses together in a coherent, supportive structure. This in turn may help form new hypotheses or place groups of hypotheses into context.

Among other facets shared by the various fields of inquiry is the conviction that the process be objective to reduce biased interpretations of the results. Another basic expectation is to document, archive and share all data and methodology so they are available for careful scrutiny by other scientists, thereby allowing other researchers the opportunity to verify results by attempting to reproduce them. This practice, called *full disclosure*, also allows statistical measures of the reliability of these data to be established.

## 1.2. Basic research

Does string theory provide physics with a grand unification theory?

The solution of the problem is the main goal of String Theory and basic research in the field [4].

Basic research (also called *fundamental* or *pure* research) has as its primary objective the advancement of knowledge and the theoretical understanding of the relations among variables (see statistics). It is *exploratory* and often driven by the researcher’s curiosity, interest, and intuition. Therefore, it is sometimes conducted without any practical end in mind, although it may have unexpected results pointing to practical applications. The terms “basic” or “fundamental” indicate that, through theory generation, basic research provides the foundation for further, sometimes applied research. As there is no guarantee of short-term practical gain, researchers may find it difficult to obtain funding for basic research.

Traditionally, basic research was considered as an activity that preceded applied research, which in turn preceded development into practical applications. Recently, these distinctions have become much less clear-cut, and it is sometimes the case that all stages will intermix. This is particularly the case in fields such as biotechnology and electronics, where fundamental discoveries may be made alongside work intended to develop new products, and in areas where public and private sector partners collaborate in order to develop greater insight into key areas of interest. For this reason, some now prefer the term *frontier* research.

### 1.2.1. Publishing

Academic publishing describes a system that is necessary in order for academic scholars to peer review the work and make it available for a wider audience [21-24,26]. The 'system', which is probably disorganised enough not to merit the title, varies widely by field, and is also always changing, if often slowly. Most academic work is published in journal article or book form. In publishing, STM publishing is an abbreviation for academic publications in science, technology, and medicine.

## 1.3. Information supply of scientific research.

Scientist’s bibliographic activity includes: organization, technology, control.

**Information retrieval systems and Internet.**

It is very important now to have lot’s of possibilities to have access to different kind of information. There are several ways. Indicate two of them and consider more carefully more modern: 1) go to library or 2) use Internet. As indicate many students: “I think that it is not difficult to understand why Internet is more preferable for me.” So, let as consider how works the best nowadays’s web-search Google and how a student can find article “A mathematical theory of communication” by C.E. Shannon.

### 1.3.1. How does Google work.

Google runs on a distributed network of thousands of low-cost computers and can therefore carry out fast parallel processing. Parallel processing is a method of computation in which many calculations can be performed simultaneously, significantly speeding up data processing. Google has three distinct parts:

Googlebot, a web crawler that finds and fetches web pages.

The indexer that sorts every word on every page and stores the resulting index of words in a huge database.

The query processor, which compares your search query to the index and recommends the documents that it considers most relevant.

Let’s take a closer look at each part.

### 1.3.2. Googlebot, Google’s Web Crawler

Googlebot is Google’s web crawling robot, which finds and retrieves pages on the web and hands them off to the Google indexer. It’s easy to imagine Googlebot as a little spider scurrying across the strands of cyberspace, but in reality Googlebot doesn’t traverse the web at all. It functions much like your web browser, by sending a request to a web server for a web page, downloading the entire page, then handing it off to Google’s indexer.

Googlebot consists of many computers requesting and fetching pages much more quickly than you can with your web browser. In fact, Googlebot can request thousands of different pages simultaneously. To avoid overwhelming web servers, or crowding out requests from human users, Googlebot deliberately makes requests of each individual web server more slowly than it's capable of doing.

Googlebot finds pages in two ways: through an add URL form, www.google.com/addurl.html, and through finding links by crawling the web.

Unfortunately, spammers figured out how to create automated bots that bombarded the add URL form with millions of URLs pointing to commercial propaganda. Google rejects those URLs submitted through its Add URL form that it suspects are trying to deceive users by employing tactics such as including hidden text or links on a page, stuffing a page with irrelevant words, cloaking (aka bait and switch), using sneaky redirects, creating doorways, domains, or sub-domains with substantially similar content, sending automated queries to Google, and linking to bad neighbors. So now the Add URL form also has a test: it displays some squiggly letters designed to fool automated "letter-guessers"; it asks you to enter the letters you see — something like an eye-chart test to stop spambots.

When Googlebot fetches a page, it culls all the links appearing on the page and adds them to a queue for subsequent crawling. Googlebot tends to encounter little spam because most web authors link only to what they believe are high-quality pages. By harvesting links from every page it encounters, Googlebot can quickly build a list of links that can cover broad reaches of the web. This technique, known as deep crawling, also allows Googlebot to probe deep within individual sites. Because of their massive scale, deep crawls can reach almost every page in the web. Because the web is vast, this can take some time, so some pages may be crawled only once a month.

Although its function is simple, Googlebot must be programmed to handle several challenges. First, since Googlebot sends out simultaneous requests for thousands of pages, the queue of "visit soon" URLs must be constantly examined and compared with URLs already in Google's index. Duplicates in the queue must be eliminated to prevent Googlebot from fetching the same page again. Googlebot must determine how often to revisit a page. On the one hand, it's a waste of resources to re-index an unchanged page. On the other hand, Google wants to re-index changed pages to deliver up-to-date results.

To keep the index current, Google continuously recrawls popular frequently changing web pages at a rate roughly proportional to how often the pages change. Such crawls keep an index current and are known as fresh crawls. Newspaper pages are downloaded daily, pages with stock quotes are downloaded much more frequently. Of course, fresh crawls return fewer pages than the deep crawl. The combination of the two types of crawls allows Google to both make efficient use of its resources and keep its index reasonably current.

### 1.3.3. Google's Indexer

Googlebot gives the indexer the full text of the pages it finds. These pages are stored in Google's index database. This index is sorted alphabetically by search term, with each index entry storing a list of documents in which the term appears and the location within the text where it occurs. This data structure allows rapid access to documents that contain user query terms.

To improve search performance, Google ignores (doesn't index) common words called stop words (such as the, is, on, or, of, how, why, as well as certain single digits and single letters). Stop words are so common that they do little to narrow a search, and therefore they can safely be discarded. The indexer also ignores some punctuation and multiple spaces, as well as converting all letters to lowercase, to improve Google's performance.

### 1.3.4. Google's Query Processor

The query processor has several parts, including the user interface (search box), the "engine" that evaluates queries and matches them to relevant documents, and the results formatter.

PageRank is Google's system for ranking web pages. A page with a higher PageRank is deemed more important and is more likely to be listed above a page with a lower PageRank.

Google considers over a hundred factors in computing a PageRank and determining which documents are most relevant to a query, including the popularity of the page, the position and size of the search terms within the page, and the proximity of the search terms to one another on the page. A patent application discusses other factors that Google considers when ranking a page. Visit SEOmoz.org's report for an interpretation of the concepts and the practical applications contained in Google's patent application.

Google also applies machine-learning techniques to improve its performance automatically by learning relationships and associations within the stored data. For example, the spelling-correcting system uses such techniques to figure out likely alternative spellings. Google closely guards the formulas it uses to calculate relevance; they're tweaked to improve quality and performance, and to outwit the latest devious techniques used by spammers.

Indexing the full text of the web allows Google to go beyond simply matching single search terms. Google gives more priority to pages that have search terms near each other and in the same order as the query. Google can also match multi-word phrases and sentences. Since Google indexes HTML code in addition to the text on the page, users can restrict searches on the basis of where query words appear, e.g., in the title, in the URL, in the body, and in links to the page, options offered by Google's Advanced Search Form and Using Search Operators (Advanced Operators).

## 2. ONTOLOGIES AND UPPER ONTOLOGIES

There are several definitions of the notion of ontology [10-13]. By T. R. Gruber (Gruber, 1992) “An ontology is a specification of a conceptualization”. By B. Smith and his colleagues, (Smith, 2004) “an ontology is a representational artefact whose representational units are intended to designate universals in reality and the relations between them”. By our opinion the definitions reflect critical goals of ontologies in computer science. For our purposes we will use more specific definition of ontology: concepts with relations and rules define ontology (Gruber, 1992; Ontology, 2008; Wikipedia, 2009 ).

Ontology Development aims at building reusable semantic structures that can be informal vocabularies, catalogs, glossaries as well as more complex finite formal structures representing the entities within a domain and the relationships between those entities. Ontologies, have been gaining interest and acceptance in computational audiences: formal ontologies are a form of software, thus software development methodologies can be adapted to serve ontology development. A wide range of applications is emerging, especially given the current web emphasis, including library science, ontology-enhanced search, e-commerce and configuration. Knowledge Engineering (KE) and Ontology Development (OD) aims at becoming a major meeting point for researchers and practitioners interested in the study and development of methodologies and technologies for Knowledge Engineering and Ontology Development.

There are next relations among concepts:

associative
partial order
higher
subordinate

subsumption relation (is a, is subtype of, is subclass of)
part-of relation.

More generally, we may use Description Logic (DL) [5] for constructing consepts and knowledge base (Franz Baader, Werner Nutt. Basic Description Logics). See the section: **Scientific research in Artificial Intelligence**

Different spaces are used in aforementioned courses. In our framework we treat ontology of spaces and ontology of symmetries as upper ontologies.

## 2.1. Concepts of Foundations of Research Activities

Foundations of Research Activities Concepts:

(a) Scientific Method.

(b) Ethics of Research Activity.

(c) Embedded Technology and Engineering.

(d) Communication of Results (Dublin Core).

In the section we consider briefly (a).

Investigative processes, which are assumed to operate iteratively, involved in the research method are the follows:

(i) Hypothesis, Low, Assumption, Generalization;

(ii) Deduction;

(iii) Observation, Confirmation;

(iv) Induction.

Indicate some related concepts: Problem. Class of Scientific Data. Scientific Theory. Formalization. Interpretation. Analyzing and Studying of Classic Scientific Problems. Investigation. Fundamental (pure) Research. Formulation of a Working Hypothesis to Guide Research. Developing Procedures to Testing a Hypothesis. Analysis of Data. Evaluation of Data.

## 2.2. Ontology components

Contemporary ontologies share many structural similarities, regardless of the language in which they are expressed. As mentioned above, most ontologies describe individuals (instances), classes (concepts), attributes, and relations. In this subsection each of these components is discussed in turn.

Common components of ontologies include:

- Individuals: instances or objects (the basic or "ground level" objects)
- Classes: sets, collections, concepts, types of objects, or kinds of things.[10]
- Attributes: aspects, properties, features, characteristics, or parameters that objects (and classes) can have
- Relations: ways in which classes and individuals can be related to one another
- Function terms: complex structures formed from certain relations that can be used in place of an individual term in a statement
- Restrictions: formally stated descriptions of what must be true in order for some assertion to be accepted as input
- Rules: statements in the form of an if-then (antecedent-consequent) sentence that describe the logical inferences that can be drawn from an assertion in a particular form
- Axioms: assertions (including rules) in a logical form that together comprise the overall theory that the ontology describes in its domain of application. This definition differs from that of "axioms" in generative grammar and formal logic. In those disciplines, axioms include only statements asserted as *a priori* knowledge. As used here, "axioms" also include the theory derived from axiomatic statements.
- Events: the changing of attributes or relations.

## 2.3. Ontology for the visualization of a lecture

Upper ontology: visualization. Visualization of the text (white text against the dark background) is subclass of visualization.

Visible page, data visualization, flow visualization, image visualization, spatial visualization, surface rendering, two-dimensional field visualization, three-dimensional field visualization, video content.

# 3. ONTOLOGIES OF OBJECT DOMAINS

## 3.1 Elements of the ontology of spaces and symmetries

There is the well known from mathematics

*space* - *ring_of_functions_on_the_space*

duality. In the subsection we only mention some concepts, relations and rules of the ontology of spaces and symmetries.

Two main concepts are space and symmetry.

3-dimensional real space $R^3$; Linear group $GL(3, R)$ of automorphisms of $R^3$;

Classical physical world has three spatial dimensions, so electric and magnetic fields are 3-component vectors defined at every point of space.

Minkowski space-time $M^{1,3}$ is a 4-dimensional real manifold with a pseudoriemannian metric $t^2 - x^2 - y^2 - z^2$. From $M^{1,3}$ it is possible to pass to $R^4$ by means of the substitution $t \to iu$ and an overall sign-change in the metric. A compactification of $R^4$ by means of a stereographic projection gives $S^4$. 2D space, 2D object, 3D space, 3D object.

Additiona material for advanced students:

Let *SO(1,3)* be the pseudoortogonal group. The moving frame in $M^{1,3}$ is a section of the trivial bundle $M^{1,3} \times SO(1,3)$. A complex vector bundle $M^{1,3} \times \mathrm{C}^2$ is associated with the frame bundle by the representation *SL(2,C)* of the Lorentz group *SO(1,3)*.

The space-time *M* in which strings are propagating must have many dimensions (10, 26. …) . The ten-dimensional space-time is locally a product $M = M^{1,3} \times K$ of macroscopic four-dimensional space-time and a compact six-dimensional Calabi-Yau manifold *K* whose size is on the order of the Planck length.

Principle bundle over space-time, structure group, associated vector bundle, connection, connection one-form, curvature, curvature form, norm of the curvature.

Foregoing concepts with relations and rules define elements of the domain ontology of spaces.

### 3.1.1 Concepts of Electrodynamics and Classical Gauge Theory

Preliminarities: electricity and magnetism. This subsection contains additiona material for advanced students.

Short history: Schwarzchild action, Hermann Weyl, F. London, Yang-Mills equations.

Quantum Electrodynamics is regarded as physical gauge theory. The set of possible gauge transformations of the entire configuration of a given gauge theory also forms a group, the gauge group of the theory. An element of the gauge group can be parameterized by a smoothly varying function from the points of space-time to the (finite-dimensional) Lie group, whose value at each point represents the action of the gauge transformation on the fiber over that point.

Concepts: Gauge group as a (possibly trivial) principle bundle over space-time, gauge, classical field, gauge potential.

# 4. EXAMPLES OF RESEARCH ACTIVITY

## 4.1. Scientific activity in arithmetics, informatics and discrete mathematics

Discrete mathematics becomes now not only a part of mathematics, but also a common language for various fields of cybernetics, computer science, informatics and their applications. Discrete mathematics studies discrete structures, operations with these structures and functions and mappings on the structures.

Examples of discrete structures are:

finite sets (**FSets**);

sets: **N** – natural numbers;

**Z –** integer numbers;

**Q –** rational numbers;

algebras of matrices over finite, rational and complex fields.

Operations with discrete structures:

$\cup$ - union; $\cap$ - intersection; $A\backslash B$ – set difference and others.

Operations with elements of discrete structures:

+ - addition, * - multiplication, scalar product and others.

Recall some facts about integer and natural numbers. Sum, difference and product of integer numbers are integers, but the quotient under the division of an integer number *a* by the integer number *b* (if *b* is not equal to zero) maybe as an integer as well as not the integer. In the case when *b* divides *a* we will denote it as $b|\ a$. From school program we know that any integer *a* is represented uniquely by the positive integer *b* in the form

$$a = bq + r; \quad 0 \leq r < b.$$

The number *r* is called the residues of *a* under the division by *b*. We will study in section 3, that residues under the division of all natural numbers on a natural *n* form the ring ***Z****/n****Z***. Below we will consider *positive divisors* only. Any integer that divides simultaneously integers *a, b, c,…m,* is called their *common divisor*. The largest from common divisors is called the *greatest common divisor* and is denoted by *(a, b, c,…m)*.

If $(a, b, c,...m) = 1$, then $a, b, c,...m$ are called *coprime*. The number *1* has only one positive divisor. Any integer, greater than *1*, has not less than two divisors, namely *1* and itself the integer. If an integer has exactly two positive divisors then the integer is called a *prime*.

Recall now two functions of integer and natural arguments: the Mobius function $\mu(a)$ and Euler's function $\varphi(n)$. *The Mobius function* $\mu(a)$ is defined for all positive integers $a$ : $\mu(a) = 0$ if $a$ is divided by a square that is not the unit; $\mu(a) = (-1)^k$ where $a$ is not divided by a square that is not the unit, $k$ is the number of prime divisors of $a$; $\mu(1) = 1$.

Examples of values of the Mobius function: $\mu(1)=1, \mu(2)=-1, \mu(3)=-1, \mu(4)=0, \mu(5)=-1, \mu(6)=1$. *Euler's function* $\varphi(n)$ is defined for any natural number $n$. $\varphi(n)$ is the quantity of numbers

$$0 \quad 1 \quad 2 \ldots n-1$$

that are coprime with $n$. Examples of values of Euler's function: $\varphi(1)=1, \varphi(2)=1, \varphi(3)=2, \varphi(4)=2, \varphi(5)=4, \varphi(6)=2$.

## 4.2. Algebra of logic and functions of the algebra of logic

The area of Algebra of logic and functions of the algebra of logic connects with mathematical logic and computer science. Boolean algebra is a part of the Algebra of logic.

Boolean algebra, an abstract mathematical system primarily used in computer science and in expressing the relationships between sets groups of objects or concepts). The notational system was developed by the English mathematician George Boole to permit an algebraic manipulation of logical statements. Such manipulation can demonstrate whether or not a statement is true and show how a complicated statement can be rephrased in a simpler, more convenient form without changing its meaning.

Let $p_1$, $p_2$, ...$p_n$ be propositional variables where each variable can take value 0 or 1. The logical operations (connectives) are $\wedge, \vee, \neg, \equiv, \rightarrow$. Their names are:

conjunction (AND - function), disjunction (OR - function), negation, equivalence, implication.

**Definition**. A *propositional formula* is defined inductively as:

1. Each variable $p_i$ is a formula.
2. If *A* and *B* are formulas, then $(A \vee B), (A \wedge B), (\neg A), (A \equiv B), (A \rightarrow B)$ are formulas.
3. *A* is a formula iff it follows from 1 and 2.

**Remark**. The operation of negation has several equivalent notations: ¬, ~ or ‘ (please see below).

### 4.3. Function of the algebra of logic

Let *n* be the number of Boolean variables. Let $P_n(0,1)$ be the set of Boolean functions in *n* variables. With respect to a fixed order of all $2^n$ possible arguments, every such function $f:\{0,1\}^n \rightarrow \{0,1\}$ is uniquely represented by its *truth table*, which is a vector of length $2^n$ listing the values of *f* in that order. Boolean functions of one variable ($n = 1$):

| x | 0 | 1 | x | ~x |
|---|---|---|---|---|
| 0 | 0 | 1 | 0 | 1 |
| 1 | 0 | 1 | 1 | 0 |

Boolean functions of two variables (*n = 2):*

| $x_1 \mid x_2$ | $x_1 \& x_2$ | $x_1 \cup x_2$ | $x_1 \to x_2$ | $x_1 + x_2$ | $x_1 / x_2$ |
|---|---|---|---|---|---|
| 0 \| 0 | 0 | 0 | 1 | 0 | 1 |
| 0 \| 1 | 0 | 1 | 1 | 1 | 1 |
| 1 \| 0 | 0 | 1 | 0 | 1 | 1 |
| 1 \| 1 | 1 | 1 | 1 | 0 | 0 |

$0, 1, x, \sim x, x_1 \& x_2,, x_1 \vee x_2, x_1 \to x_2, x_1 + x_2, x_1/x_2$- elementary functions ($\sim x$ (as $!x$) is the negation of $x$).

Let $p \to q$ be the implication. The implication $p \to q$ is equivalent to $\sim p \vee q$. The Venn diagram for $\sim p \vee q$ is represented as $\neg P \cup Q$, where $\neg P$ is the supplement of the set $P$ in a universal set, $Q$ is a set that corresponds to the variable $q$, $\neg P \cup Q$ is the union of $\neg P$ and $Q$.

The case of $n$ variables: $E_2 = Z/2 = \{0,1\} = \boldsymbol{F}_2$

$F: E_2^n \to E_2$

| | $x_1$ $x_2$ $x_3 \ldots x_{n-1}$ $x_n$ | $F(x_1 \ldots x_n)$ |
|---|---|---|
| 0 | 0 0 0 … 0 0 | F(0,0,…,0) |
| 1 | 0 0 0 … 0 1 | F(0,0,…,1) |
| 2 | 0 0 0 … 1 0 | F(0,…,1,0) |
| . | …………………….. | ………… |
| $2^n$ | 1 1 1 … 1 1 | F(1,1,…,1) |

***Theorem.*** Let $s = 2^n$. Then the number $\#P_n(0,1)$ of Boolean function of $n$ Boolean variables is equal to:

$\#P_n(0,1)=2^s$.

**Example.** If $n = 3$ then $\#P_3(0,1) = 2^8=256$.

# 5. SOME NOTIONS OF THE THEORY OF FINITE AND DISCRETE SETS

## 5.1. Sets, subsets, the characteristic function of a subset

If $S$ is a finite set we will put $\#S$ for the *cardinality* (the number of elements) of $S$. Let $B$ be a subset of a set $S$. A function $I_S: S \rightarrow \{0,1\}$ is called *the characteristic function of the subset B* if $I_S(s) = 1$ if s in $B$; otherwise $I_S(s) = 0$.

### 5.1.1. Set theoretic interpretation of Boolean functions.

In his 1881 treatise, *Symbolic Logic,* the English logician and mathematician John Venn interpreted Boole's work and introduced a new method of diagramming Boole's notation. This method is now know as the Venn diagram. When used in set theory, Boolean notation can demonstrate the relationship between groups, indicating what is in each set alone, what is jointly contained in both, and what is contained in neither. Boolean algebra is of significance in the study of information theory, the theory of probability, and the geometry of sets. The expression of electrical networks in Boolean notation has aided the development of switching theory and the design of computers.

Consider a Boolean algebra of subsets $2^{\#A}$ (where $\#A$ is the cardinality of a finite set $A$) generated by a set $A$, which is the set of subsets of $A$ that can be obtained by means of a finite number of the set operations union, intersection, and complementation. Then each element of the set $2^{\#A}$ defines the characteristic function of the element and this characteristic function is called a Boolean function geberated by $A$. So there are $2^{2^{\#A}}$ inequivalent Boolean functions for a set $A$ with cardinality $\#A$. This gives a set theoretic interpretation of Boolean functions.

### 5.2. Sets and their maps

Let $f : X \to Y$ be a map. By definition, the *image* of $f$ is defined by

$$Im(f) = Im\, X_f = \{y \in Y \mid \exists x \in X, f(x) = y\}.$$

A map $f$ is the **Injection** if

$Im(f) \subseteq Y$ and for $x_1, x_2 \in X, x_1 \neq x_2 \Rightarrow f(x_1) \neq f(x_2)$.

A map $f$ is the **Surjection** if

$Im(f) = Y$ and for $\forall y \in Y \exists x \in X$ such that $f(x) = y$.

A map $f$ is the **Bijection** if

$f$ is the bijection, if $f$ is the injection and $f$ is the surjection.

Let $Im(f) = Y$. Then by definition the *preimage* is defined by

$$PreIm\, Y_f = \{x \in X \mid f(x) = y\}.$$

# 6. ALGEBRAIC OPERATIONS AND ALGEBRAIC STRUCTURES

## 6.1. Elements of group theory

### 6.1.1. Groups

A group $G$ is a finite or infinite set of elements together with a binary operation which together satisfy the four fundamental properties of closure, associativity, the identity property, and the inverse property. The operation with respect to which a group is defined is often called the "group operation," and a set is said to be a group "under" this operation. Elements $A$, $B$, $C$, ... with binary operation between $A$ and $B$ denoted $AB$ form a group if

1. Closure: If $A$ and $B$ are two elements in $G$, then the product $AB$ is also in $G$.

2. Associativity: The defined multiplication is associative, i.e., for all $A, B, C \in G$. $(AB)C = A(BC)$.

3. Identity: There is an identity element $I$ (a.k.a. $1$, $E$, $0$, or $e$) such that $IA = AI = A$ for every element $A \in G$.

4. Inverse: There must be an inverse or reciprocal of each element. Therefore, the set must contain an element $B = A^{-1}$ such that $AA^{-1} = A^{-1}A = I$ for each element of $G$.

A group is a monoid each of whose elements is invertible.

A group must contain at least one element, with the unique (up to isomorphism) single-element group known as the trivial group.

The study of groups is known as group theory. If there are a finite number of elements, the group is called a finite group and the number of elements is called the group order of the group. A subset of a group that is closed under the group operation and the inverse operation is called a subgroup. Subgroups are also groups, and many commonly encountered groups are in fact special subgroups of some more general larger group.

A basic example of a finite group is the symmetric group $S_n$, which is the group of permutations (or "under permutation") of $n$ objects. The simplest infinite group is the set of integers under usual addition. Next example:

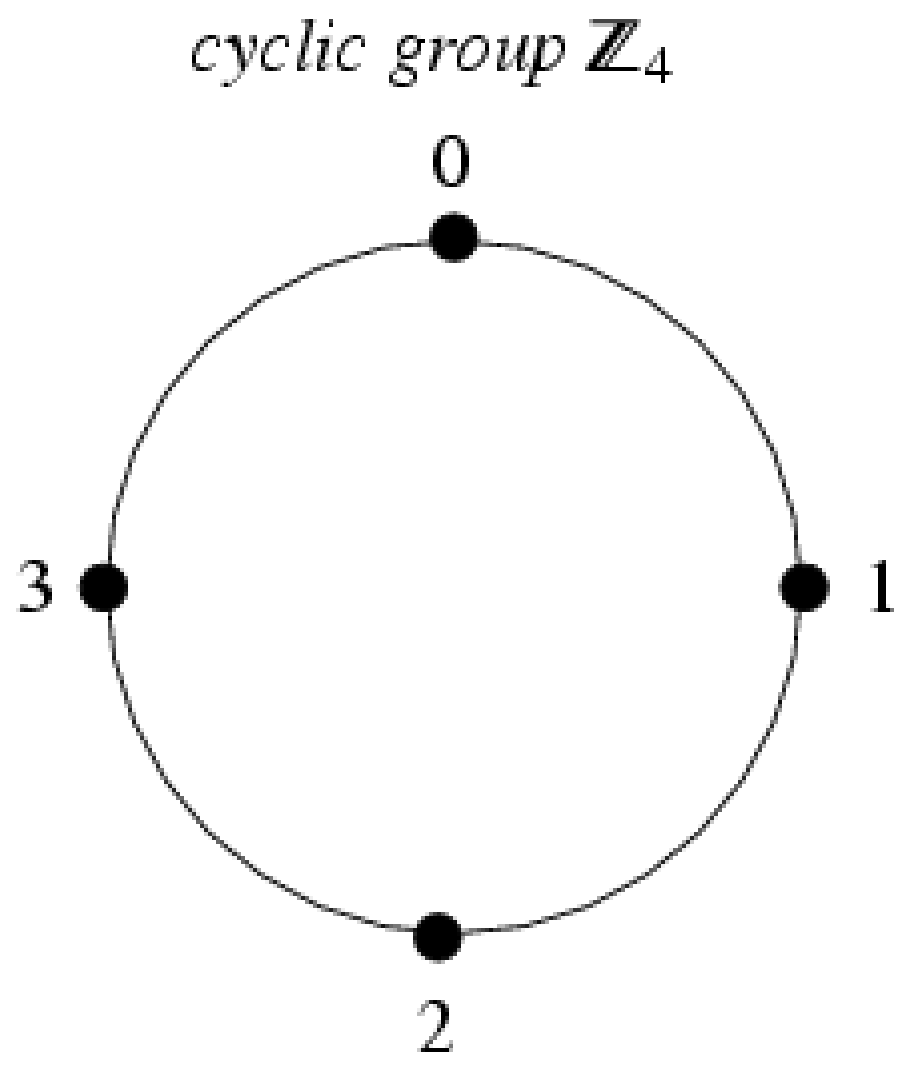

One very common type of group is the cyclic groups. This group is isomorphic to the group of integers (modulo *n*), is denoted $Z_n$, , or $\mathbb{Z}/n\mathbb{Z}$, and is defined for every integer *n* > 1. It is closed under addition, associative, and has unique inverses. The numbers from $0$ to $n-1$ represent its elements, with the identity element represented by $0$, and the inverse of *i* is represented by $n-i$.

A map between two groups which preserves the identity and the group operation is called a homomorphism. If a homomorphism has an inverse which is also a homomorphism, then it is called an isomorphism and the two groups are called isomorphic. Two groups which are isomorphic to each other are considered to be "the same" when viewed as abstract groups. For example, the group of rotations of a square, illustrated below, is the cyclic group $\mathbb{Z}_4$.

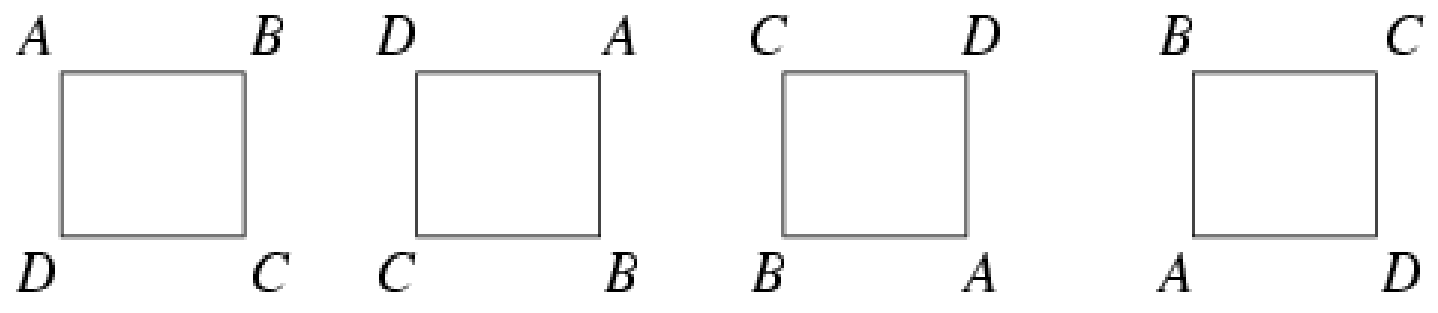

In general, a group action is when a group acts on a set, permuting its elements, so that the map from the group to the permutation group of the set is a homomorphism. For example, the rotations of a square are a subgroup of the permutations of its corners. One important group action for any group *G* is its action on itself by conjugation. These are just some of the possible group automorphisms. Another important kind of group action is a group representation, where the group acts on a vector space by invertible linear maps. When the field of the vector space is the complex numbers, sometimes a representation is called a C*G* module. Group actions, and in particular representations, are very important in applications, not only to group theory, but also to physics and aviation. Since a group can be thought of as an abstract mathematical object, the same group may arise in different contexts. It is therefore useful to think of a representation of the group as one particular incarnation of the group, which may also have other representations. An irreducible representation of a group is a representation for which there exists no unitary transformation which will transform the representation matrix into block diagonal form. The irreducible representations have a number of remarkable properties, as formalized in the group orthogonality theorem.

### 6.1.1.1. Homomorphisms of groups

Let $G$ be a group: $G = \{g \in G \mid g, g_1, g_2 \in G, \exists e : eg = ge,$
$\forall g \in G\ \exists g^{-1}$ such that $g^{-1} = g^{-1} g = e, g\,(g_1 g_2) = (g g_1)\, g_2\,.\}$
A *homomorphism* $\alpha : G \to H$ is the map $\alpha$ from $G$ to $H$ such that (s.t.) $\alpha(g_1 g_2) = \alpha(g_1)\alpha(g_2)$. It gives $\alpha(e_g) = e_h$

Let $G, H, L$ be groups with sets of homomorphisms

$Hom(G, H)$ ,$Hom(H, L)$, $Hom(G, L)$

$G \overset{\alpha}{\dashrightarrow} H \overset{\beta}{\dashrightarrow} L.$

So, there is the map:

$$Hom(G, H) \times Hom(H, L) \to Hom(G, L)$$

The *kernel* of a homomorphism $f$ is defined by (below $0$ is the identity element)

$$Ker\, f = \{g \in G \mid f(g) = 0\}.$$

*Epimorphis* **:**

$$f : G \text{ ----> } H : \forall h \in H\ \exists g \in G \text{ s.t. } f(g) = h$$

*Monomorphism* **:** $f : G \rightarrow H$; homomorphism $f$ s.t.

$$\forall g \neq e_G\ ,\ f(g) \neq e_H$$

### 6.1.2. Abelian groups (AG)

Recall once more the definition of a group:

A g*roup G* is the set of elements with binary operation * such that for any two elements $a, b \in G$: $a * b \in G$.
The operation satisfies 3 axioms:

1. $(a*b)*c = a*(b*c) - associativity$
2. The element $e$ (unit of the group) with conditions $e*a = a*e = a$ exists.
3. For every $a \in G$ exists $a^{-1} \in G$; such that $a \cdot a^{-1} = e$;

In abstract algebra, an **abelian group** is a group ($G$, *) that is commutative, i.e., in which $a * b = b * a$ holds for all elements $a$ and $b$ in $G$. Abelian groups are named after Niels Henrik Abel.

If a group is abelian, we usually write the operation as + instead of *, the identity element as 0 (often called the *zero element* in this context) and the inverse of the element $a$ as $-a$.

Examples of abelian groups include all cyclic groups such as the integers **Z** (with addition) and the integers modulo $n$ $\mathbf{Z}_n$ (also with addition). Every field gives rise to two abelian groups in the same fashion. Another important example is the factor group **Q**/**Z**, an injective cogenerator.

If $n$ is a natural number and $x$ is an element of an abelian group $G$, then $nx$ can be defined as $x + x + ... + x$ ($n$ summands) and $(-n)x = -(nx)$. In this way, $G$ becomes a module over the ring **Z** of integers. In fact, the modules over **Z** can be identified with the abelian groups. Theorems about abelian groups can often be generalized to theorems about modules over principal ideal domains. An example is the classification of finitely generated abelian groups.

Any subgroup of an abelian group is normal, and hence factor groups can be formed freely. Subgroups, factor groups, products and direct sums of abelian groups are again abelian. If $f, g : G \to H$ are two group homomorphisms between abelian groups, then their sum $f+g$, defined by $(f+g)(x) = f(x) + g(x)$, is again a homomorphism. (This is not true if $H$ is a non-abelian group). The set Hom($G$, $H$) of all group homomorphisms from $G$ to $H$ thus turns into an abelian group in its own right.

The abelian groups, together with group homomorphisms, form a category, the prototype of an abelian category.

Somewhat akin to the dimension of vector spaces, every abelian group has a *rank*. It is defined as the cardinality of the largest set of linearly independent elements of the group. The integers and the rational numbers have rank one, as well as every subgroup of the rationals. While the rank one torsion-free abelian groups are well understood, even finite-rank abelian groups are not well understood. Infinite-rank abelian groups can be extremely complex and many open questions exist, often intimately connected to questions of set theory.

### 6.1.3. Discrete abelian groups with finite number of generators (DAGFNG)

**Examples**:

$F_2 = \{0,1\} \quad 1+1=0$;

$\boldsymbol{Z}_n^m = (\boldsymbol{Z}/n\boldsymbol{Z})^m,\ m > 0$;

***Q***- rational numbers (has no finite number of generators).

Free discrete abelian groups with finite number of generators (a lattice in the geometric interpretation).

**Example**:

$Z^n, n \geq 1$

$Z^2 = Z \times Z, e = (0,0)$

$\overline{a}_1 = (1,0)$ and $\overline{a_2} = (0,1)$ are generators of the lattice.

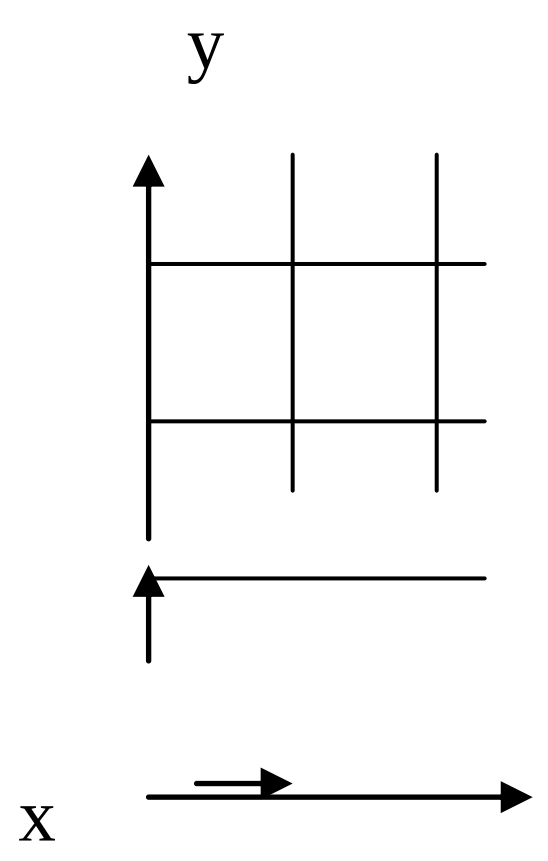

Let $R^n$ be the real space, $\overline{a_1}, \overline{a_2}, .. \overline{a_n}$, its basis (below we will sometimes denote the basis as $a_1, \ldots, a_n$). Then the sets

$$\Lambda = \{\lambda_1 a_1 + \lambda_2 a_2 + .. + \lambda_n a_n, \lambda_i \in Z\}$$
$$\Lambda = \{\lambda_1 a_1 + \lambda_2 a_2 \mid \lambda_1, \lambda_2 \in Z\}$$

are $n$-dimensional and $2$-dimensional lattices. By definition, $\det \Lambda = \left|\det(\overline{a_1}, \ldots, \overline{a_n})\right|$ is the determinant of the lattice $\Lambda$.

Example : $\Lambda = \{\lambda_1 a_1 + \lambda_2 a_2 \mid a_1 = (1,0), a_2 = (0,1)\}$

$$\det \Lambda = \begin{vmatrix} 1 & 0 \\ 0 & 1 \end{vmatrix} = 1$$

$\overline{a_1} = (3,0)$ $\quad$ $\begin{vmatrix} 3 & 0 \\ 1 & 2 \end{vmatrix} = 6$

$\overline{a_2} = (1,2)$

$$\Lambda_2 = \{\lambda_1 \overline{a}_1 + \lambda_2 \overline{a}_2\}$$
$$\det \Lambda_2 = 6$$

### 6.1.3.1. Admissible lattice

**Definition**: A lattice $\Lambda$ is said to be *admissible* for a body $R$ in $xy$ plane if it contains no points of $R$ other than perchance the origin if (0,0) is a point of $R$.

$$\Lambda = \{\lambda_1 a_1 + \lambda_2 a_2\},\ R \cap \Lambda = (0,0).$$

### 6.1.4. Symmetric group

The symmetric group $S_n$ which is the group of all $n!$ permutations $\sigma$ of $\{1,2,\ldots,n\}$ is non-abelian iff $n$ is greater or equal to $3$.

**Exercise**. Compute all elements of the group $S_3$ and their products.

The symmetric group $S_n$ is generated by all *transpositions* of $n$ consecutive numbers, $S_n = ((1,2),(2,3),\ldots,(n-1,n))$.

### 6.1.5. Subgroups

A group $H$ is called a *subgroup* of G ($G > H$) if $H$ is a subset of $G$ and for any two elements $h_1$ and $h_2$ from $H$, $h_1 * h_2 \in H$. Formally:

1) $H < G$;

2) $h_1, h_2 \in H; h_1 * h_2 \in H$;

3) $\exists\ \mathbf{e}_H,\ \mathbf{e}_H * h = h$;

4) $\forall h \in H\ \exists\ h^{-1},\ h * h^{-1} = \mathbf{e}_H$;

## 6.2. Elements of field theory in discrete mathematics

### 6.2.1. Finite fields

Elements of field theory considers sets, such as the rational number line, on which all the usual arithmetic properties hold --- those governing addition, subtraction, multiplication and division. The study of multiple fields through Galois theory is important for the study of polynomial equations, and thus has applications to number theory and group theory. Tools in this area can also be used to show certain Euclidean geometry constructions are impossible.

Specifically, a field is a commutative ring in which every nonzero element is assumed to have a multiplicative inverse. Examples include the rational numbers **Q**, finite fields (the Galois fields with $p^n$ elements for some prime $p$), and various fields of functions. All these examples are of characteristic zero except the finite fields (if there is a finite set of 1's which add to zero, the cardinality of the smallest such set is the characteristic, a prime).

Several constructions allow the creation of more fields, and in particular generate field extensions $K/F$ (i.e. nested pairs of fields $F < K$). Algebraic extensions are those in which every element of $K$ is the root of a polynomial with coefficients in $F$; elements of $K$ which are not algebraic are transcendental over $F$ (e.g. $\pi$ is transcendental over ***Q***). The field of rational functions $F(x)$ in one variable is transcendental over $F$. If $P(x)$ is an irreducible polynomial in the ring $F[x]$, then the quotient ring $K = F[x]/( P(x)$ is an algebraic extension of $F$ in which $P$ has a root.

Some themes of field theory are then immediately apparent.

First, the study of fields is the appropriate venue for the consideration of some topics in number theory. For example, one approach to Fermat's Last Theorem (that the equation $x^n + y^n = z^n$ has no solutions in positive integers when $n$ is greater than 2) suggests factoring numbers arising from a putative solution: $x^n+y^n=(x+y)(x+ry)...(x+sy)$ where $r,...,s$ are roots of unity lying in an extension field of ***Q***. This analysis centers upon the ring of integers in the extension field (which is a well-defined subring of number fields) more so than the extension field itself; thus this discussion is more appropriately considered part of Number Theory than Field Theory, but certain one uses tools from Field Theory -- the norm and trace mappings, the structure of the group of units, and so on.

$\boldsymbol{F}_p$, p – is a prime number.

$p = 5$, elements: *0, 1, 2, 3, 4.*

$\boldsymbol{F}_5$ – field of 5 elements

| + | 0 | 1 | 2 | 3 | 4 |
|---|---|---|---|---|---|
| 0 | 0 | 1 | 2 | 3 | 4 |
| 1 | 1 | 2 | 3 | 4 | 0 |
| 2 | 2 | 3 | 4 | 0 | 1 |
| 3 | 3 | 4 | 0 | 1 | 2 |
| 4 | 4 | 0 | 1 | 2 | 3 |

| * | 0 | 1 | 2 | 3 | 4 |
|---|---|---|---|---|---|
| 0 | 0 | 0 | 0 | 0 | 0 |
| 1 | 0 | 1 | 2 | 3 | 4 |
| 2 | 0 | 2 | 4 | 1 | 3 |
| 3 | 0 | 3 | 1 | 4 | 2 |
| 4 | 0 | 4 | 3 | 2 | 1 |

$\boldsymbol{N_0} = \boldsymbol{N}$ U{0}

$1019 \equiv 4 (\mathrm{mod}\, 5)$

mod 5 $\quad 6 \equiv 1 (\mathrm{mod}\, 5)$

$N_0$ → {0, 1, 2, 3, 4}

1 + 4 = 0

## 6.3. Algebras

Formally, an algebra is a vector space $V$, over a field $F$ with a multiplication which turns it into a ring defined such that, if $f \in F$ and $x, y \in V$, then

$$f(xy) = (fx)y = x(fy).$$

Examples of algebras include the algebra of real numbers **R**, vectors and matrices, tensors, complex numbers **C**, and quaternions. In the section about discrete Fourier transforms we will use finite dimensional **C**-algebras. Let $G$ be a finite group and **C**$G$ its group algebra. The group algebra is the set of all complex-valued functions of the finite group $G$ with pointwise addition and scalar multiplication. The dimension of **C**$G$ equals the order of $G$. Identifying each group element with its characteristic function, **C**$G$ can be viewed as the **C**-linear span of $G$. The algebra can be represented as a matrix algebra over **C** and as a block-diagonal matrix algebra.

Their *discrete analogues* are algebras over *rational numbers* **Q** and over *finite fields*.

## 6.4. Relations

### 6.4.1. Binary Relations

For any two sets $X$, $Y$ any subset $R \subset X \times Y$ is called *the binary relation* between $X$ and $Y$.

An $R$ relation is *reflexive* if $xRx$ for all $x$ from $X$.

An $R$ relation is *transitive* if $xRx'$ , $x'Rx''$ imply $xRx''$ for all $x, x', x''$ from $X$.

An $R$ relation is *symmetric* if $xRx'$ imply $x'Rx$ for all $x, x'$ from $X$.

### 6.4.2. A partial order relation

A relation > is a *partial order relation* on $X$ if > is reflexive and transitive.

Let $A$ and $B$ be ordering sets. A *homomorphism* $f: A \to B$ is called *monotone* if it serves the orderings of $A$ and $B$: $x \leq y \to xf \leq yf$.

### 6.4.3.Ordering.

Let $S$ be a set, $2^S$ – all subsets of $S$. $S_1, S_2, S_3$ lie in $2^S$

$S_1 < S_2 \Leftrightarrow S_1$ lies in $S_2$;

$S = ( 1\ 2\ 3 );$

$S_1 = ( 1\ 2 ); S_2 = ( 2\ 3 ); S_3 = ( 1\ 2\ 3 ); S_1$ lies in $S_3;$

$S = ( 1\ 2\ 3\ 4 );$

$( S_1 = ( 1\ 2 ); S_2 = ( 2\ 3 ));$

$( S_1 , S_2 )$ lies in $2^S * 2^S$.

Algebraic structures are connected with mathematical logic.

## 6.5. Coding theory

The goal of coding theory is to allow the receiver to detect and/or correct most transmission errors, reducing the error probability to an acceptable size [14]. The basic idea is to transmit redundant information.

The main problem in coding theory is to find codes which are well-balanced with respect to these conflicting goals of good error correction capabilities, small degree of redundancy and easy encoding and decoding.

Let *A* be an alphabet. A word in a finite alphabet $A=\{a_1 \ldots a_n\}$ is a finite set of symbols from *A*. Let *S(A)* be the set of all words of alphabet *A*. For any subset $S_1$ which belongs to *S(A)* any object or device that generates words from $S_1$ is called a message talker.

Words from $S_1$ are called communications. There are several methods of describing the talker:

a) *set theoretic description*: we fix power(coordinately) characteristics (for instance the length of the words);

b) *statistical description*: we fix some probabilistic (for instance the probability of *a* after $a_1$);

c) *logical description*: we generate $S_1$ as the automatic "language".

These are 3 main descriptions of the talker.

Let us continue to describe a setting which led to the development of coding theory. A transmitter *T* sends a message to the receiver *R*. We imagine there is only a finite number of possible messages and that *T* and *R* know beforehand the set of all possible messages. For instance, if there are 16 possible messages, then each message can be described as a binary quadruple. Each message has the form 0000 or 0110 or 1011 and so on. We say 4 bits have to be sent. Messages may of course be very much longer. Let us assume that we want to send a black and white picture doodle into pixels where each pixel has one of the $256=2^8$ shades (for example, 00000000 for white ,11111111 for black , in that case we would need to send 80 000 bits in order to transmit the picture.

We image *T* and *R* separated in space or time. So, the transmission error may occur (for example if 0110 is sent, then an error occurs in the third coordinate, then 0100 will be received). Moreover we treat these errors as being randomly generated. A popular and particularly simple model of a channel is the *binary symmetric channel (BSC)*. A channel is called *binary* as it transmits only one of two symbols (0 or 1) at a time. The channel is called *symmetric* as the probability of an error is independent on the symbol sent.

Let *A* be a finite set of *q* elements (the alphabet). A *q-ary block code C* of length *n* is a family of *n*-tuples with entries in *A.*

Let $a = (a_1, \dots a_n)$, $b = (b_1, \dots b_n)$ from $A^n$. The Hamming distance between *a* and *b* is

$$d(a,b) = \sum_{i=1}^{n} (1 - \delta(a_i, b_i))$$

where $\delta == \begin{cases} 1, if\ a_i = b_i, \\ 0, if\ a_i \neq b_i. \end{cases}$

Let *C* be a *block code* of length *n* in *A. The code distance d(C)* is

$$d(C) = d = min\{d(a,b),\ a \neq b\}.$$

The minimum distance is important because it is closely connected to the error correction and detection capabilities of a code.

Let us put $K(C) = K = log_q \#C$.

**Lemma.** For every $q$-ary block code $C$ of length $n$ we may compute values $K$ and $d$.

Let $C$ be a $q$-ary block code of length $n$. The values $q, n, \#C, d, K$ and some arithmetic expressions with these quantities are called *the parameters of code C.*

### 6.5.2. Finite Fields

Let $p$ be a prime number ($p=2,3,5,7,11,\ldots$), $\mathbf{F}_p$ the prime finite field. Let $p=2$. $\mathbf{F}_2$ :$1+1=0$;

In the field $\mathbf{F}_p$ the following condition holds: $(a+b)^p = a^p+b^p$.

Let $p=5$. In the field $\mathbf{F}_5$ the polynomials $X^2 - 3=0; X^2-2=0,$ are irreducible. If we add the roots of the polynomials to $\mathbf{F}_5$ we obtain the algebraic extensions of finite field $\mathbf{F}_5$.

There is the construction that builds *algebraic closure* of the given finite field.

In the field $\mathbf{F}_p$ $(a+b)^p=a^p+b^p$.

**Theorem**. Every finite field has $p^n$ elements for some prime $p$. The subfield generated by the element 1 is $\mathbf{F}_p = \mathbf{Z}/p\mathbf{Z}$.

In order to generate finite fields we may use irreducible polynomials. So let $f(x) \in \mathbf{F}_p[x]$ be an irreducible polynomial of degree $n$.

Let $f$ to be monic. The factor ring at the polynomial ring over the ideal generated by $f(x)$, is a field with $p^n$ elements.

### 6.5.3. Linear Codes. Bounds On Codes. Cyclic And Hamming Codes

Let $n$ be a natural number, $\boldsymbol{F}_q$ - finite field with $q$ elements, $\boldsymbol{F}_q^n$ the vector space over $\boldsymbol{F}_q$. Denote $\boldsymbol{F}_q^n$ by $V$.

**Definition**. Any linear subspace of dimension $k$ of $V$ is called *the k-dimensional linear code of length n over* $\boldsymbol{F}_q$.

The $k$-dimensional linear code of length $n$ over $\boldsymbol{F}_q$ with the distance $d$ will be denoted as $[n, k, d]$ code.

Let now $C$ be a binary linear code. As $C$ is a vector space, the minimum distance $d$ equals the minimum weight of a nonzero codeword. A code with minimum distance $\geq 2t + 1$ can correct up to $t$ errors. A code with minimum distance $t + 1$ can detect up to $t$ errors.

Let $a = (a_1, \ldots a_n) = a_1 \ldots a_n$ be a *codeword*.

It is convenient to represent a codeword $a_1 \ldots a_n$ by the polynomial $a(x) = a_1 x^{n-1} + \ldots + a_{n-1}x + a_n$.

### 6.5.4. Bounds On Codes

Some bounds on codes are very easily obtained.

**Theorem (**Singleton bound). Let $C$ be a $q$-ary block code of length $n$. Then the following holds:

$$\#C \leq q^{n-d+1}.$$

For any linear code $d \leq n - k + 1$.

**Proof**. Fix some $n - (d - 1)$ coordinates of a code. By the definition of minimal distance after projection to these coordinates different codewords remain different.

### 6.5.5.Cyclic and Hamming Codes

A code is *cyclic* if whenever $a_1 \ldots a_n$ is a codeword so is $a_2, \ldots a_n a_1$. Let $C$ be an $[n, k, d]$ cyclic code. Let $q = 2$. The *[7, 4, 3] Hamming code* $H_7$ is the binary cyclic code of length 7 with generator polynomials $g(x) = x^3 + x + 1$.

### 6.5.6. Reed-Muller codes

Below in this subsection we will consider binary linear codes. Let $P_m(0,1)$ be a set of Boolean functions in $m$ variables. With respect to a fixed order of all $2^m$ possible arguments, every such function $f:\{0,1\}^m \rightarrow \{0,1\}$ is uniquely represented by its *truth table,* which is a vector of length $2^m$ listing the values of $f$ in that order. Here we will identify functions and their truth tables.

The first-order Reed-Muller code $R(m)$ consists of the truth tables of all affine linear functions of $m$ variables:

$$R(m) = \{\sum_{i=1}^{m} a_i x_i + a_{m+1} \mid a_1, \ldots, a_{m+1} \in \mathbf{F}_2\}.$$

This is a linear code of length $2^m$ and dimension $m + 1$. Its minimum distance is $2^{m-1}$ From the minimum distance we know that $R(m)$ can correct up to $2^{m-2} - 1$ errors. Encoding $R(m)$ is ease:

$$(a_1, \ldots a_{m+1}) \mapsto \sum_{i=1}^{m} a_i x_i + a_{m+1} .$$

An efficient decoding procedure is based on generalized Fourier transform.

In the case the generalized Fourier transform is called Walsh-Hadamar transform (WHT). The WHT matrix $H_m$ is the $2^m$-square matrix. The evaluation of $H_m$ can be reduced to two evaluations of $H_{m-1}$ followed by $2^{m-1}$ additions and $2^{m-1}$ subtractions. The fast WHT algorithm allows an efficient decoding of first-order Reed-Muller codes.

# 7. ELEMENTS OF THE THEORY OF GRAPHS AND NETS

## 7.1. Base notions of graph theory

A *graph G* is a pair *(V,E)*, where *V* is a finite set of elements called *vertices* and *E* is a finite set of elements called *edges* **s**uch that each edge meets exactly 2 vertices called the *end-points* of the edge. Edge is said to join its end-points.

We will denote the number of vertices in a graph by *#V* and the number of edges by *#E*.

If an edge *e* has *v* as a point, then it is said that *e* is the *incident* with *v*. If the end-points of *e* are $v_1$ and $v_2$ then we write $(v_1, v_2)$ or $(v_2, v_1)$.

**Example 1.**

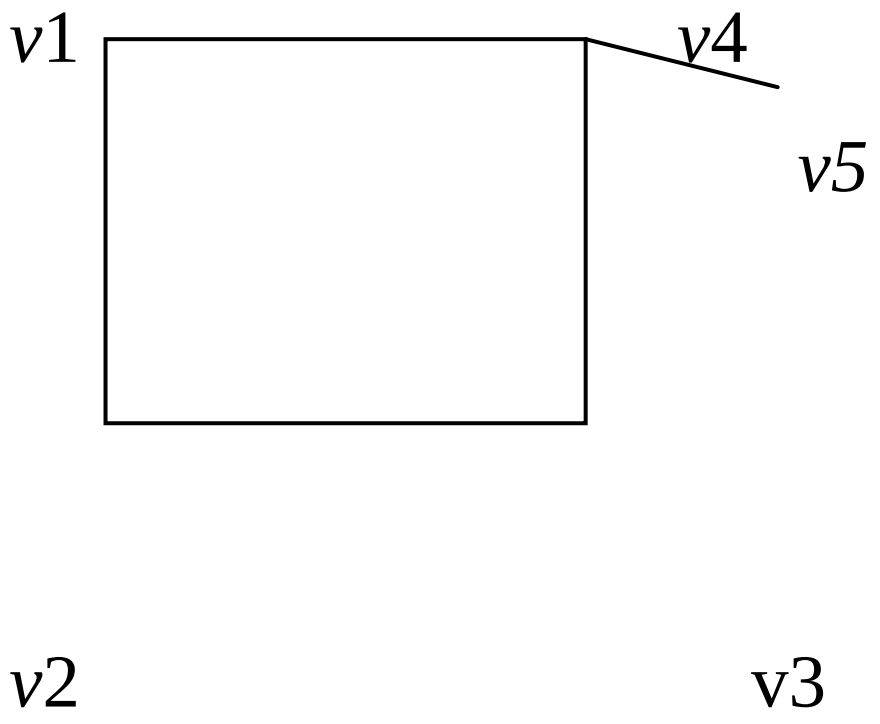

$V = \{ v_1, v_2, v_3, v_4, v_5 \}$

$E = \{ v_1v_2, v_1v_4, v_4v_5, v_2v_3, v_4v_3 \}$

$G = (V,E)$, $\#V = 5$

$\#E = 5$

## 7.1.2. The degree of a vertex *v*

*The degree of a vertex v,* written *d(v)* is the number of edges incident with *v*. A vertex *v* for which *d(v)=0* is called an *isolated vertex.*

Example 2.

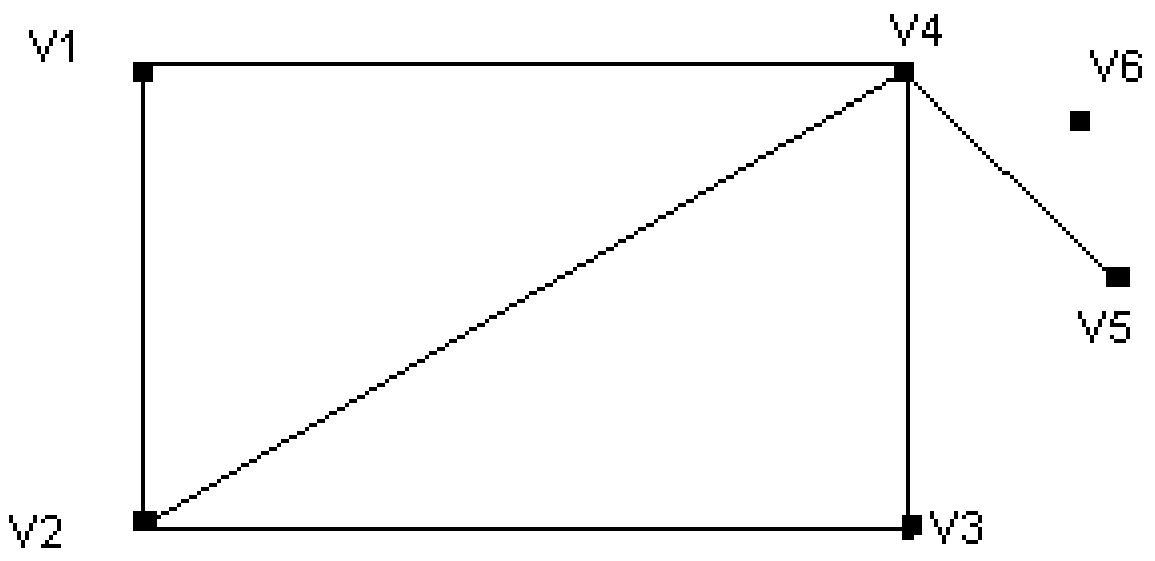

***d(v1)=2, d(v2)=3, d(v3)=2, d(v4)=4, d(v5)=1, d(v6)=0. #V = 6.***

*#E=6*

A *loop* is an edge *(u,v)* for which *u = v*.

## 7.1.3. Multiple edges

***A* multiple edge *is another edge which connects* u *and* v.**

**Example 3.**

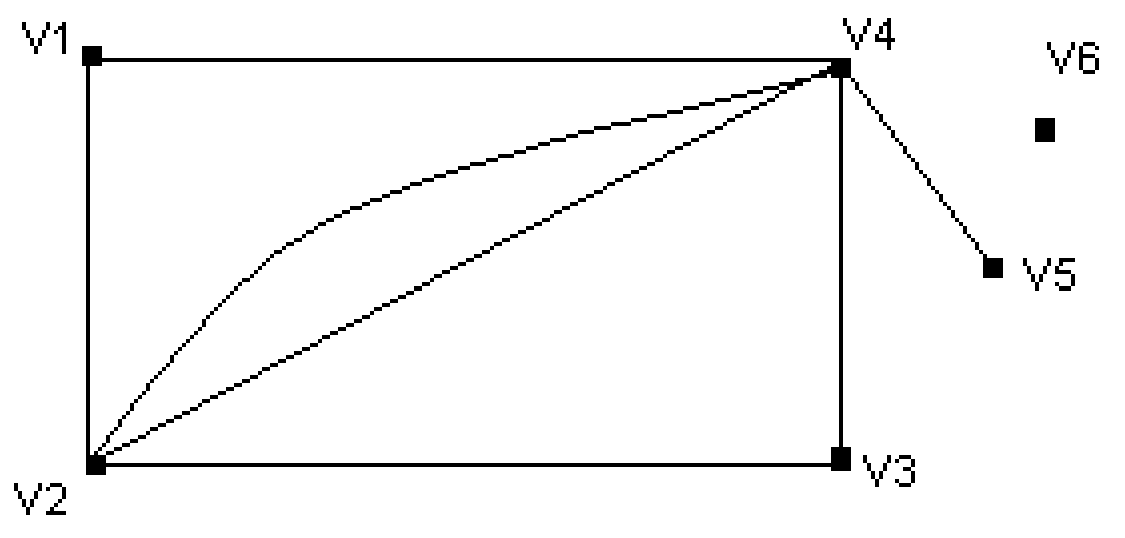

*#V=6*

*#E=7*

$G_3=\{V,E\}$

A graph is called *simple* if it contains no loops and multiple edges.

**Example 4.**

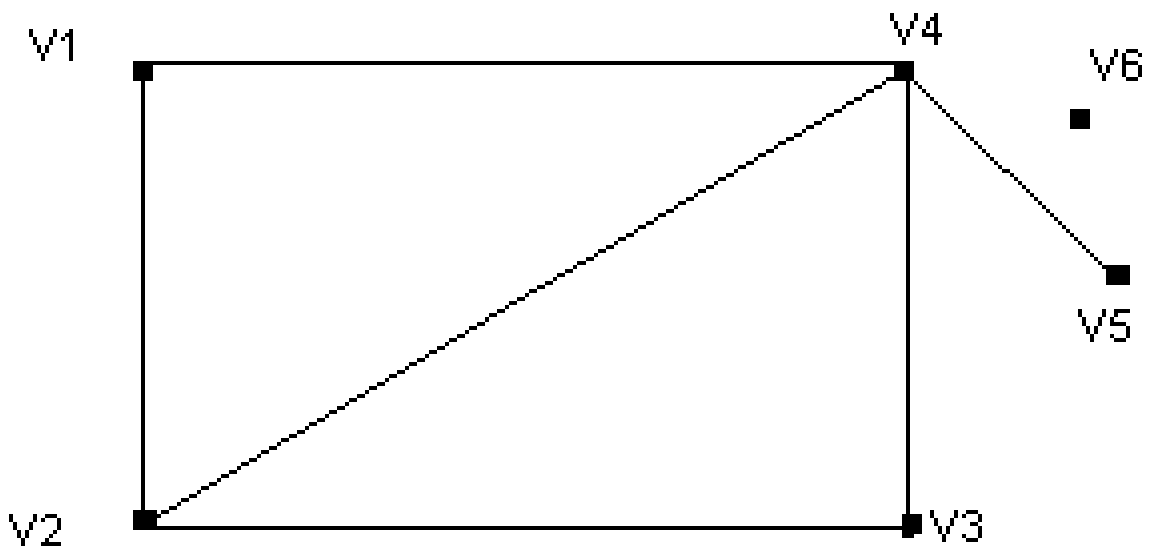

G1= {V,E}

*#V=6*

*#E=6*

A graph is called *regular* if every vertex has the same degree. If this is *k*, then the graph is called *k-regular*.

**Example 5.**

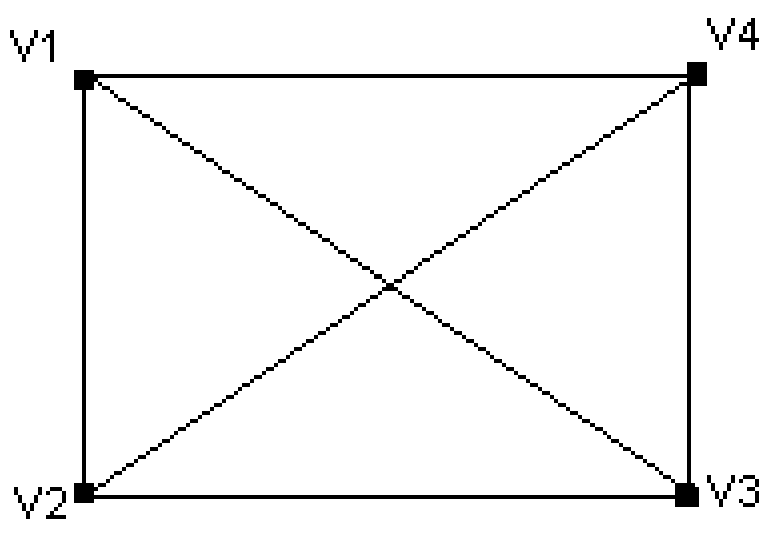

*d(v1)=3*

*d(v2)=3*

*d(v3)=3*

*d(v4)=3*

3-regular graph

### 7.1.4. Complete graph

A graph for which every pair of distinct vertex defines an edge is called a *complete graph*. A complete graph with *n* vertices is denoted by $K_n$.

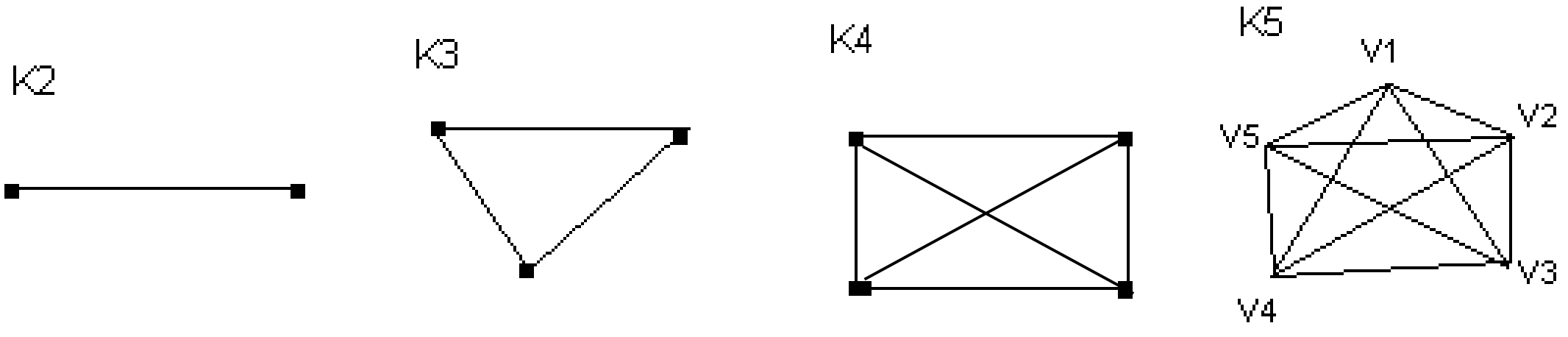

A (proper) *sub graph* of *G* is a graph obtainable by the removal of a (nonzero) number of edges and/or vertices of *G*. The removal of a vertex necessarily implies the removal of every edge incident with it, whereas the removal of an edge does not remove a vertex although it may result in one (or even two) isolated vertices. If we remove an edge *e* or a vertex *v* from *G*, then the resulting graphs are respectively denoted by *(G – e)* and *(G - v)*. If *H* is a subgraph of *G*, then *G* is called a *supergraph* of *H* and we write $H \subseteq G$. A subgraph *induced* by a subset of its vertices $V' \subset V$ is the graph consisting of $V'$ and those edges of *G* with both end-points in $V'$.

A *spanning* sub graph of a connected graph G is a sub graph of G obtained by removing edges only and such that any pair of vertices remain connected.

### 7.1.5. Paths

A *path* from $v_1$ to $v_i$ is a sequence $p=v_1, e_1, v_2, e_2, e_{i-1}, v_i$ of alternating vertices and edges such that for $1 \leq j < i$, $e_j$ is incident with $v_j$ and $v_{j+1}$. If $v_1=v_i$ then *p* is said to be a *cycle* or circuit. In a simple graph a path or a cycle $v_1,e_1,v_2,e_2,\ldots,e_{i-1},v_i$ can be more simply specified by the sequences of vertices $v_1,v_2,\ldots,v_i$. If in a path each vertex appears only once, then the sequence is called a *simple* path. If each vertex appears once except $v_1=v_i$, then *p* is a *simple* path or a cycle. Two paths are *edge-disjoint* if they do not have an edge in common.

**Example 6.**

A path is a sequence of edges such that the previous edge is connected with the next one.

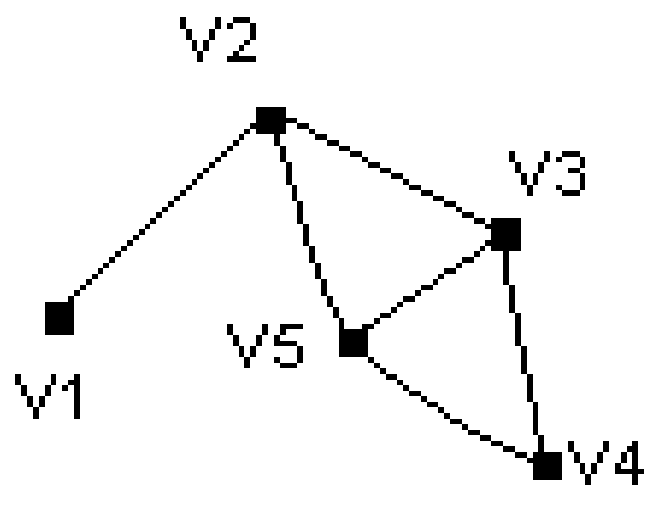

path: *((v1v2) (v2v3))*

*(v1v2) (v2v3) (v3v5);*

*(v2v3) (v3v5) (v5v2)* cycle;

### 7.1.6. Eulerian Cycles

**Definition.** A path or cycle that includes every edge of a graph exactly once is called **Eulerian**. A graph that contains an Eulerian cycle is called an **Eulerian graph**.

The problem of characterizing the graphs that contain these paths was formulated by L. Euler in 1736.

Two vertices $v_i$ and $v_j$ are *connected* if there is a path from $v_i$ to $v_j$. By convention, every vertex is connected to itself. Connection is an equivalence relation on the vertex set of a graph which partition it into subset $V_1, V_2, .., V_k$. A pair of vertices are connected if and only if they belong to the same subset of the partition. The subgraphs induced in turn by the subsets $V_1, V_2, .., V_k$, are called the *components* of the graph. A *connected* graph has only one component, otherwise it is disconnected.

**Theorem** (Euler). A connected graph is Eulerian if and only if every vertex has even degree.

### 7.1.7. Hamiltonian Cycles

**Definition.** In a graph $G$ with $\#V$ vertices, a path (or cycle) that contains exactly $\#V - 1$ (respectively. $\#V$) edges and spans $G$ is called **Hamiltonian**. A graph is called **Hamiltonian** if it contains an **Hamiltonian** cycle.

**Theorem.** If $G$ has $\#V \geq 3$ vertices and every vertex has degree at least $\#V/2$, then $G$ is Hamiltonian.

## 7.2. Operations with graphs

### 7.2.1. Join

Let $G$ and $H$ be graphs. The *join* of $G$ and $H$, denoted by $G + H$, is the graph formed by joining every vertex of $G$ to each vertex of $H$.

*G*(*v1,v2,v3*)

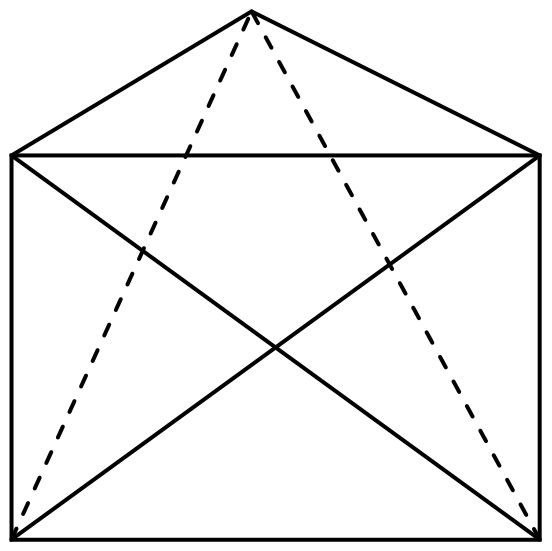

*H(u1,u2)*

When *H* is a vertex, say *x*, the graph *G + x* is formed by joining the component vertex x to every vertex of G. We say that G is suspended from *x*, and call *G + x* a suspended graph. When *G* is a chain (a tree with (nodes = vertices) of (degrees = valencies) 1 and 2 only) *G + x* is called a *fan*. When *G* is a cycle (circuit) the suspended graph *G + x* is called a *wheel*.

### 7.2.2. Product of graphs

Let *Gi = (Vi, Ei) (i=1,2)* be two graphs. The product $G = G_1 * G_2$ is a graph, for which $VG = V_1 * V_2$ (the Cartesian product of vertices) and *EG* is defined by the following way: *(u, v)* is adjacent with ($u_1$, $v_2$) if $u = u_1$ and *v* is adjacent with $v_1$ in $G_2$ or $v = v_1$ and *u* is adjacent with $u_1$ in $G_1$:

## 7.4. Blocks

Let *H* be a connected graph or a component. If the removal of a vertex *v* disconnected *H*, then *v* is said to be an *articulation point*

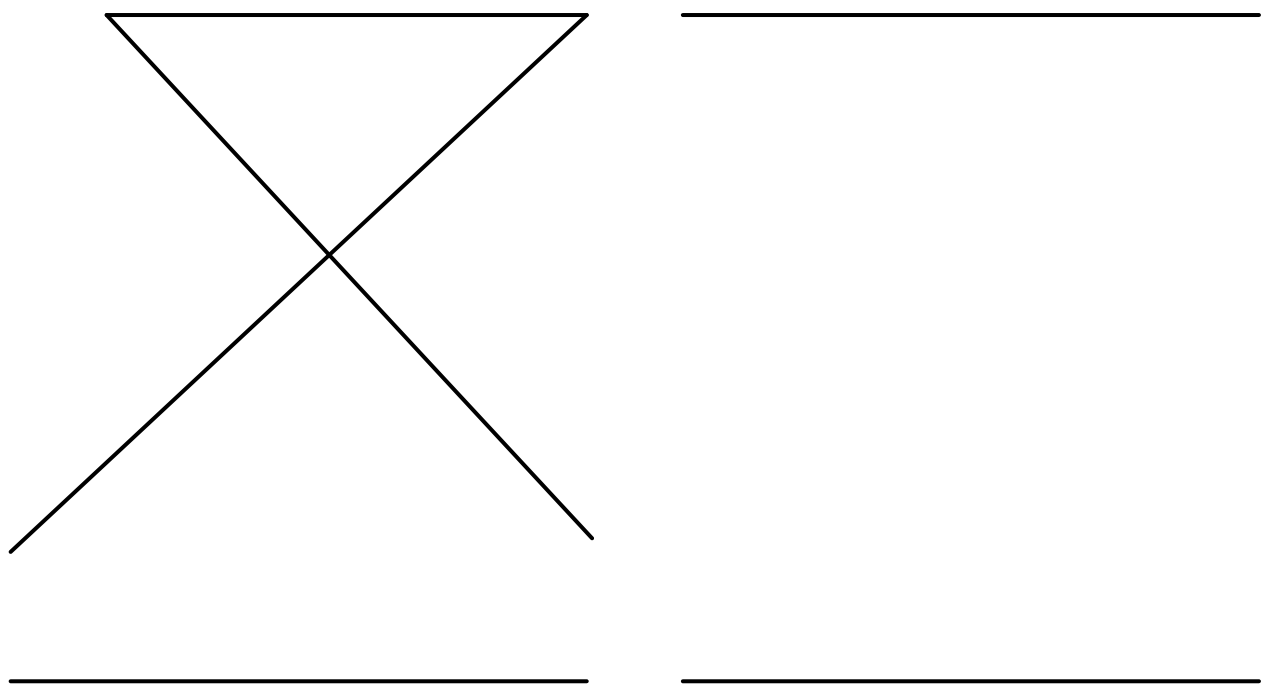

If *H* contains no articulation point then *H* is a *block*, sometimes called a *2 connected* graph or component.

If *H* contains an edge *e*, such that its removal will disconnect *H*, then *e* is said to be a *cut-edge*.

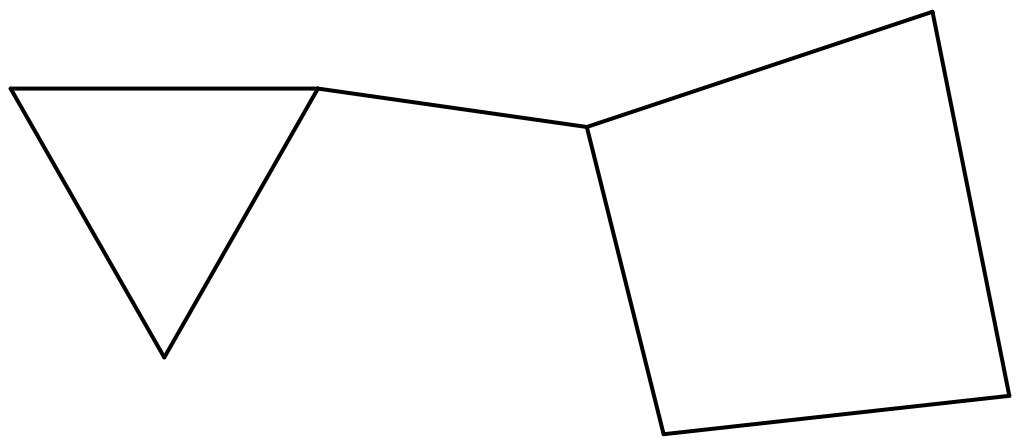

The end-points of a cut-edge are usually articulation points.

## 7.3. Trees

**Definition-Theorem.**

If *T* is a *tree* with *n* vertices, then

a) any two vertices of *T* are connected by precisely one path.
b) for any edge *e* , not in *T*, but connecting two vertices of *T*, the graph *(T+e)* contains exactly one circuit.
c) *T* has *(n-1)* edges.

The depth or level of a vertex in a rooted-tree is the number of edges in the path from the root to that vertex. If *(u, v)* is an edge of a rooted-tree such that U lies on the path from the root to *v*, then *u* is said to be the *father* of *v* and *v* is the *son* of *u*. An *ancestor* of *u* is any vertex of the path from *u* to the root of the tree. A *proper ancestor* of *u* is any ancestor of *u* including *u*. Similarly, if *u* is an ancestor of *v*, then *v* is a *descendant* of *u*. A proper descendant of *u* excludes *u*. Finally, a *binary tree* is a rooted-tree in which every vertex, unless it is a leaf, has two sons.

An *m-star* (m>=0) is a tree, consisting of a node of degree (valency) *m*, called the center, joined to m nodes (vertices) of valency one, called tips.
Let *G* be a graph. A *star-cover* of *G* is a spanning subgraph of *G*, in which every component is a star.

## 7.4. Directed graphs

In some applications it is natural to assign a *direction* to each edge of a graph. Thus in a diagram of the graph each edge is represented by an arrow. A graph augment in this way called a ***directed*** graph or a *digraph.*

If $e=(v_i,v_j)$ is an edge of a digraph, then the order of $v_i$ and $v_j$ becomes significant. The edge *e* is understood to be directed from the first vector $v_i$ to the second vector $v_j$. Thus if a digraphs contains the edge ($v_i$, $v_j$) then it may or it may not contain the edge ($v_j$, $v_i$). The directed edge ($v_i$, $v_j$) is said to be ***incident from*** $v_i$ and ***incident to*** $v_j$. For the vertex *v*, the ***out-degree*** $d^{+}(v)$ and the ***in-degree*** $d^{-}(v)$ are, respectively, the number of edges incident from v and the number of edges incident to v. A ***symmetric*** digraph is a digraph in which for every edge ($v_i$, $v_j$) there is an edge ($v_j$, $v_i$). A digraph is ***balanced*** if for every vertex *v*, $d^{+}(v)=d^{-}(v)$.

Of course, every digraph has an ***underlying (undirected simple) graph*** obtained by deleting the edge directions.
As defined earlier, a path (or circuit) in a corresponding undirected graph is a sequence $S=v_1, e_1, v_2, e_2, \ldots, v_{i+1}, e_i$, of vertices and edges. In the associated digraph this sequence may be such that for all $j$, $1 \leq j < i$, $e_j$ is incident from $v_j$ and incident to $v_{j+1}$. In this case *S* is said to be a ***directed*** path (or circuit). Otherwise it is an ***undirected***

**Example 7.**

A digraph is a graph such that each of its edges has a direction. So, in a digraph there are two edges: in-edge, out-edge.

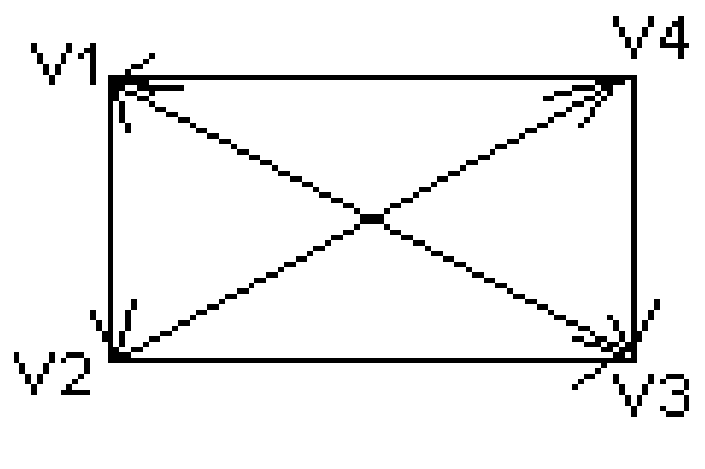

$G_4=\{V,E\}$

$ind(V1)=2$ $oud(V1)=1$

$ind(V2)=1$ $oud(V2)=2$

$ind(V3)=2$ $oud(V3)=0$

$ind(V4)=1$ $oud(V4)=2$

$$d(V)=ind(V)+oud(V)$$

## 7.5.1. Trees and formulas

A tree is a connected (directed) graph with the following properties:

1) $r \in V(T)$ $ind(r)=0$;
2) $\forall v' \neq r,\ v' \in V(T),\ ind(v')=1$.

### 7.5.1.1. Graphical representation of formulas

Let us consider examples of trees:

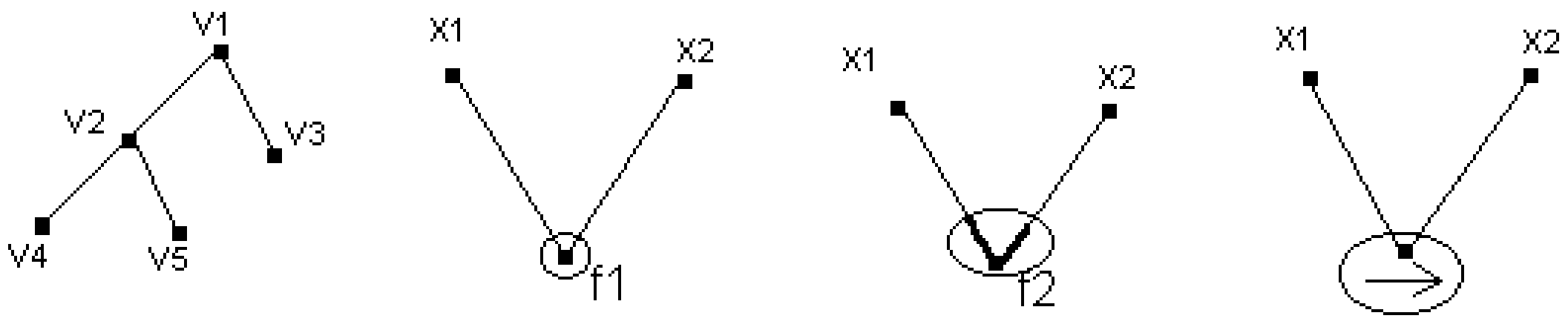

*f(x1,x2)* | *f1 + f2*

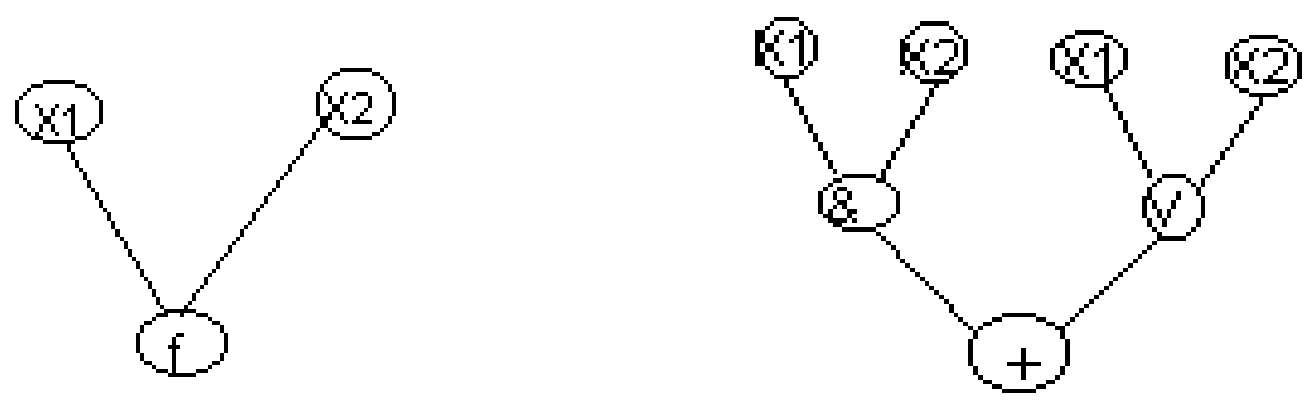

### 7.6. Graphs and Directed graphs and their edges

$V, E \subset V^{(2)}$, ($V^{(2)}$ is the set of all two element subsets of $V$ for undirected graphs.

$E \subset V^2$ (Cartesian Product for directed graphs)

### 7.7. Depth – First Search (DFS) algorithm

Recursive version is proposed by R. Tarjan 1972 y.

*Backtracking*, or depth-first search is a technique, which has been widely used for finding solutions to problems in combinational theory, and AI (Artificial Intelligence).

Suppose, *G* is a graph, which we wish to explore. Initially, all the vertices of *G* are unexplored. We start from some vertex of *G* and choose an edge to follow. Traversing the edge leads to a new vertex. We continue in this way, at each step we select an unexplored edge leading from a vertex already reached and traverse this edge. The edge leads to some vertex, either new or already reached. Whenever we run out of edges leading from old vertices, we choose some unreached vertex, if any exists, and begin a new exploration from this point.

Eventually, we will traverse all the edges of *G*, each exactly once. Such a process is called **a** *search of G***.**

There are many ways of searching a graph, depending upon the way in which edges to search are selected.

Consider the following choice rule:

When selecting an edge to traverse, always choose an edge emanating from the vertex most recently reached which still has unexplored edges.

A search, which uses this rule is called a *depth-first search***.** The set of all vertices with possibly unexplored edges may be stored on a stack.

Thus, a **depth-first search** is very easy to program either iteratively or recursively, provided we have a suitable computer representation of the graph.

**Definition.**

Let $G = (V, E)$ be a graph. For each vertex $u \in V$ we may construct a list containing all vertices, such, that $(u, w) \in E$. Such a list is called an adjacency list for vertex u.

A set of such lists, one for each vertex in *G* is called an *adjacency structure* of *G*.

If the graph *G* is undirected, each edge *(u, w)* is represented twice in an adjacency structure; once for *u*, and once for w. If *G* is directed, each edge *(u, w)* is represented once: vertex *w* appears in the adjacency list of vertex *u*.

## 7.8. Nets

Electrical, communication, transportation, computer, and neural networks are special kinds of nets. Designing these networks demands sophisticated mathematical models for their analysis. The theory of nets brings together elements of abstract graph theory and circuit analysis to network problems. Let us consider very shortly one important class of nets the Petri nets. Petri Nets are a formal and graphical appealing language which is appropriate for modeling systems with concurrency. Petri Nets have been under development since the beginning of the 60'ies, when Carl Adam Petri defined the language. It was the first time a general theory for discrete parallel systems was formulated. The language is a generalization of automata theory such that the concept of concurrently occurring events can be expressed. Petri Nets are a general graphical language which can be used for a wide variety of purposes, and hence not just specifically related with network communication technology. However, Petri Nets have proved useful for modeling, analyzing, and verifying protocols typically used in networks.

Some fundamental notions of the theory of Petri nets can be described in Mizar formalism. The Mizar Language is a formal language derived from the mathematical vernacular. The principle idea of its author was to design a language readable for mathematicians and, simultaneously, sufficiently rigorous to enable processing and verifying by computer software. A Petri net is defined as a triple of the form $\langle {\rm places},\,{\rm transitions},\,{\rm flow} \rangle$ with places and transitions being disjoint sets and flow being a relation included in ${\rm places} \times {\rm transitions}$.

Here we mention two applications of Petri nets.

### 7.8.1. Hardware design and Petri nets

Petri Nets have proved to be useful in the modelling, analysis and synthesis of digital systems, both synchronous (with global clocks) and asynchronous (self-timed). They also appear to be very promising in hardware-software co-design, for example as a formal model that may potentially be the mathematical foundation for the solution of many important co-design problems such as scheduling, timing analysis, verification, synthesis of hardware, etc., and thus enable the construction of concurrent and reactive real-time embedded systems.

### 7.8.2. Applications of Petri nets to intelligent system development

The application of Petri Nets to the development of Intelligent Systems (IS) has been receiving an increasing attention, since there are many fundamental problems addressed by Artificial Intelligence for which Petri Net formalism has played or will play an important role. Petri nets have already found their way into the development of different types of intelligent systems. Tasks addressed by these systems include: planning, scheduling, diagnosis, supervision and monitoring of discrete systems, uncertain or vague knowledge representation, knowledge base design and verification, distributed agents cooperation. Petri nets offer a very effective framework in dealing with various issues of intelligent system development for two main reasons:

- the net model can be interpreted as a clear and well-defined knowledge representation language for the task at hand;
- analysis techniques can be exploited as reasoning mechanisms to perform the given task.

Finally, recent advances in the development of intelligent agent systems in the INTERNET/INTRANET and WEB environments open new opportunities for using Petri Nets techniques in both the design and the analysis of such cutting-edge applications.

# 8. SCIENTIFIC ACTIVITY ON THE EXAMPLE “INFORMATION AND ITS INVESTIGATION”.

We are beginning our considerations in the titled framework and will continue it in next sections on the example of the history of scientific research on the field “Information and its investigation”. The scientific research was motivated by investigation of communication systems and consider the communication of information in information channel. At first recall the notion of the communication system.

## 8.2. System Diagram of data transmission or storage system

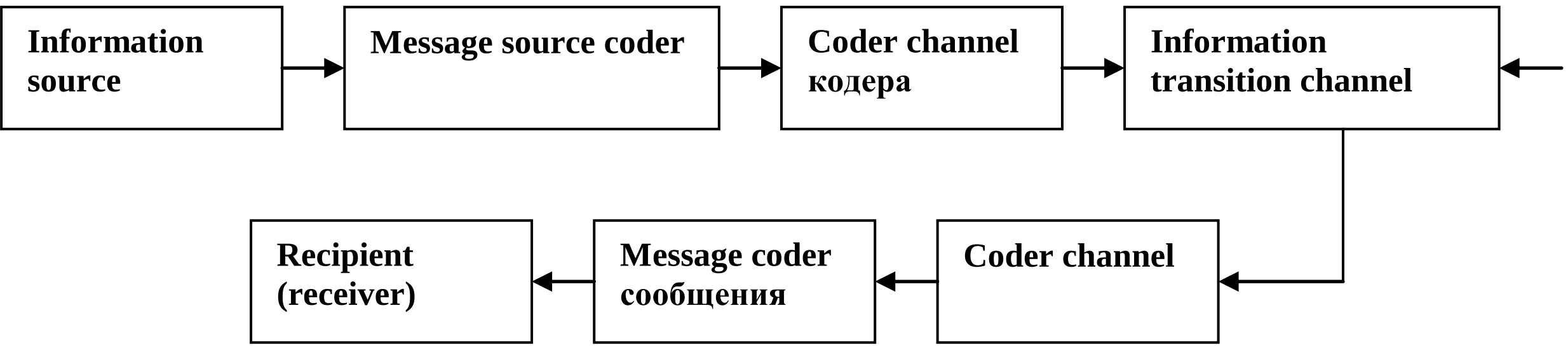

Message generates by information source, like a talking men, or computer. Source coder transforms this massage in digital form. This process consists of quantization or transformation of analog date to digital. On channel coder output are quantized date from coder and changes them into a coding signals (cods), which transmits to channel or writes to memory unit. Data transformation (message) transmits through a channel (phone cable, radio wave package or memory channel). There is often a noise in channel, that’s why received message could be differ from sent. Channel encoder perceives received signal (message code) and makes (correct) estimation of that code. Source encoder transforms decoder output channel to initial state.

H. Nyquist, R. Hartley and C. Shannon have investigated communication system with more simple structure. Here we follow to C. Shannon's paper "*A mathematical theory of communication*". A system related to:

INFORMATION SOURCE

MESSAGE TRANSMITTER

SIGNAL RECEIVED

SIGNAL RECEIVER

MESSAGE DESTINATION

NOISE SOURCE

It consists of essentially five parts:

1. An *information source* which produces a message or sequence of messages to be communicated to the receiving terminal. The message may be of various types: (a) A sequence of letters as in a telegraph of teletype system; (b) A single function of time $f(t)$ as in radio or telephony; (c) A function of time and other variables as in black and white television — here the message may be thought of as a function $f(x;y;\ t)$ of two space coordinates and time, the light intensity at point $(x;y)$ and time $t$ on a pickup tube plate; (d) Two or more functions of time, say $f(t)$, $g(t)$, $h(t)$—this is the case in "threedimensional" sound transmission or if the system is intended to service several individual channels in multiplex; (e) Several functions of several variables—in color television themessage consists of three functions $f(x;y;\ t)$, $g(x;y;\ t)$, $h(x;y;\ t)$ defined in a three-dimensional continuum—we may also think of these three functions as components of a vector field defined in the region — similarly, several black and white television sources would produce “messages” consisting of a number of functions of three variables; (f) Various combinations also occur, for example in television with an associated audio channel.

2. A *transmitter* which operates on the message in some way to produce a signal suitable for transmission over the channel. In telephony this operation consists merely of changing sound pressure into a proportional electrical current. In telegraphy we have an encoding operation which produces a sequence of dots, dashes and spaces on the channel corresponding to the message. In a multiplex PCM system the different speech functions must be sampled, compressed, quantized and encoded, and finally interleaved properly to construct the signal. Vocoder systems, television and frequency modulation are other examples of complex operations applied to the message to obtain the signal.

3. The *channel* is merely the medium used to transmit the signal from transmitter to receiver. It may be a pair of wires, a coaxial cable, a band of radio frequencies, a beam of light, etc.

4. The *receiver* ordinarily performs the inverse operation of that done by the transmitter, reconstructing the message from the signal.

5. The *destination* is the person (or thing) for whom the message is intended.

By a discrete system we will mean one in which both the message and the signal are a sequence of discrete symbols.

A continuous system is one in which the message and signal are both treated as continuous functions, e.g., radio or television. A mixed system is one in which both discrete and continuous variables appear, e.g., PCM transmission of speech.

We first consider the discrete case (Computers, Phone,…).

Generally, a discrete channel will mean a system whereby a sequence of choices from a finite set of elementary symbols $S_1,\ldots, S_n$ can be transmitted from one point to another. Each of the symbols *Si* is assumed to have a certain duration in time $t_i$ seconds (not necessarily the same for different *Si*, for example the dots and dashes in telegraphy). It is not required that all possible sequences of the *Si* be capable of transmission on the system; certain sequences only may be allowed. These will be possible signals for the channel. Thus in telegraphy suppose the symbols are: (1) A dot, consisting of line closure for a unit of time and then line open for a unit of time; (2) A dash, consisting of three time units of closure and one unit open; (3) A letter space consisting of, say, three units of line open; (4) A word space of six units of line open. We might place the restriction on allowable sequences that no spaces follow each other (for if two letter spaces are adjacent, it is identical with a word space). The question we now consider is how one can measure the capacity of such a channel to transmit information.

## . 8.3. Measures of information

There are several information measures or quantities:

- Combined quantity of information (Hartley)
- Probabilistic information quantity (Shannon)
- Algorithmic information quantity (Kolmogorov)

The choice of a logarithmic base corresponds to the choice of a unit for measuring information. If the base 2 is used the resulting units may be called binary digits, or more briefly *bits,* a word suggested by J. W. Tukey. A device with two stable positions, such as a relay or a flip-flop **circuit**, can store one bit of information. *N* such devices can store *N* bits, since the total number of possible states is 2*N* and log2 2*N* =*N*. If the base 10 is used the units may be called decimal digits. Since

$$\log_2 M = \log_{10} M / \log_{10} 2 = 3{,}32 \log_{10} M$$

**Definition**: The capacity *C* of a discrete channel is given by

$$C = \lim (\log N(T)) / T$$

under *T* tends to infinity and where *N*(*T*) is the number of allowed signals of duration *T*.

## 9. SCIENTIFIC RESEARCH IN ARTIFICIAL INTELLIGENCE

Basic premise of artificial intelligence (AI) is that knowledge of something is the ability to form a mental model that accurately represents the thing as well as the actions than can be performed by it and on it.

Knowledge-based systems are computer systems that deal with complex problems by making use of knowledge. This knowledge may be acquired from humans or automatically derived with abductive, deductive, and inductive techniques. This knowledge is mainly represented declaratively rather than encoded using complex algorithms. This declarative representation of knowledge economizes the development and maintenance process of these systems and improves their understandability. Therefore, knowledge-based systems originally used simple and generic inference mechanisms, like inheritance and forward or backward resolution, to compute solutions for given problems. This approach, however, turned out to become infeasible for many real-world tasks. Indeed, it also contrasted with human experts who exploited knowledge about the dynamics of the problem-solving *process* in order so solve their problems.

Recall what is Knowledge Acquisition.

- Objective: Construct a model to perform defined task
- Participants: Collaboration between problem expert(s) and modeling expert(s)
- Process: Iterate until done
  - Define task objective
  - Construct model
  - Evaluate model

For knowledge engineers with a background in artificial intelligence, knowledge domain concepts and the relations established between them can only be represented by using a formal language.

### 9.1. Description Logic

Description logic (DL) is the most resent name for a family of knowledge representation (KR) formal reason that represent the knowledge of an application domain ("the world") by fast defining the relevant concepts of the domain (it's terminology), and than using these concepts to specify properties of objects and individuals or carrying in the domain (the world description) [5]. As the name DL indicates one of the characteristics of these languages is that, unlike some their predecessors, they are equipped with a formal, logic-based semantics. Another distinguished feature is the emphasizes on reasoning as central service: reasoning allows one-to-infinitively represented knowledge from the knowledge that is explicitly contained in the knowledge base.

DL supports inference patterns that occur in many applications or intelligent information processing systems and which are also used by humans to structure and understand the world: classification of concepts and individuals.

Definition of the Basic Formalism.

An artificial intelligence system based on DL provide facilities to set up knowledge bases, to reason about their concepts and to manipulate them. Next figure sketches the architecture of such a system.

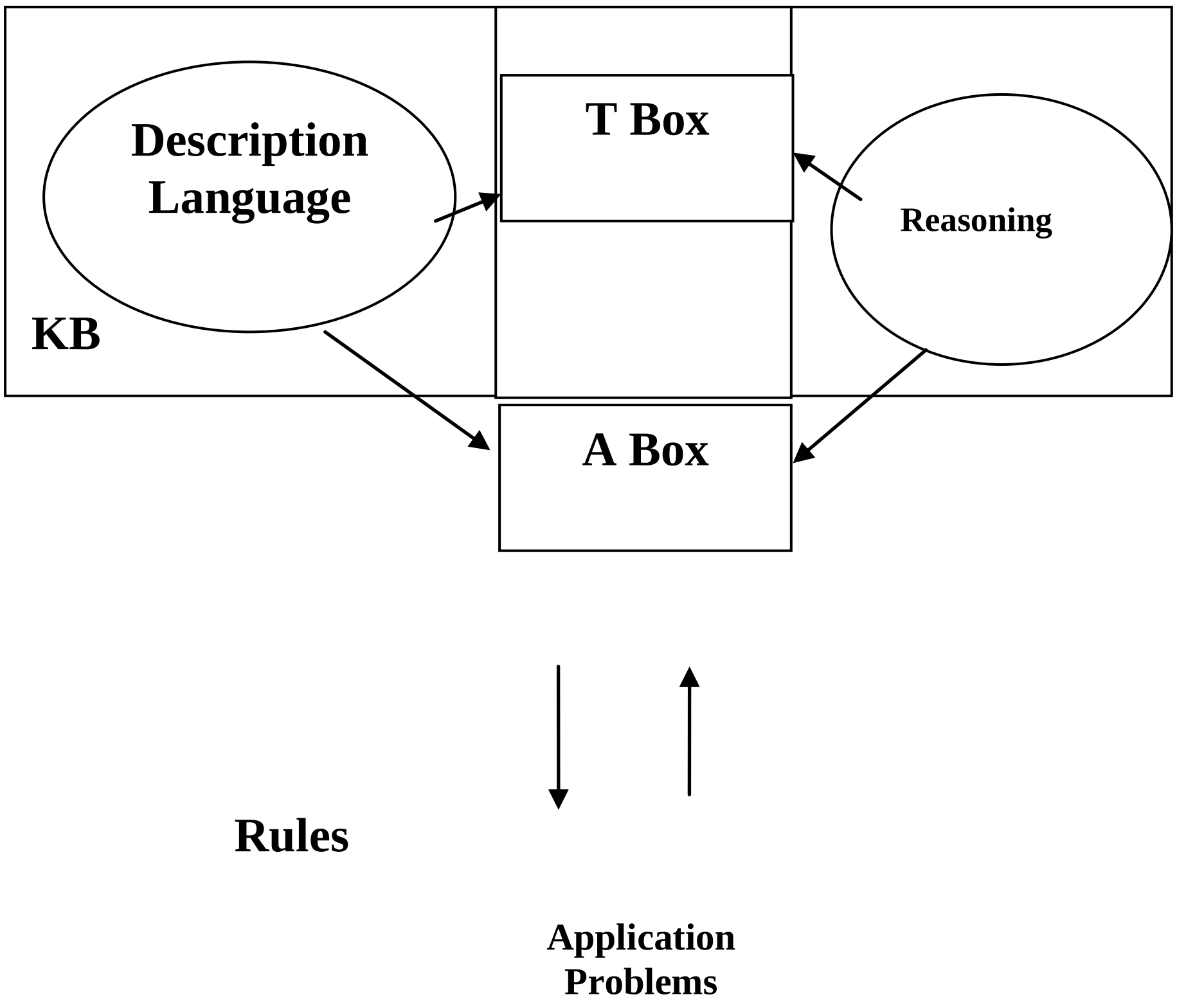

A knowledge base comprises two components: the TBox and the ABox. The TBox introduces the terminology, i.e. the vocabulary of an application domain, while the ABox contains accessions about named individuals in terms of this vocabulary. The vocabulary consists of concepts which denote sets of individuals and rows which denote binary relationships between individuals. In addition to atomic concept and rows (concept and row names) allow all DL systems to all their users to build complex description of concept and rows.

TBox can be used to assign names to complex descriptions. The language of such complex description is characteristic of each DL system, and different systems are distinguished by their description languages. The Description Language has a model theoretic semantics. Thus statements in the TBox can be identified with formulae in the $1^{st}$ order logic or in some cases in slide extension of it.

DL system not only stored terminology and accessions but also offer services that reason about them.

Typical reasoning tasks for a terminology are to determine whether a description is satisfaible (i.e. non contradictory), or whether one description is more general than another one, that is whether the first subsumes the second. Important problem for an ABox is to find out whether its set of accessions is consistent, that is whether it has a model, and whether the accessions in the ABox inhale that a particular individual is an instance of a given concept description.

Description Languages.

Elementary descriptions are atomic concept and atomic rows. Complex descriptions can be build from them inductively with concept constructions. An Abstract notation, we use letter A and B for atomic concepts, R – for atomic rows, C and D – for concept descriptions.

Below we'll consider so-called attribute – Alive Language (AL).

The Basic Descriptive Language AL.

Concept description in AL are formed according to the following syntax rules:

$C,D \rightarrow A \mid$ (atomic concept)

$T \mid$ (universal concept)

$\bot \mid$ (bottom concept)

$\neg A \mid$ (atomic negation)

$C \sqcap D \mid$ (intersection)

$\forall R.C \mid$ (value restriction)

$\exists R.T \mid$ (limited existential quantification)

Below we introduce terminological axioms, which make statements about how concepts or rows are related to each other. Than we single out definitions as specific axioms and identify terminologies as sets of definitions by which we can introduce atomic concepts as abriviations or names for complex concepts.

Terminology Axioms.

In the most general case, terminological axioms have the following:

$$C \equiv D \qquad (R \equiv S)$$

Axioms of the first kind are called inclusions, while axioms of the second kind are called equalities.

Definitions.

An equality whose left-hand-side is an atomic concept is an definition. A definitions are used to introduce symbolic names for complex description.

For instance for the axiom

$$Mother \equiv Woman\Pi\exists hasChild, Person$$

We associate to the description of the right-hand-side sign the name *Mother*.

Symbolic names may be used as abbreviation in other descriptions. I for example we have define Father in analogy to Mother we can define as Parent.

$$Parent \equiv Mother \cup Father$$

$$Woman \equiv Person\,\Pi\,Female$$

$$Man \equiv Person\Pi\neg(Person\,\Pi\,Female)$$

$$Mother \equiv (Person\,\Pi\,Female)\Pi\exists hasChild.Person$$

$$Father \equiv \big(Person\,\Pi\,\neg(Person\,\Pi\,Female)\big)\Pi\exists hasChild.Person$$

Extractions from John McCarthy' paper "Programs with common sense" (John McCarthy, Computer Science Department, Stanford University).

## 9.2. The Dr. McCarthy's paper on Artificial Intellegence and its discussion

The paper “Programs with common sense” was published by John McCarthy in 1959 year (see mcc59.pdf in Google).

Prof. Y. Bar-hillel: Dr. McCarthy's paper belongs in the Journal of Half-Baked Ideas, the creation of which was recently proposed by Dr. I. J. Good. Dr. McCarthy will probably be the first to admit this. Before he goes on to bake his ideas fully, it might be well to give him some advice and raise some objections. He himself mentions some possible objections, but I do not think that he treats them with the full consideration they deserve; there are others he does not mention.

For lack of time, I shall not go into the first part of his paper, although I think that it contains a lot of highly unclear philosophical, or pseudo-philosophical assumptions. I shall rather spend my time in commenting on the example he works out in his paper at some length. Before I start, let me voice my protest against the general assumption of Dr. McCarthy - slightly caricatured - that a machine, if only its program is specifed with a sufficient degree of carelessness, will be able to carry out satisfactory even rather discult tasks.

Consider the assumption that the relation he designates by at is transitive (page 7). However, since he takes both \at(I; desk)" and \at(desk; home)" as premises, I presume { though this is never made quite clear { that at means something like being-a-physical-part or in-the-immediate-spatial-neighborhood-of. But then the relation is clearly not transitive. If A is in the immediate spatial neighborhood of B and B in the immediate spatial neighborhood of C then A need not be in the immediate spatial neighborhood of C. Otherwise, everything would turn out to be in the immediate spatial neighborhood of everything, which is surely not Dr. McCarthy's intention. Of course, start-ing from false premises, one can still arrive at right conclusions. We do such things quite often, and a machine could do it. But it would probably be bad advice to allow a machine to do such things consistently.

Many of the other 23 steps in Dr. McCarthy's argument are equally or more questionable, but I don't think we should spend our time showing this in detail. My major question is the following: On page 9 McCarthy states that a machine which has a competence of human order in _nding its way around will have almost all the premises of the argument stored in its memory. I am
at a complete loss to understand the point of this remark. If Dr. McCarthy wants to say no more than that a machine, in order to behave like a human being, must have the knowledge of a human being, then this is surely not a very important remark to make. But if not, what was the intention of this remark?
The decisive question how a machine, even assuming that it will have somehow countless millions of facts stored in its memory, will be able to pick out those facts which will serve as premises for its deduction is promised to receive its treatment in another paper, which is quite right for a half-baked idea. It sounds rather incredible that the machine could have arrived at its conclusion | which, in plain English, is \Walk from your desk to your car!" | by sound deduction. This conclusion surely could not possibly follow from the premise in any serious sense. Might it not be occasionally cheaper to call a taxi and have it take you over to the airport? Couldn't you decide to cancel your ight or to do a hundred other things? I don't think it would be wise to develop a programming language so powerful as to make a machine arrive at the conclusion Dr. McCarthy apparently intends it to make. Let me also point out that in the example the time factor has never been mentioned, probably for the sake of simplicity. But clearly this factor is here so important that it could not possibly be disregarded without distorting the whole argument. Does not the solution depend, among thousands of other things, also upon the time of my being at my desk, the time at which I have to be at the airport, the distance from the airport, the speed of my car, etc. To make the argument deductively sound, its complexity will have to
be increased by many orders of magnitude. So long as this is not realized, any discussions of machines able to perform the deductive | and inductive! | operations necessary for treating problems of the kind brought forward by Dr. McCarthy is totally pointless. The gap between Dr. McCarthy's general programme (with which I have little quarrel, after discounting its \philosophical" features) and its execution even in such a simple case as the one discussed seems to me so enormous that much more has to be done to persuade me that even the _rst step in bridging this gap has already been taken.

Dr. O. G. Selfridge: I have a question which I think applies to this. It seems to me in much of that work, the old absolutist Prof. Bar-Hillel has really put his _nger on something; he is really worried about the deduction actually made. He seemed really to worry that that system is not consistent, and he made a remark that conclusions should not be drawn from false premises. In my experience those are the only conclusions that have ever been drawn. I have never yet heard of someone drawing correct conclusions from correct premises. I mean this seriously. This, I think is Dr. Minsky's point this morning. What this leads to is that the notion of deductive logic being something sitting there sacred which you can borrow for particularly sacred uses and producing inviolable results is a lot of nonsense. Deductive logic is inferrred as much as anything else. Most women have never inferred it, but they get along pretty well, marrying happy husbands, raising happy children, without ever using deductive logic at all. My feeling is that my criticism of Dr. McCarthy is the other way. He assumes deductive logic, whereas in fact that is something to be concocted. This is another important point which I think Prof. Bar-Hillel ignores in this, the criticism of the programme should not be as to whether it is logically consistent, but only will he be able to wave it around saying \this in fact works the way I want it". Dr. McCarthy would be the _rst to admit that his proramme is not now working, so it has to be changed. Then can you make the changes in the programme to make it work? That has nothing to do with logic. Can he amend it in such a way that it includes the logic as well as the little details of the programme? Can he manage in such a way that it works the way he does? He said at the begining of his talk that when he makes an arbitrary change in the programme it will not work usually, evidence, to me at least, that small changes in his programme will not obviously make the programme work and might even improve it. His next point is whether he can make small changes that in fact make it work. That is what we do not know yet.

Prof. Y. Bar-hillel: May I ask whether you could thrash this out with Dr. McCarthy? It was my impression that Dr. McCarthy's advice taker was meant to be able, among other things, to arrive at a certain conclusion from appropriate premises by faultless deductive reasoning. If this is still his programme, then I think your defence is totally beside the point.

Dr. O. G. Selfridge: I am not defending his programme, I am only defending him.

Dr. J. McCarthy: Are you using the word `programme' in the technical sense of a bunch of cards or in the sense of a project that you get money for?

Prof. Y. Bar-hillel: When I uttered my doubts that a machine working under the programme outlined by Dr. McCarthy would be able to do what he expects it to do, I was using `programme' in the technical sense.

Dr. O. G. Selfridge: In that case your criticisms are not so much philo-sophical as technical.

Prof. Y. Bar-hillel: They are purely technical. I said that I shall not
make any philosophical criticisms, for lack of time.

Dr. O. G. Selfridge: A technical objection does not make ideas half-baked.

Prof. Y. Bar-hillel: A deductive argument, where you have first to
and out what are the relevant premises, is something which many humans are not always able to carry out successfully. I do not see the slightest reason to believe that at present machines should be able to perform things that humans and trouble in doing. I do not think there could possibly exist a programme which would, given any problem, divide all facts in the universe into those which are and those which are not relevant for that problem. Developing such a programme seems to me by 1010 orders of magnitude more discult than, say, the Newell-Simon problem of developing a heuristic for deduction in the propositional calculus. This cavalier way of jumping over orders of magnitude only tends to becloud the issue and throw doubt on ways of thinking for which I have a great deal of respect. By developing a powerful programme language you may have paved the way for the first step in solving problems of the kind treated in your example, but the claim of being well on the way towards their solution is a gross exaggeration. This was the major point of my objections.

Dr. McCarthy (in reply): Prof. Bar-Hillel has correctly observed that my paper is based on unstated philosophical assumptions although what he means by \pseudo-philosophical" is unclear. Whenever we program a computer to learn from experience we build into the program a sort of epis-temology. It might be argued that this epistemology should be made explicit before one writes the programme, but epistemology is in a foggier state than computer programming even in the present half-baked state of the latter. I hope that once we have succeeded in making computer programs reason about the world, we will be able to reformulate epistemology as a branch of applied mathematics no more mysterious or controversial than physics. On re-reading my paper I can't see how Prof. Bar-Hillel could see in it a proposal to specify a computer program carelessly. Since other people have proposed this as a device for achieving \creativity", I can only conclude that he has some other paper in mind. In his criticism of my use of the symbol \at", Prof. Bar-Hillel seems to have misunderstood the intent of the example. First of all, I was not trying to formalize the sentence form, A is at B, as it is used in English. \at" merely was intended to serve as a convenient mnemonic for the relation between a place and a sub-place. Second, I was not proposing a practical problem for the program to solve but rather an example intended to allow us to think about the kinds of reasoning involved and how a machine may be made to perform them. Prof. Bar-Hillel's major point concerns my statement that the premises listed could be assumed to be in memory. The intention of this statement is to explain why I have not included formalizations of statements like, \it is possible to drive from my home to the airport" among my premises. If there were n known places in the county there would be

$$n(n - 1)/2$$

such sentences and, since we are quite sure that we do not have each of them in our memories, it would be cheating to allow the machine to start with them. The rest of Prof. Bar-Hillel's criticisms concern ways in which the model mentioned does not reect the real world; I have already explained that this was not my intention. He is certainly right that the complexity of the model will have to be increased for it to deal with practical problems. What we disagree on is my contention that the conceptual disculties arise at the present level of complexity and that solving them will allow us to increase the complexity of the model easily. With regard to the discussion between Prof. Bar-Hillel and Oliver Self-ridge | the logic is intended to be faultless although its premises cannot be guaranteed. The intended conclusion is \do(go(desk; car;walking))"|not, of course, \at(I; airport)". The model oversimpli_es but is not intended to oversimplify to the extent of allowing one to deduce one's way to the airport.

## 10. COMPILERS AND COMPILATION

Here we follow to [15-16] and references theirin. Software for early computers was primarily written in assembly language for many years. Higher level programming languages were not invented until the benefits of being able to reuse software on different kinds of CPUs started to become significantly greater than the cost of writing a compiler. The very limited memory capacity of early computers also created many technical problems when implementing a compiler.

Towards the end of the 1950s, machine-independent programming languages were first proposed. Subsequently, several experimental compilers were developed. The first compiler was written by Grace Hopper, in 1952, for the A-0 programming language. The FORTRAN team led by John Backus at IBM is generally credited as having introduced the first complete compiler in 1957. COBOL was an early language to be compiled on multiple architectures, in 1960.

In many application domains the idea of using a higher level language quickly caught on. Because of the expanding functionality supported by newer programming languages and the increasing complexity of computer architectures, compilers have become more and more complex.

Early compilers were written in assembly language. The first *self-hosting* compiler — capable of compiling its own source code in a high-level language — was created for Lisp by Tim Hart and Mike Levin at MIT in 1962. Since the 1970s it has become common practice to implement a compiler in the language it compiles, although both Pascal and C have been popular choices for implementation language. Building a self-hosting compiler is a bootstrapping problem—the first such compiler for a language must be compiled either by a compiler written in a different language, or (as in Hart and Levin's Lisp compiler) compiled by running the compiler in an interpreter.

## 10.1. Purposes, problems and concepts of compilers and compilation

Programming language is “a set of conventions for communicating an algorithm” (Horowitz).

Purposes of a programming languages (PL):

1. Specifying algorithm and data.
2. Communication to other people.
3. Establishing correctness.

Levels of PL:

- Machine language;
- Assemble language;
- High-level language (C/C++, Ada, Java, Fortran).

Translation – the process of converting a program written in a high-level language into machine language or into an intermediate PL.

## 10.2. Compilation

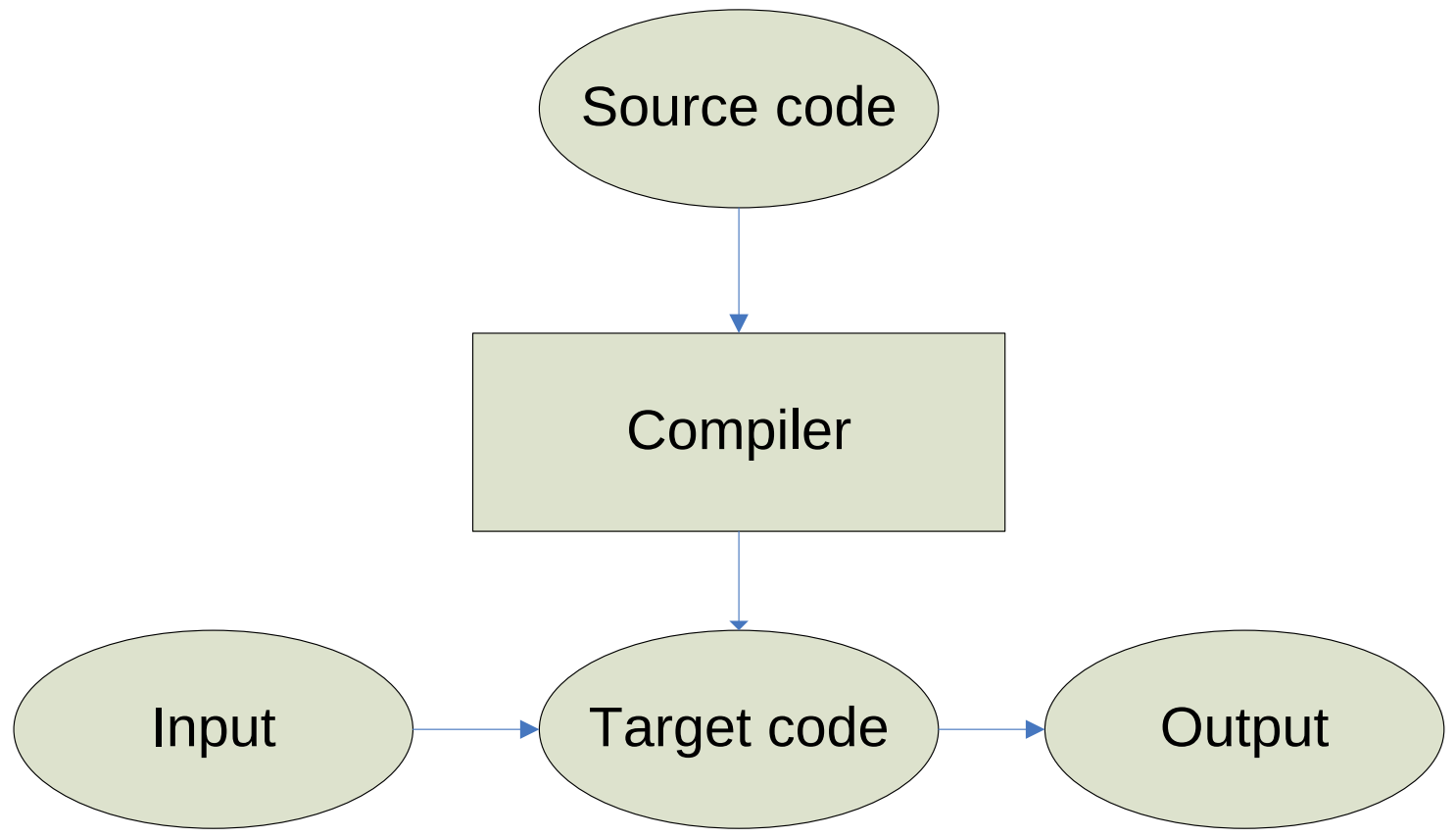

## 10.3. Interpretation

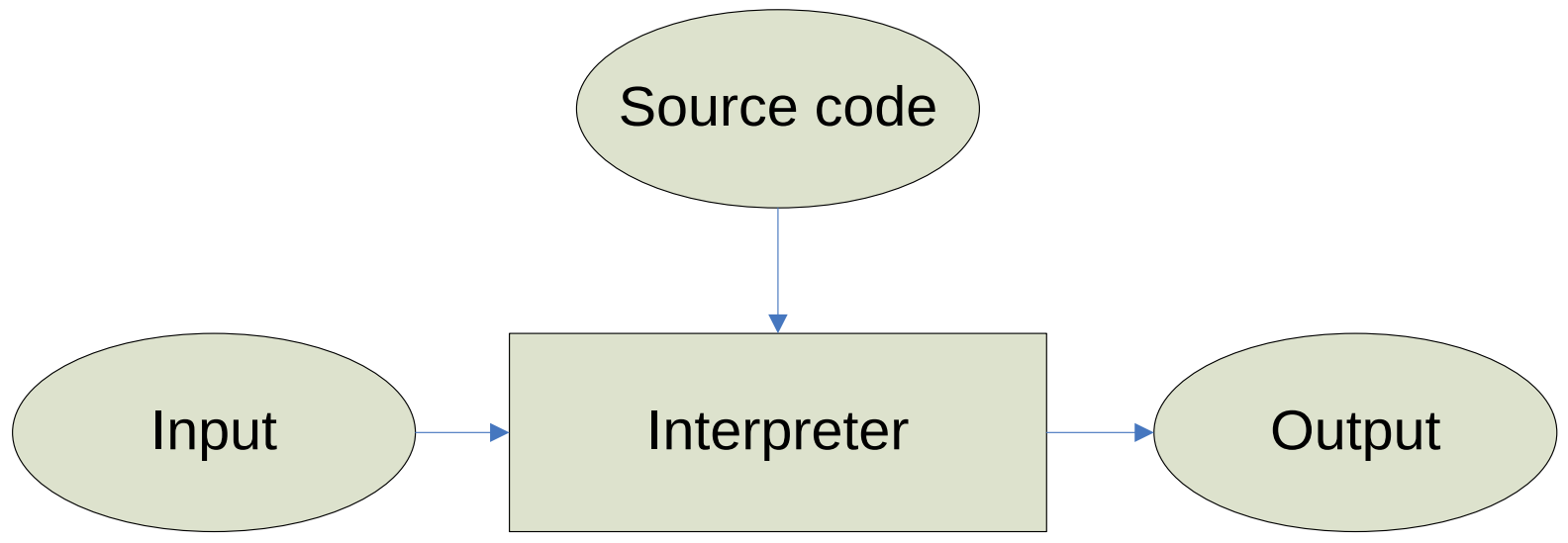

Compilation – the whole program is translated before execution.

Interpretation – translate and execute one statement at a time.

## 10.4. Pseudo-compilation

A hybrid of compilation and interpretation.

A compiler translates the whole program before execution into intermediate code. The intermediate code can be executed on any machine that has an interpreter for the intermediate code.

Java used initially this hybrid strategy.

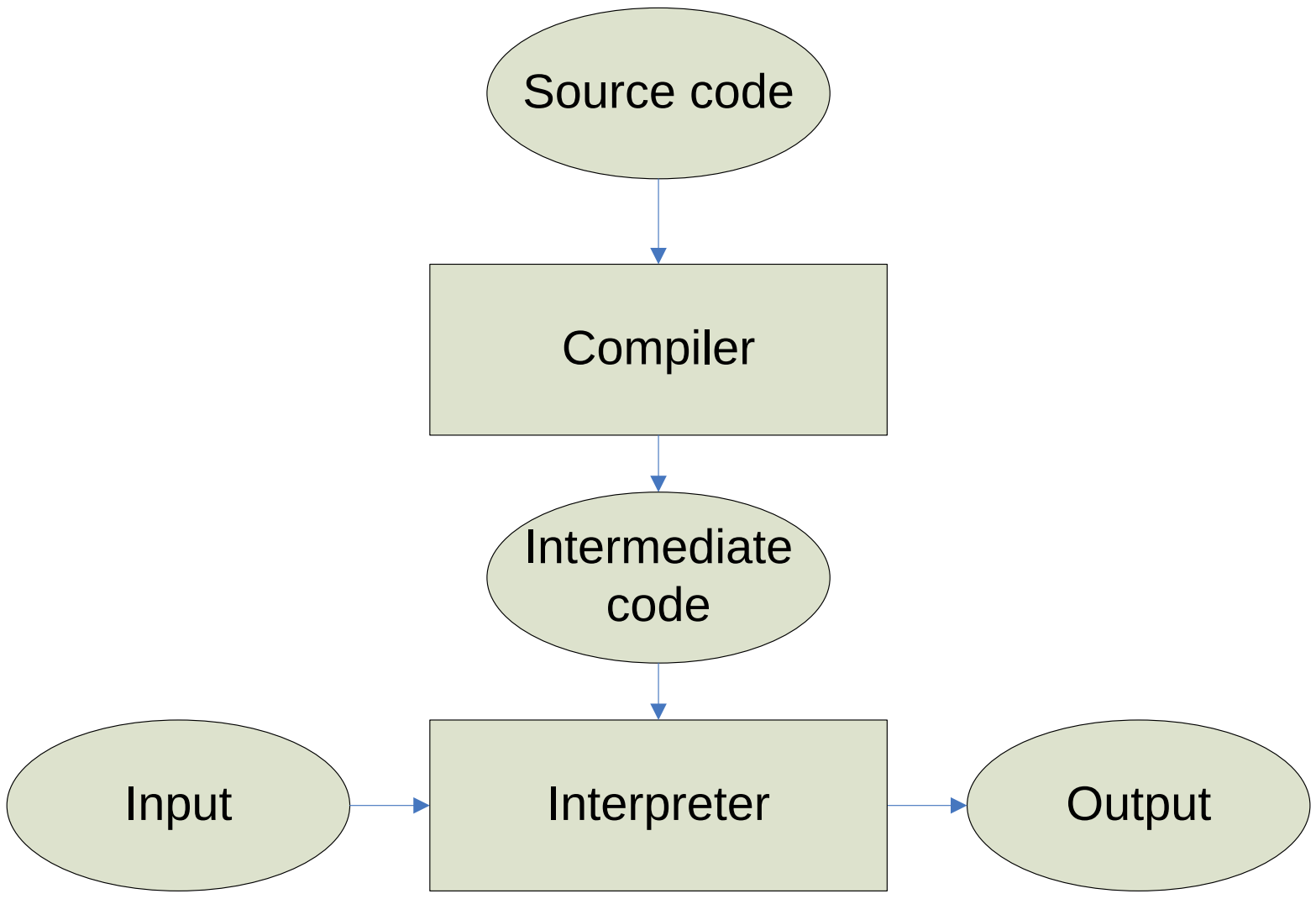

This intermediate code is called the byte-code. An interpreter translates and executes the intermediate code one statement at a time.

### 10.5. Advantages and disadvantages

| *Compilation* | *Interpretation* |
| --- | --- |
| Brings the program down to the level of the machine | Brings the machine up to the level of the program |
| Can execute translated program many times because the entire translation is produced | Must be retranslated for every execution |
| Program execution is much faster because the translator can do optimization | Program execution is much slower |
| Harder to provide useful feedback when debugging because executing the target code | Easier to provide useful feedback when debugging because executing the source code |
| | Flexibly supports rapid prototyping (Matlab) |

### 10.6. Paradigms for programming languages

1. Imperative languages
   - Program statements are commands (Ex. ADD 12 to X)
   - Key operations: assignment, looping.

These languages fit the von Neumann architecture closely: Fortran, Pascal, C, Turing.

2. Functional languages.
   - Program statement describes the value of expressions, using “essentially noun phrases”.
   - Key operation: expression evaluation by applying a function.

Examples: LISP, Scheme.

3. OO languages

   The idea behind object-oriented programming is that a computer program may be seen as composed of a collection of individual units, or *objects*, that act on each other, as opposed to a traditional view in which a program may be seen as a collection of functions or procedures, or simply as a list of instructions to the computer. Each object is capable of receiving messages, processing data, and sending messages to other objects.

   Examples: C++ Java C#

4. Logical languages

   Example: Prolog

Recall concepts of OO languages. Object-oriented programming (OOP) emphasizes the following concepts:

- Class — the unit of definition of data and behavior (functionality) for some kind-of-thing. For example, the 'class of Dogs' might be a set which includes the various breeds of dogs. A class is the basis of modularity and structure in an object-oriented computer program. *A class should typically be recognizable to a non-programmer familiar with the problem domain,* and the code for a class should be (relatively) self-contained and independent (as should the code for any good pre-OOP function). With such modularity, the structure of a program will correspond to the aspects of the problem that the program is intended to solve. This simplifies the mapping to and from the problem and program.
- Object — an instance of a class, an object (for example, "Lassie" the Dog) is the run-time manifestation (*instantiation*) of a *particular exemplar* of a class. (For the class of dogs which contains breed types, an acceptable exemplar would only be the ***subclass*** 'collie'; "Lassie" would then be an object in that subclass.) Each object has its own data, though the code within a class (or a subclass or an object) may be shared for economy. (Thus, OOLs must be capable of being reentrant.)
- Encapsulation — a type of privacy applied to the data and some of the methods (that is, functions or subroutines) in a class, encapsulation ensures that an object can be changed only through established channels (namely, the class's public methods). Encapsulation means wrapping up of data and associated functions into a single unit(called class).Each object exposes an *interface* — those public methods, which specify how other objects may read or modify it. An interface can prevent, for example, any caller from adding a list of children to a Dog when the Dog is less than one year old.
- Inheritance — a mechanism for creating subclasses, inheritance provides a way to define a (sub)class as a specialization or subtype or extension of a more general class (as Dog is a subclass of Canidae, and Collie is a subclass of the (sub)class Dog); a subclass acquires all the data and methods of all of its superclasses, but it can add or change data or methods as the programmer chooses. Inheritance is the "is-a" relationship: a Dog is-a Canidae. This is in contrast to composition, the "has-a" relationship, which user-defined datatypes brought to computer science: a Dog has-a mother (another Dog) and has-a father, etc.

- Abstraction — the ability of a program to ignore the details of an object's (sub)class and work at a more generic level when appropriate; For example, "Lassie" the Dog may be treated as a Dog much of the time, but when appropriate she is abstracted to the level of Canidae (superclass of Dog) or Carnivora (superclass of Canidae), and so on.

Polymorphism — polymorphism is behavior that varies depending on the class in which the behavior is invoked, that is, two or more classes can react *differently* to the *same message*.

## 10.7. Compiler optimization

**Compiler optimization** is the process of tuning the output of a compiler to minimize or maximize some attribute of an executable computer program. The most common requirement is to minimize the time taken to execute a program; a less common one is to minimize the amount of memory occupied. The growth of portable computers has created a market for minimizing the power consumed by a program. Compiler optimization is generally implemented using a sequence of *optimizing transformations*, algorithms which take a program and transform it to produce an output program that uses less resources.

It has been shown that some code optimization problems are NP-complete, or even undecidable[*citation needed*] . In practice, factors such as the programmer's willingness to wait for the compiler to complete its task place upper limits on the optimizations that a compiler implementor might provide. (Optimization is generally a very CPU- and memory-intensive process.) In the past, computer memory limitations were also a major factor in limiting which optimizations could be performed. Because of all these factors, optimization rarely produces "optimal" output in any sense, and in fact an "optimization" may impede performance in some cases; rather, they are heuristic methods for improving resource usage in typical programs.

**Building a control-flow graph.** The control-flow graph (CFG) is a fundamental data structure needed by almost all the techniques that compilers use to find opportunities for optimization and to prove the safety of those optimizations. Such analysis includes global data-flow analysis, the construction of a static single assignment form ssa-graph [8], and data-dependence analysis. Other techniques use the CFG to guide a more local analysis and replacement phase. These techniques all assume the existence of a CFG.

### 10.7.1. Types of optimizations

Techniques used in optimization can be broken up among various *scopes* which can affect anything from a single statement to the entire program. Generally speaking, locally scoped techniques are easier to implement than global ones but result in smaller gains. Some examples of scopes include:

- Peephole optimizations: Usually performed late in the compilation process after machine code has been generated. This form of optimization examines a few adjacent instructions (like "looking through a peephole" at the code) to see whether they can be replaced by a single instruction or a shorter sequence of instructions. For instance, a multiplication of a value by 2 might be more efficiently executed by left-shifting the value or by adding the value to itself. (This example is also an instance of strength reduction.)

- Local optimizations: These only consider information local to a function definition. This reduces the amount of analysis that needs to be performed (saving time and reducing storage requirements) but means that worst case assumptions have to be made when function calls occur or global variables are accessed (because little information about them is available).

- *Interprocedural* or whole-program optimization: These analyze all of a program's source code. The greater quantity of information extracted means that optimizations can be more effective compared to when they only have access to local information (i.e., within a single function). This kind of optimization can also allow new techniques to be performed. For instance function inlining, where a call to a function is replaced by a copy of the function body.

- Loop optimizations: These act on the statements which make up a loop, such as a *for* loop (eg, loop-invariant code motion). Loop optimizations can have a significant impact because many programs spend a large percentage of their time inside loops.

In addition to scoped optimizations there are two further general categories of optimization:

- *Programming language-independent vs. language-dependent*: Most high-level languages share common programming constructs and abstractions — decision (if, switch, case), looping (for, while, repeat.. until, do.. while), encapsulation (structures, objects). Thus similar optimization techniques can be used across languages. However, certain language features make some kinds of optimizations difficult. For instance, the existence of pointers in C and C++ makes it difficult to optimize array accesses (see Alias analysis). However, languages such as PL/1 (that also supports pointers) nevertheless have available sophisticated optimizing compilers to achieve better performance in various other ways. Conversely, some language features make certain optimizations easier. For example, in some languages functions are not permitted to have "side effects". Therefore, if a program makes several calls to the same function with the same arguments, the compiler can immediately infer that the function's result need be computed only once.

- *Machine independent vs. machine dependent*: Many optimizations that operate on abstract programming concepts (loops, objects, structures) are independent of the machine targeted by the compiler, but many of the most effective optimizations are those that best exploit special features of the target platform.

The following is an instance of a local machine dependent optimization. To set a register to 0, the obvious way is to use the constant '0' in an instruction that sets a register value to a constant. A less obvious way is to XOR a register with itself. It is up to the compiler to know which instruction variant to use. On many RISC machines, both instructions would be equally appropriate, since they would both be the same length and take the same time. On many other microprocessors such as the Intel x86 family, it turns out that the XOR variant is shorter and probably faster, as there will be no need to decode an immediate operand, nor use the internal "immediate operand register". (A potential problem with this is that XOR may introduce a data dependency on the previous value of the register, causing a pipeline stall. However, processors often have XOR of a register with itself as a special case that doesn't cause stalls.)

### 10.7.2. Factors affecting optimization

Many of the choices about which optimizations can and should be done depend on the characteristics of the target machine. It is sometimes possible to parameterize some of these machine dependent factors, so that a single piece of compiler code can be used to optimize different machines just by altering the machine description parameters. GCC is a compiler which exemplifies this approach.

*The architecture of the target CPU*

- Number of CPU registers: To a certain extent, the more registers, the easier it is to optimize for performance. Local variables can be allocated in the registers and not on the stack. Temporary/intermediate results can be left in registers without writing to and reading back from memory.

- RISC vs. CISC: CISC instruction sets often have variable instruction lengths, often have a larger number of possible instructions that can be used, and each instruction could take differing amounts of time. RISC instruction sets attempt to limit the variability in each of these: instruction sets are usually constant length, with few exceptions, there are usually fewer combinations of registers and memory operations, and the instruction issue rate (the number of instructions completed per time period, usually an integer multiple of the clock cycle) is usually constant in cases where memory latency is not a factor. There may be several ways of carrying out a certain task, with CISC usually offering more alternatives than RISC. Compilers have to know the relative costs among the various instructions and choose the best instruction sequence (see instruction selection).
- Pipelines: A pipeline is essentially an ALU broken up into an assembly line. It allows use of parts of the ALU for different instructions by breaking up the execution of instructions into various stages: instruction decode, address decode, memory fetch, register fetch, compute, register store, etc. One instruction could be in the register store stage, while another could be in the register fetch stage. Pipeline conflicts occur when an instruction in one stage of the pipeline depends on the result of another instruction ahead of it in the pipeline but not yet completed. Pipeline conflicts can lead to pipeline stalls: where the CPU wastes cycles waiting for a conflict to resolve.

  Compilers can *schedule*, or reorder, instructions so that pipeline stalls occur less frequently.

- Number of functional units: Some CPUs have several ALUs and FPUs. This allows them to execute multiple instructions simultaneously. There may be restrictions on which instructions can pair with which other instructions ("pairing" is the simultaneous execution of two or more instructions), and which functional unit can execute which instruction. They also have issues similar to pipeline conflicts.

Here again, instructions have to be scheduled so that the various functional units are fully fed with instructions to execute.

### 10.7.3. The architecture of the machine

- Cache size (256kB-12MB) and type (direct mapped, 2-/4-/8-/16-way associative, fully associative): Techniques such as inline expansion and loop unrolling may increase the size of the generated code and reduce code locality. The program may slow down drastically if a highly utilized section of code (like inner loops in various algorithms) suddenly cannot fit in the cache. Also, caches which are not fully associative have higher chances of cache collisions even in an unfilled cache.
- Cache/Memory transfer rates: These give the compiler an indication of the penalty for cache misses. This is used mainly in specialized applications.

### 10,7.4. Intended use of the generated code

- Debugging: When a programmer is still writing an application, he or she will recompile and test often, and so compilation must be fast. This is one reason most optimizations are deliberately avoided during the test/debugging phase. Also, program code is usually "stepped through" (see Program animation) using a symbolic debugger, and optimizing transformations, particularly those that reorder code, can make it difficult to relate the output code with the line numbers in the original source code. This can confuse both the debugging tools and the programmers using them.

- General purpose use: Prepackaged software is very often expected to be executed on a variety of machines and CPUs that may share the same instruction set, but have different timing, cache or memory characteristics. So, the code may not be tuned to any particular CPU, or may be tuned to work best on the most popular CPU and yet still work acceptably well on other CPUs.

- Special-purpose use: If the software is compiled to be used on one or a few very similar machines, with known characteristics, then the compiler can heavily tune the generated code to those specific machines (if such options are available). Important special cases include code designed for parallel and vector processors, for which special parallelizing compilers are employed.
  - Embedded systems: These are a "common" case of special-purpose use. Embedded software can be tightly tuned to an exact CPU and memory size. Also, system cost or reliability may be more important than the code's speed. So, for example, compilers for embedded software usually offer options that reduce code size at the expense of speed, because memory is the main cost of an embedded computer. The code's timing may need to be predictable, rather than "as fast as possible," so code caching might be disabled, along with compiler optimizations that require it.

### 10.7.5. Optimization techniques

#### Common themes

To a large extent, compiler optimization techniques have the following themes, which sometimes conflict.

*Optimize the common case*

> The common case may have unique properties that allow a *fast path* at the expense of a *slow path*. If the fast path is taken most often, the result is better over-all performance.

*Avoid redundancy*

> Reuse results that are already computed and store them for use later, instead of recomputing them.

*Less code*

Remove unnecessary computations and intermediate values. Less work for the CPU, cache, and memory usually results in faster execution. Alternatively, in embedded systems, less code brings a lower product cost.

*Fewer jumps* by using *straight line code*, also called *branch-free code*

Less complicated code. Jumps (conditional or unconditional branches) interfere with the prefetching of instructions, thus slowing down code. Using inlining or loop unrolling can reduce branching, at the cost of increasing binary file size by the length of the repeated code. This tends to merge several basic blocks into one. Even when it makes performance worse, programmers may eliminate certain kinds of jumps in order to defend against side-channel attacks.

*Locality*

Code and data that are accessed closely together in time should be placed close together in memory to increase spatial locality of reference.

*Exploit the memory hierarchy*

Accesses to memory are increasingly more expensive for each level of the memory hierarchy, so place the most commonly used items in registers first, then caches, then main memory, before going to disk.

*Parallelize*

Reorder operations to allow multiple computations to happen in parallel, either at the instruction, memory, or thread level.

*More precise information is better*

The more precise the information the compiler has, the better it can employ any or all of these optimization techniques.

*Runtime metrics can help*

Information gathered during a test run can be used in profile-guided optimization. Information gathered at runtime (ideally with minimal overhead) can be used by a JIT compiler to dynamically improve optimization.

*Strength reduction*

Replace complex or difficult or expensive operations with simpler ones. For example, replacing division by a constant with multiplication by its reciprocal, or using induction variable analysis to replace multiplication by a loop index with addition.

### 10.7.5.1. Optimization techniques (continuation)

**Loop optimizations**

*Main article: Loop optimization*

Some optimization techniques primarily designed to operate on loops include:

Induction variable analysis

Roughly, if a variable in a loop is a simple function of the index variable, such as j:= 4*i+1, it can be updated appropriately each time the loop variable is changed. This is a strength reduction, and also may allow the index variable's definitions to become dead code. This information is also useful for bounds-checking elimination and dependence analysis, among other things.

Loop fission or *loop distribution*

Loop fission attempts to break a loop into multiple loops over the same index range but each taking only a part of the loop's body. This can improve locality of reference, both of the data being accessed in the loop and the code in the loop's body.

Loop fusion or *loop combining*

Another technique which attempts to reduce loop overhead. When two adjacent loops would iterate the same number of times (whether or not that number is known at compile time), their bodies can be combined as long as they make no reference to each other's data.

Loop inversion

This technique changes a standard *while* loop into a *do/while* (also known as *repeat/until*) loop wrapped in an *if* conditional, reducing the number of jumps by two, for cases when the loop is executed. Doing so duplicates the condition check (increasing the size of the code) but is more efficient because jumps usually cause a pipeline stall. Additionally, if the initial condition is known at compile-time and is known to be side-effect-free, the *if* guard can be skipped.

Loop interchange

These optimizations exchange inner loops with outer loops. When the loop variables index into an array, such a transformation can improve locality of reference, depending on the array's layout.

Loop-invariant code motion

If a quantity is computed inside a loop during every iteration, and its value is the same for each iteration, it can vastly improve efficiency to hoist it outside the loop and compute its value just once before the loop begins. This is particularly important with the address-calculation expressions generated by loops over arrays. For correct implementation, this technique must be used with loop inversion, because not all code is safe to be hoisted outside the loop.

Loop nest optimization

Some pervasive algorithms such as matrix multiplication have very poor cache behavior and excessive memory accesses. Loop nest optimization increases the number of cache hits by performing the operation over small blocks and by using a loop interchange.

Loop reversal

Loop reversal reverses the order in which values are assigned to the index variable. This is a subtle optimization which can help eliminate dependencies and thus enable other optimizations.

Loop unrolling

Unrolling duplicates the body of the loop multiple times, in order to decrease the number of times the loop condition is tested and the number of jumps, which hurt performance by impairing the instruction pipeline. A "fewer jumps" optimization. Completely unrolling a loop eliminates all overhead, but requires that the number of iterations be known at compile time.

Loop splitting

Loop splitting attempts to simplify a loop or eliminate dependencies by breaking it into multiple loops which have the same bodies but iterate over different contiguous portions of the index range. A useful special case is *loop peeling*, which can simplify a loop with a problematic first iteration by performing that iteration separately before entering the loop.

Loop unswitching

Unswitching moves a conditional from inside a loop to outside the loop by duplicating the loop's body inside each of the if and else clauses of the conditional.

Software pipelining

The loop is restructured in such a way that work done in an iteration is split into several parts and done over several iterations. In a tight loop this technique hides the latency between loading and using values.

Automatic parallelization

A loop is converted into multi-threaded or vectorized (or even both) code in order to utilize multiple processors simultaneously in a shared-memory multiprocessor (SMP) machine, including multi-core machines.

**Data-flow optimizations**

Data flow optimizations, based on Data-flow analysis, primarily depend on how certain properties of data are propagated by control edges in the control flow graph. Some of these include:

Common subexpression elimination

In the expression "(a+b)-(a+b)/4", "common subexpression" refers to the duplicated "(a+b)". Compilers implementing this technique realize that "(a+b)" won't change, and as such, only calculate its value once.

Constant folding and propagation

replacing expressions consisting of constants (e.g. "3 + 5") with their final value ("8") at compile time, rather than doing the calculation in run-time. Used in most modern languages.

Induction variable recognition and elimination see discussion above about *induction variable analysis*.

Alias classification and pointer analysis in the presence of pointers, it is difficult to make any optimizations at all, since potentially any variable can have been changed when a memory location is assigned to. By specifying which pointers can alias which variables, unrelated pointers can be ignored.

Dead store elimination

removal of assignments to variables that are not subsequently read, either because the lifetime of the variable ends or because of a subsequent assignment that will overwrite the first value.

## SSA-based optimizations

These optimizations are intended to be done after transforming the program into a special form called static single assignment, in which every variable is assigned in only one place. Although some function without SSA, they are most effective with SSA. Many optimizations listed in other sections also benefit with no special changes, such as register allocation.

**Global value numbering**

GVN eliminates redundancy by constructing a value graph of the program, and then determining which values are computed by equivalent expressions. GVN is able to identify some redundancy that common subexpression elimination cannot, and vice versa.

**Sparse conditional constant propagation**

Effectively equivalent to iteratively performing constant propagation, constant folding, and dead code elimination until there is no change, but is much more efficient. This optimization symbolically executes the program, simultaneously propagating constant values and eliminating portions of the control flow graph that this makes unreachable.

## Code generator optimizations

**register allocation**

The most frequently used variables should be kept in processor registers for fastest access. To find which variables to put in registers an interference-graph is created. Each variable is a vertex and when two variables are used at the same time (have an intersecting liverange) they have an edge between them. This graph is colored using for example Chaitin's algorithm using the same number of colors as there are registers. If the coloring fails one variable is "spilled" to memory and the coloring is retried.

**instruction selection**

Most architectures, particularly CISC architectures and those with many addressing modes, offer several different ways of performing a particular operation, using entirely different sequences of instructions. The job of the instruction selector is to do a good job overall of choosing which instructions to implement which operators in the low-level intermediate representation with. For example, on many processors in the 68000 family and on the x86 architecture, complex addressing modes can be used in statements like "lea 25(a1,d5*4), a0", allowing a single instruction to perform a significant amount of arithmetic with less storage.

**instruction scheduling**

Instruction scheduling is an important optimization for modern pipelined processors, which avoids stalls or bubbles in the pipeline by clustering instructions with no dependencies together, while being careful to preserve the original semantics.

**rematerialization**

Rematerialization recalculates a value instead of loading it from memory, preventing a memory access. This is performed in tandem with register allocation to avoid spills.

**code factoring**

If several sequences of code are identical, or can be parameterized or reordered to be identical, they can be replaced with calls to a shared subroutine. This can often share code for subroutine set-up and sometimes tail-recursion.[1]

**trampolines**

Many CPUs have smaller subroutine call instructions to access low memory. A compiler can save space by using these small calls in the main body of code. Jump instructions in low memory can access the routines at any address. This multiplies space savings from code factoring.[2]

**reordering computations**

Based on integer linear programming, restructuring compilers enhance data locality and expose more parallelism by reordering computations. Space-optimizing compilers may reorder code to lengthen sequences that can be factored into subroutines.

## Functional language optimizations

Although many of these also apply to non-functional languages, they either originate in, are most easily implemented in, or are particularly critical in functional languages such as Lisp and ML.

Removing recursion

Recursion is often expensive, as a function call consumes stack space and involves some overhead related to parameter passing and flushing the instruction cache. Tail recursive algorithms can be converted to iteration, which does not have call overhead and uses a constant amount of stack space, through a process called tail recursion elimination or tail call optimization. Some functional languages, e.g. Scheme, mandate that tail calls be optimized by a conforming implementation, due to their prevalence in these languages.

Data structure fusion

Because of the high level nature by which data structures are specified in functional languages such as Haskell, it is possible to combine several recursive functions which produce and consume some temporary data structure so that the data is passed directly without wasting time constructing the data structure.

## Other optimizations

Bounds-checking elimination

Many languages, for example Java, enforce bounds-checking of all array accesses. This is a severe performance bottleneck on certain applications such as scientific code. Bounds-checking elimination allows the compiler to safely remove bounds-checking in many situations where it can determine that the index must fall within valid bounds, for example if it is a simple loop variable.

Branch offset optimization (machine independent)

Choose the shortest branch displacement that reaches target

Code-block reordering

Code-block reordering alters the order of the basic blocks in a program in order to reduce conditional branches and improve locality of reference.

Dead code elimination

Removes instructions that will not affect the behaviour of the program, for example definitions which have no uses, called dead code. This reduces code size and eliminates unnecessary computation.

Factoring out of invariants

If an expression is carried out both when a condition is met and is not met, it can be written just once outside of the conditional statement. Similarly, if certain types of expressions (e.g. the assignment of a constant into a variable) appear inside a loop, they can be moved out of it because their effect will be the same no matter if they're executed many times or just once. Also known as total redundancy elimination. A more powerful optimization is Partial redundancy elimination (PRE).

Inline expansion or macro expansion

When some code invokes a procedure, it is possible to directly insert the body of the procedure inside the calling code rather than transferring control to it. This saves the overhead related to procedure calls, as well as providing great opportunity for many different parameter-specific optimizations, but comes at the cost of space; the procedure body is duplicated each time the procedure is called inline. Generally, inlining is useful in performance-critical code that makes a large number of calls to small procedures. A "fewer jumps" optimization.

Jump threading

In this pass, conditional jumps in the code that branch to identical or inverse tests are detected, and can be "threaded" through a second conditional test.

Reduction of cache collisions

(e.g. by disrupting alignment within a page)

Stack height reduction

Rearrange expression tree to minimize resources needed for expression evaluation.

Test reordering

If we have two tests that are the condition for something, we can first deal with the simpler tests (e.g. comparing a variable to something) and only then with the complex tests (e.g. those that require a function call). This technique complements lazy evaluation, but can be used only when the tests are not dependent on one another. Short-circuiting semantics can make this difficult.

## Interprocedural optimizations

Interprocedural optimization works on the entire program, across procedure and file boundaries. It works tightly with intraprocedural counterparts, carried out with the cooperation of a local part and global part. Typical interprocedural optimizations are: procedure inlining, interprocedural dead code elimination, interprocedural constant propagation, and procedure reordering. As usual, the compiler needs to perform interprocedural analysis before its actual optimizations. Interprocedural analyses include alias analysis, array access analysis, and the construction of a call graph.

Interprocedural optimization is common in modern commercial compilers from SGI, Intel, Microsoft, and Sun Microsystems. For a long time the open source GCC was criticized for a lack of powerful interprocedural analysis and optimizations, though this is now improving.

Due to the extra time and space required by interprocedural analysis, most compilers do not perform it by default. Users must use compiler options explicitly to tell the compiler to enable interprocedural analysis and other expensive optimizations.

**Problems with optimization** (from [16])

Early in the history of compilers, compiler optimizations were not as good as hand-written ones. As compiler technologies have improved, good compilers can often generate better code than human programmers — and good post pass optimizers can improve highly hand-optimized code even further. For RISC CPU architectures, and even more so for VLIW hardware, compiler optimization is the key for obtaining efficient code, because RISC instruction sets are so compact that it is hard for a human to manually schedule or combine small instructions to get efficient results. Indeed, these architectures were designed to rely on compiler writers for adequate performance.

However, optimizing compilers are by no means perfect. There is no way that a compiler can guarantee that, for all program source code, the fastest (or smallest) possible equivalent compiled program is output; such a compiler is fundamentally impossible because it would solve the halting problem.

This may be proven by considering a call to a function, foo(). This function returns nothing and does not have side effects (no I/O, does not modify global variables and "live" data structures, etc.). The fastest possible equivalent program would be simply to eliminate the function call. However, if the function foo() in fact does *not* return, then the program with the call to foo() would be different from the program without the call; the optimizing compiler will then have to determine this by solving the halting problem.

Additionally, there are a number of other more practical issues with optimizing compiler technology:

- Optimizing compilers focus on relatively shallow "constant-factor" performance improvements and do not typically improve the algorithmic complexity of a solution. For example, a compiler will not change an implementation of bubble sort to use quicksort instead.
- Compilers usually have to support a variety of conflicting objectives, such as cost of implementation, compilation speed and quality of generated code.
- A compiler typically only deals with a part of a program at a time, often the code contained within a single file or module; the result is that it is unable to consider contextual information that can only be obtained by processing the other files.
- The overhead of compiler optimization: Any extra work takes time; whole-program optimization is time consuming for large programs.
- The often complex interaction between optimization phases makes it difficult to find an optimal sequence in which to execute the different optimization phases.

Work to improve optimization technology continues. One approach is the use of so-called "post pass" optimizers (some commercial versions of which date back to mainframe software of the late 1970s). These tools take the executable output by an "optimizing" compiler and optimize it even further. Post pass optimizers usually work on the assembly language or machine code level (contrast with compilers that optimize intermediate representations of programs). The performance of post pass compilers are limited by the fact that much of the information available in the original source code is not always available to them.

As processor performance continues to improve at a rapid pace, while memory bandwidth improves more slowly, optimizations that reduce memory bandwidth (even at the cost of making the processor execute "extra" instructions) will become more useful. Examples of this, already mentioned above, include loop nest optimization and rematerialization.

Verification of compilers

## 11. OBJECTIVE, CONCEPTS AND HISTORY OF COMPUTER SECURITY

Computer security is a branch of computer technology known as information security as applied to computers and networks. The objective of computer security includes protection of information and property from theft, corruption, or natural disaster, while allowing the information and property to remain accessible and productive to its intended users. The term computer system security means the collective processes and mechanisms by which sensitive and valuable information and services are protected from publication, tampering or collapse by unauthorized activities or untrustworthy individuals and unplanned events respectively. The strategies and methodologies of computer security often differ from most other computer technologies because of its somewhat elusive objective of preventing unwanted computer behavior instead of enabling wanted computer behavior.

Give following to the Open Security Archetecture http://www.opensecurityarchitecture.org/cms/index.php

Some definitions.

Security provided by IT Systems can be defined as the IT system's ability to being able to protect confidentiality and integrity of processed data, provide availability of the system and data, accountability for transactions processed, and assurance that the system will continue to perform to its design goals.

Security Architecture. The design artifacts that describe how the security controls (= security countermeasures) are positioned, and how they relate to the overall IT Architecture. These controls serve the purpose to maintain the system's quality attributes, among them confidentiality, integrity, availability, accountability and assurance.

Security Control. A technical countermeasure, an organizational setup or a process, that helps to maintain an IT systems security-quality properties.

Security Incident. In IT Security: A violation or imminent threat of violation of computer security policies, acceptable use policies, or standard computer security practices.

**Concepts**

Security; Computer security model; Security Architecture; Security Incident; Authentication techniques; Automated theorem proving; verification tools; Capability and access control list; Chain of trust techniques; Cryptographic techniques; Firewalls microkernel; Endpoint Security software;

## 11.1. Computer security model

A computer security model is a scheme for specifying and enforcing security policies. A security model may be founded upon a formal model of access rights, a model of computation, a model of distributed computing, or no particular theoretical grounding at all.

## 11.2. Enterprise security architecture

### Security by design

The technologies of computer security are based on logic. As security is not necessarily the primary goal of most computer applications, designing a program with security in mind often imposes restrictions on that program's behavior.

There are 4 approaches to security in computing, sometimes a combination of approaches is valid:

1. Trust all the software to abide by a security policy but the software is not trustworthy (this is computer insecurity).
2. Trust all the software to abide by a security policy and the software is validated as trustworthy (by tedious branch and path analysis for example).
3. Trust no software but enforce a security policy with mechanisms that are not trustworthy (again this is computer insecurity).
4. Trust no software but enforce a security policy with trustworthy hardware mechanisms.

Computers consist of software executing atop hardware, and a "computer system" is, by frank definition, a combination of hardware, software (and, arguably, firmware, should one choose so separately to categorize it) that provides specific functionality, to include either an explicitly expressed or (as is more often the case) implicitly carried along security policy. Indeed, citing the Department of Defense Trusted Computer System Evaluation Criteria (the TCSEC, or Orange Book)—archaic though that may be —the inclusion of specially designed hardware features, to include such approaches as tagged architectures and (to particularly address "stack smashing" attacks of recent notoriety) restriction of executable text to specific memory regions and/or register groups, was a *sine qua non* of the higher evaluation classes, to wit, B2 and above.)

Many systems have unintentionally resulted in the first possibility. Since approach two is expensive and non-deterministic, its use is very limited. Approaches one and three lead to failure. Because approach number four is often based on hardware mechanisms and avoids abstractions and a multiplicity of degrees of freedom, it is more practical. Combinations of approaches two and four are often used in a layered architecture with thin layers of two and thick layers of four.

There are various strategies and techniques used to design security systems. However there are few, if any, effective strategies to enhance security after design. One technique enforces the principle of least privilege to great extent, where an entity has only the privileges that are needed for its function. That way even if an attacker gains access to one part of the system, fine-grained security ensures that it is just as difficult for them to access the rest.

Furthermore, by breaking the system up into smaller components, the complexity of individual components is reduced, opening up the possibility of using techniques such as automated theorem proving to prove the correctness of crucial software subsystems. This enables a closed form solution to security that works well when only a single well-characterized property can be isolated as critical, and that property is also assessible to math. Not surprisingly, it is impractical for generalized correctness, which probably cannot even be defined, much less proven. Where formal correctness proofs are not possible, rigorous use of code review and unit testing represent a best-effort approach to make modules secure.

The design should use "defense in depth", where more than one subsystem needs to be violated to compromise the integrity of the system and the information it holds. Defense in depth works when the breaching of one security measure does not provide a platform to facilitate subverting another. Also, the cascading principle acknowledges that several low hurdles does not make a high hurdle. So cascading several weak mechanisms does not provide the safety of a single stronger mechanism.

Subsystems should default to secure settings, and wherever possible should be designed to "fail secure" rather than "fail insecure" (see fail-safe for the equivalent in safety engineering). Ideally, a secure system should require a deliberate, conscious, knowledgeable and free decision on the part of legitimate authorities in order to make it insecure.

In addition, security should not be an all or nothing issue. The designers and operators of systems should assume that security breaches are inevitable. Full audit trails should be kept of system activity, so that when a security breach occurs, the mechanism and extent of the breach can be determined. Storing audit trails remotely, where they can only be appended to, can keep intruders from covering their tracks. Finally, full disclosure helps to ensure that when bugs are found the "window of vulnerability" is kept as short as possible.

### 11.3. Short history of computer security

The early Multics operating system was notable for its early emphasis on computer security by design, and Multics was possibly the very first operating system to be designed as a secure system from the ground up. In spite of this, Multics' security was broken, not once, but repeatedly. The strategy was known as 'penetrate and test' and has become widely known as a non-terminating process that fails to produce computer security. This led to further work on computer security that prefigured modern security engineering techniques producing closed form processes that terminate.

In computer science, an Access Control Matrix or Access Matrix is an abstract, formal security model of protection state in computer systems, that characterizes the rights of each subject with respect to every object in the system. It was first introduced by Butler W. Lampson in 1971. According to the model, the protection state of a computer system can be abstracted as a set of objects *O*, that is the set of entities that needs to be protected (e.g. processes, files, memory pages) and a set of subjects *S*, that consists of all active entities (e.g. users, processes). Further there exists a set of rights *R* of the form *r*(*s*,*o*), where $s \in S$, $o \in O$and $r(s,o) \subseteq R$. A right thereby specifies the kind of access a subject is allowed to process with regard to an object. Because it does not define the granularity of protection mechanisms, the Access Control Matrix can be used as a model of the static access permissions in any type of access control system. It does not model the rules by which permissions can change in any particular system, and therefore only gives an incomplete description of the system's access control security policy.

An Access Control Matrix should be thought of only as an abstract model of permissions at a given point in time; a literal implementation of it as a two-dimensional array would have excessive memory requirements. Capability-based security and access control lists are categories of concrete access control mechanisms whose static permissions can be modeled using Access Control Matrices. Although these two mechanisms have sometimes been presented (for example in Butler Lampson's *Protection* paper) as simply row-based and column-based *implementations* of the Access Control Matrix, this view has been criticized as drawing a misleading equivalence between systems that does not take into account dynamic behaviour.

The Clark-Wilson integrity model provides a foundation for specifying and analyzing an integrity policy for a computing system.

The model is primarily concerned with formalizing the notion of information integrity. Information integrity is maintained by preventing corruption of data items in a system due to either error or malicious intent. An integrity policy describes how the data items in the system should be kept valid from one state of the system to the next and specifies the capabilities of various principals in the system. The model defines enforcement rules and certification rules.

The model was described in a 1987 paper (*A Comparison of Commercial and Military Computer Security Policies*) by David D. Clark and David R. Wilson. The paper develops the model as a way to formalize the notion of information integrity, especially as compared to the requirements for multi-level security (MLS) systems described in the Orange Book. Clark and Wilson argue that the existing integrity models such as Bell-La Padula model (read-down/write-up) and Biba (read-up/write-down) were better suited to enforcing data confidentiality rather than information integrity. The Bell-La Padula and Biba models are more clearly useful in, for example, military classification systems to prevent the theft of information and the tainting of information at higher classification levels, respectively. In contrast, Clark-Wilson is more clearly applicable to business and industry processes in which the integrity of the information content is paramount at any level of classification (although the authors stress that all three models are obviously of use to both government and industry organizations).

The model's enforcement and certification rules define data items and processes that provide the basis for an integrity policy. The core of the model is based on the notion of a transaction.

- A well-formed transaction is a series of operations that transition a system from one consistent state to another consistent state.
- In this model the integrity policy addresses the integrity of the transactions.
- The principle of separation of duty requires that the certifier of a transaction and the implementer be different entities.

The model contains a number of basic constructs that represent both data items and processes that operate on those data items. The key data type in the Clark-Wilson model is a Constrained Data Item (CDI). An Integrity Verification Procedure (IVP) ensures that all CDIs in the system are valid at a certain state. Transactions that enforce the integrity policy are represented by Transformation Procedures (TPs). A TP takes as input a CDI or Unconstrained Data Item (UDI) and produces a CDI. A TP must transition the system from one valid state to another valid state. UDIs represent system input (such as that provided by a user or adversary). A TP must guarantee (via certification) that it transforms all possible values of a UDI to a “safe” CDI.

At the heart of the model is the notion of a relationship between an authenticated principal (i.e., user) and a set of programs (i.e., TPs) that operate on a set of data items (e.g., UDIs and CDIs). The components of such a relation, taken together, are referred to as a *Clark-Wilson triple.* The model must also ensure that different entities are responsible for manipulating the relationships between principals, transactions, and data items. As a short example, a user capable of certifying or creating a relation should not be able to execute the programs specified in that relation.

The model consists of two sets of rules: Certification Rules (C) and Enforcement Rules (E). The nine rules ensure the external and internal integrity of the data items. To paraphrase these:

> C1—When an IVP is executed, it must ensure the CDIs are valid.
> C2—For some associated set of CDIs, a TP must transform those CDIs from one valid state to another.

Since we must make sure that these TPs are certified to operate on a particular CDI, we must have E1 and E2.

> E1—System must maintain a list of certified relations and ensure only TPs certified to run on a CDI change that CDI.
> E2—System must associate a user with each TP and set of CDIs. The TP may access the CDI on behalf of the user if it is “legal.”

This requires keeping track of triples (user, TP, {CDIs}) called “allowed relations.”

> C3—Allowed relations must meet the requirements of “separation of duty.”

We need authentication to keep track of this.

E3—System must authenticate every user attempting a TP. Note that this is per TP request, not per login.

For security purposes, a log should be kept.

C4—All TPs must append to a log enough information to reconstruct the operation.

When information enters the system it need not be trusted or constrained (i.e. can be a UDI). We must deal with this appropriately.

C5—Any TP that takes a UDI as input may only perform valid transactions for all possible values of the UDI. The TP will either accept (convert to CDI) or reject the UDI.

Finally, to prevent people from gaining access by changing qualifications of a TP:

E4—Only the certifier of a TP may change the list of entities associated with that TP.

The Graham-Denning Model is a computer security model that shows how subjects and objects should be securely created and deleted. It also addresses how to assign specific access rights. It is mainly used in access control mechanisms for distributed systems. This model addresses the security issues associated with how to define a set of basic rights on how specific subjects can execute security functions on an object. The model has eight basic protection rules (actions) that outline:

- How to securely create an object.
- How to securely create a subject.
- How to securely delete an object.
- How to securely delete a subject.
- How to securely provide the read access right.
- How to securely provide the grant access right.

- How to securely provide the delete access right.
- How to securely provide the transfer access right.

Moreover, each object has an owner that has special rights on it, and each subject has another subject(controller) that has special rights on it.

The model is based on the Access Control Matrix model where rows correspond to subjects and columns correspond to objects and subjects, each element contains a set of rights between subject i and object j or between subject i and subject k.

For example an action A[s,o] contains the rights that subject s has on object o (example: {own, execute}).

When executing one of the 8 rules, for example creating an object, the matrix is changed: a new column is added for that object, and the subject that created it becomes its owner. Each rule is associated with a precondition.

The Brewer and Nash model was constructed to provide information security access controls that can change dynamically. This security model, also known as the Chinese wall model, was designed to provide controls that mitigate conflict of interest in commercial organizations, and is built upon an information flow model.

In the Brewer and Nash Model no information can flow between the subjects and objects in a way that would create a conflict of interest.

## 11.4. Security architecture

Security Architecture can be defined as the design artifacts that describe how the security controls (security countermeasures) are positioned, and how they relate to the overall information technology architecture. These controls serve the purpose to maintain the system's quality attributes, among them confidentiality, integrity, availability, accountability and assurance.". A security architecture is the plan that shows where security measures need to be placed. If the plan describes a specific solution then, prior to building such a plan, one would make a risk analysis. If the plan describes a generic high level design (reference architecture) then the plan should be based on a threat analysis.

### 11.4.2. Hardware mechanisms that protect computers and data

Hardware based or assisted computer security offers an alternative to software-only computer security. Devices such as dongles may be considered more secure due to the physical access required in order to be compromised.

While many software based security solutions encrypt the data to prevent data from being stolen, a malicious program or a hacker may corrupt the data in order to make it unrecoverable or unusable. Similarly, encrypted operating systems can be corrupted by a malicious program or a hacker, making the system unusable. Hardware-based security solutions can prevent read and write access to data and hence offers very strong protection against tampering and unauthorized access.

Working of hardware based security: A hardware device allows a user to login, logout and to set different privilege levels by doing manual actions. The device uses biometric technology to prevent malicious users from logging in, logging out, and changing privilege levels. The current state of a user of the device is read both by a computer and controllers in peripheral devices such as harddisks. Illegal access by a malicious user or a malicious program is interrupted based on the current state of a user by harddisk and DVD controllers making illegal access to data impossible. Hardware based access control is more secure than logging in and logging out using operating systems as operating systems are vulnerable to malicious attacks. Since software cannot manipulate the user privilege levels, it is impossible for a hacker or a malicious program to gain access to secure data protected by hardware or perform unauthorized privileged operations. The hardware protects the operating system image and file system privileges from being tampered. Therefore, a completely secure system can be created using a combination of hardware based security and secure system administration policies.

### 11.4.3. Secure operating systems

One use of the term computer security refers to technology to implement a secure operating system. Much of this technology is based on science developed in the 1980s and used to produce what may be some of the most impenetrable operating systems ever. Though still valid, the technology is in limited use today, primarily because it imposes some changes to system management and also because it is not widely understood. Such ultra-strong secure operating systems are based on operating system kernel technology that can guarantee that certain security policies are absolutely enforced in an operating environment. An example of such a Computer security policy is the Bell-La Padula model. The strategy is based on a coupling of special microprocessor hardware features, often involving the memory management unit, to a special correctly implemented operating system kernel. This forms the foundation for a secure operating system which, if certain critical parts are designed and implemented correctly, can ensure the absolute impossibility of penetration by hostile elements. This capability is enabled because the configuration not only imposes a security policy, but in theory completely protects itself from corruption. Ordinary operating systems, on the other hand, lack the features that assure this maximal level of security. The design methodology to produce such secure systems is precise, deterministic and logical.

Systems designed with such methodology represent the state of the art of computer security although products using such security are not widely known. In sharp contrast to most kinds of software, they meet specifications with verifiable certainty comparable to specifications for size, weight and power. Secure operating systems designed this way are used primarily to protect national security information, military secrets, and the data of international financial institutions. These are very powerful security tools and very few secure operating systems have been certified at the highest level (Orange Book A-1) to operate over the range of "Top Secret" to "unclassified" (including Honeywell SCOMP, USAF SACDIN, NSA Blacker and Boeing MLS LAN.) The assurance of security depends not only on the soundness of the design strategy, but also on the assurance of correctness of the implementation, and therefore there are degrees of security strength defined for COMPUSEC. The Common Criteria quantifies security strength of products in terms of two components, security functionality and assurance level (such as EAL levels), and these are specified in a Protection Profile for requirements and a Security Target for product descriptions. None of these ultra-high assurance secure general purpose operating systems have been produced for decades or certified under the Common Criteria.

In USA parlance, the term High Assurance usually suggests the system has the right security functions that are implemented robustly enough to protect DoD and DoE classified information. Medium assurance suggests it can protect less valuable information, such as income tax information. Secure operating systems designed to meet medium robustness levels of security functionality and assurance have seen wider use within both government and commercial markets. Medium robust systems may provide the same security functions as high assurance secure operating systems but do so at a lower assurance level (such as Common Criteria levels EAL4 or EAL5). Lower levels mean we can be less certain that the security functions are implemented flawlessly, and therefore less dependable. These systems are found in use on web servers, guards, database servers, and management hosts and are used not only to protect the data stored on these systems but also to provide a high level of protection for network connections and routing services.

### 11.4.4. Secure coding

If the operating environment is not based on a secure operating system capable of maintaining a domain for its own execution, and capable of protecting application code from malicious subversion, and capable of protecting the system from subverted code, then high degrees of security are understandably not possible. While such secure operating systems are possible and have been implemented, most commercial systems fall in a 'low security' category because they rely on features not supported by secure operating systems (like portability, et al.). In low security operating environments, applications must be relied on to participate in their own protection. There are 'best effort' secure coding practices that can be followed to make an application more resistant to malicious subversion.

In commercial environments, the majority of software subversion vulnerabilities result from a few known kinds of coding defects. Common software defects include buffer overflows, format string vulnerabilities, integer overflow, and code/command injection. It is to be immediately noted that all of the foregoing are specific instances of a general class of attacks, where situations in which putative "data" actually contains implicit or explicit, executable instructions are cleverly exploited.

Some common languages such as C and C++ are vulnerable to all of these defects (see Seacord, *"Secure Coding in C and C++"*). Other languages, such as Java, are more resistant to some of these defects, but are still prone to code/command injection and other software defects which facilitate subversion.

Recently another bad coding practice has come under scrutiny; dangling pointers. The first known exploit for this particular problem was presented in July 2007. Before this publication the problem was known but considered to be academic and not practically exploitable.

Unfortunately, there is no theoretical model of "secure coding" practices, nor is one practically achievable, insofar as the variety of mechanisms are too wide and the manners in which they can be exploited are too variegated. It is interesting to note, however, that such vulnerabilities often arise from archaic philosophies in which computers were assumed to be narrowly disseminated entities used by a chosen few, all of whom were likely highly educated, solidly trained academics with naught but the goodness of mankind in mind. Thus, it was considered quite harmless if, for (fictitious) example, a FORMAT string in a FORTRAN program could contain the J format specifier to mean "shut down system after printing." After all, who would use such a feature but a well-intentioned system programmer? It was simply beyond conception that software could be deployed in a destructive fashion.

It is worth noting that, in some languages, the distinction between code (ideally, read-only) and data (generally read/write) is blurred. In LISP, particularly, there is no distinction whatsoever between code and data, both taking the same form: an S-expression can be code, or data, or both, and the "user" of a LISP program who manages to insert an executable LAMBDA segment into putative "data" can achieve arbitrarily general and dangerous functionality. Even something as "modern" as Perl offers the eval() function, which enables one to generate Perl code and submit it to the interpreter, disguised as string data.

### 11.4.5. Capabilities and access control lists

Within computer systems, two security models capable of enforcing privilege separation are access control lists (ACLs) and capability-based security. The semantics of ACLs have been proven to be insecure in many situations, e.g., the confused deputy problem. It has also been shown that the promise of ACLs of giving access to an object to only one person can never be guaranteed in practice. Both of these problems are resolved by capabilities. This does not mean practical flaws exist in all ACL-based systems, but only that the designers of certain utilities must take responsibility to ensure that they do not introduce flaws.

Capabilities have been mostly restricted to research operating systems and commercial OSs still use ACLs. Capabilities can, however, also be implemented at the language level, leading to a style of programming that is essentially a refinement of standard object-oriented design. An open source project in the area is the E language.

First the Plessey System 250 and then Cambridge CAP computer demonstrated the use of capabilities, both in hardware and software, in the 1970s. A reason for the lack of adoption of capabilities may be that ACLs appeared to offer a 'quick fix' for security without pervasive redesign of the operating system and hardware.

The most secure computers are those not connected to the Internet and shielded from any interference. In the real world, the most security comes from operating systems where security is not an add-on.

### 11.4.6. Applications

Computer security is critical in almost any technology-driven industry which operates on computer systems.Computer security can also be referred to as computer safety. The issues of computer based systems and addressing their countless vulnerabilities are an integral part of maintaining an operational industry.

### Cloud computing Security

Security in the cloud is challenging, due to varied degree of security features and management schemes within the cloud entitites. In this connection one logical protocol base need to evolve so that the entire gamet of components operate synchronously and securely.

### Computer Security in Aviation

The aviation industry is especially important when analyzing computer security because the involved risks include human life, expensive equipment, cargo, and transportation infrastructure. Security can be compromised by hardware and software malpractice, human error, and faulty operating environments. Threats that exploit computer vulnerabilities can stem from sabotage, espionage, industrial competition, terrorist attack, mechanical malfunction, and human error.

The consequences of a successful deliberate or inadvertent misuse of a computer system in the aviation industry range from loss of confidentiality to loss of system integrity, which may lead to more serious concerns such as data theft or loss, network and air traffic control outages, which in turn can lead to airport closures, loss of aircraft, loss of passenger life. Military systems that control munitions can pose an even greater risk.

A proper attack does not need to be very high tech or well funded; for a power outage at an airport alone can cause repercussions worldwide.. One of the easiest and, arguably, the most difficult to trace security vulnerabilities is achievable by transmitting unauthorized communications over specific radio frequencies. These transmissions may spoof air traffic controllers or simply disrupt communications altogether. These incidents are very common, having altered flight courses of commercial aircraft and caused panic and confusion in the past.[*citation needed*] Controlling aircraft over oceans is especially dangerous because radar surveillance only extends 175 to 225 miles offshore. Beyond the radar's sight controllers must rely on periodic radio communications with a third party.

Lightning, power fluctuations, surges, brown-outs, blown fuses, and various other power outages instantly disable all computer systems, since they are dependent on an electrical source. Other accidental and intentional faults have caused significant disruption of safety critical systems throughout the last few decades and dependence on reliable communication and electrical power only jeopardizes computer safety.[*citation needed*]

## Notable system accidents

In 1994, over a hundred intrusions were made by unidentified crackers into the Rome Laboratory, the US Air Force's main command and research facility. Using trojan horse viruses, hackers were able to obtain unrestricted access to Rome's networking systems and remove traces of their activities. The intruders were able to obtain classified files, such as air tasking order systems data and furthermore able to penetrate connected networks of National Aeronautics and Space Administration's Goddard Space Flight Center, Wright-Patterson Air Force Base, some Defense contractors, and other private sector organizations, by posing as a trusted Rome center user.  Now, a technique called Ethical hack testing is used to remediate these issues.

Electromagnetic interference is another threat to computer safety and in 1989, a United States Air Force F-16 jet accidentally dropped a 230 kg bomb in West Georgia after unspecified interference caused the jet's computers to release it.

A similar telecommunications accident also happened in 1994, when two UH-60 Blackhawk helicopters were destroyed by F-15 aircraft in Iraq because the IFF system's encryption system malfunctioned.

On April 1, 2009, Senator Jay Rockefeller (D-WV) introduced the "Cybersecurity Act of 2009 - S. 773" (full text) in the Senate; the bill, co-written with Senators Evan Bayh (D-IN), Barbara Mikulski (D-MD), Bill Nelson (D-FL), and Olympia Snowe (R-ME), was referred to the Committee on Commerce, Science, and Transportation, which approved a revised version of the same bill (the "Cybersecurity Act of 2010") on March 24, 2010[8]. The bill seeks to increase collaboration between the public and the private sector on cybersecurity issues, especially those private entities that own infrastructures that are critical to national security interests (the bill quotes John Brennan, the Assistant to the President for Homeland Security and Counterterrorism: "our nation’s security and economic prosperity depend on the security, stability, and integrity of communications and information infrastructure that are largely privately-owned and globally-operated" and talks about the country's response to a "cyber-Katrina".), increase public awareness on cybersecurity issues, and foster and fund cybersecurity research. Some of the most controversial parts of the bill include Paragraph 315, which grants the President the right to "order the limitation or shutdown of Internet traffic to and from any compromised Federal Government or United States critical infrastructure information system or network[9]." The Electronic Frontier Foundation, an international non-profit digital rights advocacy and legal organization based in the United States, characterized the bill as promoting a "potentially dangerous approach that favors the dramatic over the sober response".

**International Cybercrime Reporting and Cooperation Act**

On March 25, 2010, Representative Yvette Clarke (D-NY) introduced the "International Cybercrime Reporting and Cooperation Act - H.R.4962" (full text) in the House of Representatives; the bill, co-sponsored by seven other representatives (among whom only one Republican), was referred to three House committees[11]. The bill seeks to make sure that the administration keeps Congress informed on information infrastructure, cybercrime, and end-user protection worldwide. It also "directs the President to give priority for assistance to improve legal, judicial, and enforcement capabilities with respect to cybercrime to countries with low information and communications technology levels of development or utilization in their critical infrastructure, telecommunications systems, and financial industries" as well as to develop an action plan and an annual compliance assessment for countries of "cyber concern".

## Terminology

The following terms used in engineering secure systems are explained below.

- Authentication techniques can be used to ensure that communication end-points are who they say they are.
- Automated theorem proving and other verification tools can enable critical algorithms and code used in secure systems to be mathematically proven to meet their specifications.
- Capability and access control list techniques can be used to ensure privilege separation and mandatory access control. This section discusses their use.
- Chain of trust techniques can be used to attempt to ensure that all software loaded has been certified as authentic by the system's designers.
- Cryptographic techniques can be used to defend data in transit between systems, reducing the probability that data exchanged between systems can be intercepted or modified.
- Firewalls can provide some protection from online intrusion.

- A microkernel is a carefully crafted, deliberately small corpus of software that underlies the operating system *per se* and is used solely to provide very low-level, very precisely defined primitives upon which an operating system can be developed. A simple example with considerable didactic value is the early '90s GEMSOS (Gemini Computers), which provided extremely low-level primitives, such as "segment" management, atop which an operating system could be built. The theory (in the case of "segments") was that—rather than have the operating system itself worry about mandatory access separation by means of military-style labeling—it is safer if a low-level, independently scrutinized module can be charged solely with the management of individually labeled segments, be they memory "segments" or file system "segments" or executable text "segments." If software below the visibility of the operating system is (as in this case) charged with labeling, there is no theoretically viable means for a clever hacker to subvert the labeling scheme, since the operating system *per se* does not provide mechanisms for interfering with labeling: the operating system is, essentially, a client (an "application," arguably) atop the microkernel and, as such, subject to its restrictions.

- Endpoint Security software helps networks to prevent data theft and virus infection through portable storage devices, such as USB drives.

Some of the following items may belong to the computer insecurity article*:*

- Access authorization restricts access to a computer to group of users through the use of authentication systems. These systems can protect either the whole computer – such as through an interactive logon screen – or individual services, such as an FTP server. There are many methods for identifying and authenticating users, such as passwords, identification cards, and, more recently, smart cards and biometric systems.
- Anti-virus software consists of computer programs that attempt to identify, thwart and eliminate computer viruses and other malicious software (malware).

- Applications with known security flaws should not be run. Either leave it turned off until it can be patched or otherwise fixed, or delete it and replace it with some other application. Publicly known flaws are the main entry used by worms to automatically break into a system and then spread to other systems connected to it. The security website Secunia provides a search tool for unpatched known flaws in popular products.
- Backups are a way of securing information; they are another copy of all the important computer files kept in another location. These files are kept on hard disks, CD-Rs, CD-RWs, and tapes. Suggested locations for backups are a fireproof, waterproof, and heat proof safe, or in a separate, offsite location than that in which the original files are contained. Some individuals and companies also keep their backups in safe deposit boxes inside bank vaults. There is also a fourth option, which involves using one of the file hosting services that backs up files over the Internet for both business and individuals.
    - Backups are also important for reasons other than security. Natural disasters, such as earthquakes, hurricanes, or tornadoes, may strike the building where the computer is located. The building can be on fire, or an explosion may occur. There needs to be a recent backup at an alternate secure location, in case of such kind of disaster. Further, it is recommended that the alternate location be placed where the same disaster would not affect both locations. Examples of alternate disaster recovery sites being compromised by the same disaster that affected the primary site include having had a primary site in World Trade Center I and the recovery site in 7 World Trade Center, both of which were destroyed in the 9/11 attack, and having one's primary site and recovery site in the same coastal region, which leads to both being vulnerable to hurricane damage (e.g. primary site in New Orleans and recovery site in Jefferson Parish, both of which were hit by Hurricane Katrina in 2005). The backup media should be moved between the geographic sites in a secure manner, in order to prevent them from being stolen.

- Encryption is used to protect the message from the eyes of others. It can be done in several ways by switching the characters around, replacing characters with others, and even removing characters from the message. These have to be used in combination to make the encryption secure enough, that is to say, sufficiently difficult to crack. Public key encryption is a refined and practical way of doing encryption. It allows for example anyone to write a message for a list of recipients, and only those recipients will be able to read that message.
- Firewalls are systems which help protect computers and computer networks from attack and subsequent intrusion by restricting the network traffic which can pass through them, based on a set of system administrator defined rules.
- Honey pots are computers that are either intentionally or unintentionally left vulnerable to attack by crackers. They can be used to catch crackers or fix vulnerabilities.
- Intrusion-detection systems can scan a network for people that are on the network but who should not be there or are doing things that they should not be doing, for example trying a lot of passwords to gain access to the network.
- Pinging The ping application can be used by potential crackers to find if an IP address is reachable. If a cracker finds a computer they can try a port scan to detect and attack services on that computer.
- Social engineering awareness keeps employees aware of the dangers of social engineering and/or having a policy in place to prevent social engineering can reduce successful breaches of the network and servers.
- File Integrity Monitors are tools used to detect changes in the integrity of systems and files.

## 12. METHODOLOGY AND CATEGORIAL APPARATUS OF SCIENTIFIC RESEARCH

Methodology - is the doctrine about organization of activity.

Methodology of scientific research is the doctrine about organization of scientific research.

Methodology of scientific research includes:

1. Characteristics of scientific activities:

- Peculiarities
- Principles
- Conditions
- Norms of scientific activities

2. Logical structure of scientific activities:

    - individual (person) or social group
    - Object
    - Subject
    - Forms of activity
    - Facilities
    - Methods
    - Results

3. Time structure of scientific activities:

    - Phases (design phase, technological phase, reflexive phase)
    - Stages (conceptual stage, construction of hypothesis, construction of scientific research, process engineering of scientific research, stage of experimentation, presentation of results)

- Steps of  scientific activities

**Object** of investigation (For instance: entropy and information).

**Subject** of investigation in the framework of the object  (For instance: information creation rate of deterministic and probabilistic information sources).

**Research Issue**  (For instance:  values of information creation rate of deterministic information sources and probabilistic information sources).

Plan design of research work.

Detection and using of regularities, lows and tendencies under developing of technical and scientific systems.

## 13. METHODOLOGY AND METHODS OF SCIENTIFIC RESEARCH. SCIENTIFIC PROBLEM. SCIENTIFIC RESEARCH PURPOSE. SCIENTIFIC HYPOTHESIS. RESEARCH TASKS.

More complete presentation of scientific activities.

Time structure of scientific activities:

- Design phase
    - Conceptual stage
        - revelation of  contradiction
        - formulation of the problem
        - determination of the aim of the research
        - choice of decision criteria
    - Stage of construction of hypothesis

- ♦ construction of hypothesis
    - ♦ refinement of hypothesis
  - Stage of construction of scientific research
    - ♦ Decomposition (determination of research tasks)
    - ♦ Study of conditions (resource facilities)
    - ♦ research program construction
  - Stage of process engineering of scientific research
- Technological phase
  - Stage of experimentation

      Theoretical period
      Empirical period
  - Stage of presentation of results

      Approbation of results
      Presentation of results
- Reflexive phase

## 13.1. Methods of theoretical level of research

These methods include: induction, deduction, analysis, synthesis, abstraction.

### 13.1.1. Induction.

From The Oxford English Dictionary (OED); to induce (in relation to science and logic) means "to derive by reasoning, to lead to something as a conclusion, or inference, to suggest or imply," and induction "as the process of inferring a general law or principle from observation of particular instances." Another version is the "adducing (pulling together) of a number of separate facts, particulars, etc. especially for the purpose of proving a general statement."

My 1967 edition of the Encyclopedia Britannica (E. Brit.) gives two versions by John Stuart Mill: "the operation of discovering and proving general propositions" or "that operation of the mind by which we infer that what we know to be true in a particular case or cases will be true in all cases that resemble the former in certain assignable respects."

A paraphrase of Francis Bacon's view (also from the E. Brit.) is "a selective process of elimination among a number of alternative possibilities." The E. Brit. in a separate entry defines primary induction as "the deliberate attempt to find more laws about the behavior of the thing that we can observe and so to draw the boundaries of natural possibility more narrowly" (that is, to look for a generalization about what we can observe), and secondary induction as "the attempt to incorporate the results of primary induction in an explanatory theory covering a large field of enquiry" (that is, to try to fit the generalization made by primary induction into a more comprehensive theory).

E. Mayr in his Growth of Biologic Thought offers this definition: "inductivism claims that (we) can arrive at objective unbiased conclusions only by…recording, measuring, and describing what we encounter without any root hypothesis…."

### 13.1.2. Deduction.

Sherlock Holmes' "Elementary, my dear Watson!" has made deduction common knowledge a more familiar feature than induction in problem solving. The OED definition of *to deduce* is "to show or hold a thing to be derived from etc…" or "to draw as a conclusion from something known or assumed, to infer"; *deduction* thus is "inference by reasoning from generals to particulars," or "the process of deducing from something known or assumed…"

Both terms define systems of logic the purpose of which is to solve problems, in the one case by looking for a general characteristic (generalization, conclusion, conjecture, supposition, inference, etc.) in a set or group of observations, in the other to identify a particular instance through its resemblance to a set or group of known instances or observations. Popper’s ridicule of induction was based on the premise that induction requires the observation of *every* instance of a given phenomenon for the generalization to be true—an obvious impossibility; the fact that all known crows are black, for example, doesn’t prove that no white crows exist. Of course it is ridiculous when looked at in this way, but what *really* matters is that most if not all crows *are* black, and even if a white one should show up and prove to be a crow and not another kind of bird, most crows would still be black. His argument can also be used to make deduction useless for it, too, is based on an incomplete set of known facts. Even if the identified fact resembles the members of the set, how can we be sure that *every* possible feature of either the unknown or the members of the set itself has been considered? As we will see in what follows, in many of the examples of the way science is practiced, induction is as much a part of this practice as is deduction or any system of logic that serves the purpose of advancing knowledge. Induction and deduction are two, usually different but never contradictory, approaches to problem solving. The problem must be solved by testing the validity of the conclusion or inference, etc. reached from either direction. Induction and deduction are thus valuable, often complementary, tools that facilitate problem solving.

### 13.2. Methods of empirical level of research

These includes: computational experiments, descriptions, statistics.

## 14. SCIENTIFIC IDEA AND SIGNIFICANCE OF SCIENTIFIC RESEARCH

Here we will concider: scientific idea; scientific novelty; significance of scientific research.

## 14.1. Scientific idea.

There are two interpretations of the notion of scientific idea**.**

Idea (in philosophical sense) is the highest form of cognition. This understanding of the notion of scientific idea will main for us.

The next understanding of the notion of scientific idea is connected with planning or implementing of scientific research. It appears as some intellectual insight.

Let us give an example of the notion of scientific idea in framework of A.N. Kolmogorov’s explanation. In one interview Kolmogorov answered the question: “How do you work” with the words : “… You read… ”

## 14.2. Significance (Importance) of scientific research.

**(Example):** Definition of the Subject and its Importance (by Matthew Nicol And Karl Petersen).

“Measure-preserving systems are a common model of processes which evolve in time and for which the rules governing the time evolution don’t change. For example, in Newtonian mechanics the planets in a solar system undergo motion according to Newton’s laws of motion: the planets move but the underlying rule governing the planets’ motion remains constant. The model adopted here is to consider the time evolution as a transformation (either a map in discrete time or a flow in continuous time) on a probability space or more generally a measure space. This is the setting of the subject called ergodic theory. Applications of this point of view include the areas of statistical physics, classical mechanics, number theory, population dynamics, statistics, information theory and economics. The purpose of this chapter is to present a flavor of the diverse range of examples of measure-preserving transformations which have played a role in the development and application of ergodic theory and smooth dynamical systems theory. We also present common constructions involving measure-preserving systems. Such constructions may be considered a way of putting ‘building-block’ dynamical systems together to construct examples or decomposing a complicated system into simple ‘building-blocks’ to understand it better.”

## 15. METHODOLOGY AND METHODS OF SCIENTIFIC RESEARCH: FORMS OF SCIENTIFIC KNOWLEDGE ORGANIZATION AND PRINCIPLES OF SCIENTIFIC RESEARCH

Forms of science (scientific knowledge) organization includes: scientific fact, (theoretical) proposition, notion, category, principle, law, theory (conception).

### 15.1. Forms of science (scientific knowledge)

Scientific fact (synonym: scientific event, scientific result). Scientific fact is just only the events, phenomenon, their properties, connections, relations which are locked in by a specific manner.

(Theoretical) proposition is scientific assertion, lay down (preconceived) thought. A particular cases of proposition are axiom and theorem.

Gruber and Olsen (Gruber, 1993) gave the ontology for physical quantities, units of measure and algebra for engineering models. We will use it for presentation of elements of ontology of charges and fields from the problem domain Electromagnetism.

The SI unit of quantity of electric charge is the coulomb. The charge of an electron is approximately $-1.602 \times 10^{-19}$ COULOMB.

```
(defobject COULOMB  (basic-unit COULOMB))

(defobject   NEWTON (basic-unit   NEWTON))
```

The electric field is a vector field with SI units of newtons per coulomb (NEWTON COULOMB$^{-1}$).

```
(defobject    NEWTON / COULOMB
          (=  NEWTON / COULOMB
             (unit* newton (unit^   coulomb -1))))
```

## 15.2. Basic principles of scientific research

Illustrate basic research principles follow to American Health Information Management Association's "Basic Research Principles. Chapter 13". But instead of problem domain "Health" we will consider follow to technical universities program in physics problem domains "Relativity Theory" and "Quantum Mechenics".

Research Process

Six major steps

- Define the problem
- Review the literature
- Design the research
- Collect the data
- Analyze the data
- Draw Conclusions

Recall facts from Relativity Theory.

Experimental fact: light velocity does not depends on the state of motion of radiator and absorber.

Invariance in the Relativity Theory: invariance under shifts in space and time.

Relativity relation among energy $E$, impulse $p$ and mass $m$ for a massive particle has the form

$$E^2 = p^2c^2 + m^2c^4, \qquad (8.2)$$

We can obtain wave equation for the free particle by replacing $E$, and $p$ with quantum-mechenical operators

$$E \rightarrow i\hbar\, \partial/\partial t, \quad p \rightarrow i\hbar \nabla \; = - i\hbar(\partial/\partial x + \partial/\partial y + \partial/\partial z).$$

The operators act on wave function $\psi$ and we obtain instead of equation (8.2) the equation

$$\nabla^2 \psi + (m^2c^2/\hbar^2)\psi - (1/c^2)\, \partial^2/\partial t^2\, \psi = 0.$$

Define the problem

- Forming a question regarding a topic you would like to study
  - The most important thing is to be clear about is what you want to study
- Often an analysis of historical data, or secondary information, has gone into the problem definition
  - Looking at the past to see what has been done before
  - Prevents "reinventing the wheel"
  - In previous research, the investigator may have made recommendations for future studies

For example, a short historical overview of the development of gauge theory: from Schwarzschid action principle for electrodynamics of 1903 and Hermann Weyl's 1918 paper, in which he "gauges" the scale of length, through the Yang-Mills generalization to nonabelian gauge groups of 1954, to the recent developments.

- Problems may also be broken down into sub-problems or smaller problems
- May also define the scope of the study - who or what will be included in the study

Main problems in the Relativity Theory: the problem of the measuring of space and the problem of observation.

- Review the literature
  - An investigation of all the information about a topic
  - Three reasons for Literature Review
    - Has research already been done to answer your question?
    - Are there any data sources you can use in your study?
    - May help make your hypothesis more specific
- Sources of previous work
  - Journals, books, position papers, conference presentations, videos, interviews and online subscription services
- Design the research and Collect the data
- Several types of research design
  - Exploratory
  - Historical
  - Conclusive
    - Includes descriptive and causal
  - Correlational
  - Evaluation

- Experimental

- Design the research and Collect the data
    - Exploratory research
    - Undertaken because a problem is not very clearly defined
    - Allows the researcher to study the problem and gather information about the problem
    - May generate a hypothesis
    - Generally informal and relies on literature review and informal discussions with others in order to find out more about a problem
    - This type of research may not help you answer your problem but rather may provide you with some insights to the problem
- Design the research and Collect the data
  - Historical research
    - Involves investigation and analysis of events of the past
    - This type of research does not focus solely on the past, but it also allows the researcher to apply a previous researcher's experiences and conclusions from their study into current investigation
- Design the research and Collect the data
  - Conclusive research
    - Performed in order to come to some sort of conclusion or help in decision making
    - May be done using primary research
      - Data collected specifically for your study; or

- Secondary research
    - Includes a literature review to see if previous studies can be used to answer your question
    - May also include summaries of past works
- Design the research and Collect the data

  Conclusive research
    - Descriptive research - also called statistical research
        – Provides data about the dynamics of particle or body you are studying, including the frequency that something occurs
        – The two most common collection techniques of descriptive research are observations and surveys
- Design the research and Collect the data
    – Conclusive research
        - Causal research - tries to answer questions about what causes certain things to occur
            – This type of research is difficult because there may always be an additional cause to consider
            – Uses experimentation and simulation as its data collection methods
- Design the research and Collect the data
    – Correlational research
        - Studies that try to discover a relationship between variables
            – Variable - anything under study

- The strength of the relationship is measured
            - There may be a positive relationship or a negative relationship between variables
        - This type of research only determines if there is a relationship between two or more variables
            - It does not determine the cause of those relationships
            - Uses questionnaires, observations and secondary data as its data collection methods

- Design the research and Collect the data
    - Evaluation research
        - A process used to determine what has happened during a given activity
        - The purpose of evaluation is to lead to better understanding of whether a program is effective, whether a policy is working, or if something that was agreed upon is the most cost-effective way of doing something
- Design the research and Collect the data
    - Experimental research
        - Entails manipulation of a situation in some way in order to test a hypothesis
        - Certain variables are kept constant and an independent or experimental variable is manipulated
        - Researchers select both independent and dependent variables

- Independent variables are the factors that researchers manipulate directly
- Dependent variables are the measured variables

- Statement of the Hypothesis
  - In formal research, a *hypothesis* is formed
  - A statement of the predicted relationship of what the researcher is studying
  - A proposed solution or explanation at which the researcher has arrived through the review of the literature
  - Simply – it is the tentative answer to a question
  - The statement of the hypothesis is important because it allows the researcher to think about the variables to include in the study and type of research design to use in the study
- Statement of the Hypothesis
  - The research tests the hypothesis proving it to be positive or negative (correct or incorrect)
    - If the hypothesis is rejected, that is, incorrect, it does not necessarily mean that the research is poor, but only that the results are different from what was expected
  - The formulation of the hypothesis in advance of the data-gathering process is necessary for an unbiased investigation
- Statement of the Hypothesis
  - There are two forms of the hypothesis
    - The null hypothesis

- This hypothesis states that there is no difference between the velocity (or mass) means or proportions that are being compared; or that there is no association between the two variables that are being compared
  - For example, in an experimental trial of a new condition, the null hypothesis is: The new condition is no better than the current condition

- Statement of the Hypothesis
  - The alternative hypothesis
    - This hypothesis is a statement of what the study is set up to establish
    - For example, in our experimental trial of a new condition, the alternative hypothesis is: The new condition is better than the current condition
- Collect the Data
  - Includes primary research
    - Data you get from observations, surveys and interviews
  - Includes secondary research
    - Literature review or summaries of original studies
  - Irrespective of the data collection method the research must be valid and reliable
- Collect the Data
  - Validity

- The degree to which scientific observations actually measure or record what they purport to measure

- Collect the Data
    - Reliability
        - The repeatability, including interperson replicability, of scientific observations
        - With reliability, the major question is: Can another researcher reproduce the study using a similar instrument and get similar results?
    - A study may be reliable but not valid
    - That is, the study may be able to be replicated, but still doesn't answer the research question
- Collect the Data
    - Depends on the type of research the investigator wishes to conduct
    - For example, if the researcher want to establish a causal relationship, they should conduct one of the experimental studies
    - If they are breaking new ground in a poorly understood area of practice, they may want to consider an exploratory study in a qualitative design
- Collect the Data
    - Surveys
        - Gathers data from a relatively large number of cases at a particular time
        - Can include interviews and questionnaire surveys

- The questions should be well thought out to be sure to answer the question of the research study
      - The questions can be restricted - closed ended
        - The investigator only wants certain answers
        - Provide for unanticipated response by providing an “other” category that permits respondents to indicate other thoughts

- Collect the Data
  - Surveys
    - Questionnaire surveys
      - The questions can be unrestricted - open ended
        - This allows the participant to express a freer response in his own words
        - These questionnaires can be difficult to tabulate, however, they are easier to write
- Collect the Data
  - Observation
    - Instead of asking a participant questions, the investigator observes the participant
      - Non-participant observation, the examiner is a neutral observer who does not interact with the participants
      - Participant observation, the researcher may also participant in the actions being observed, however they try to maintain their objectivity

- Collect the Data
  - Observation
    - Ethnography or Naturalistic Inquiry
      - The researcher observes, listens to, and sometimes converses with the subjects in as free and natural an atmosphere as possible
      - The assumption is that the most important behavior of individuals in groups is a dynamic process of complex interactions and consists of more than a set of facts, statistics, or even discrete incidents
      - A position of neutrality is important in this type of research
- Collect the Data
  - Experimental study
    - This provides a logical, systematic way to answer the question, "If this is done under carefully controlled conditions, what will happen?"
    - Experimentation is a sophisticated technique for collection of data
- Selecting an Instrument
  - The instrument, also called a tool, is a consistent way to collect data
  - There are many different types of instruments to use
  - Should fit the purpose of the research
  - Researchers should not develop an instrument until they have established that one does not already exist

- If you decide to develop your own instrument take your time in developing the questions
    - It would be too costly and time consuming to have to repeat a survey because you neglected a question or two

- Samples
  - The selection of subjects for the study
- Samples
  - Types of Samples
    - Two types of sampling techniques
      - Probability sampling
      - Non-probability sampling
- Probability Samples
  - Simple Random Sample (SRS)
    - Statistics books also include a table of random numbers that can be used for the selection of random samples
    - Systematic random sampling
      - A systematic pattern is used with random sampling
- Probability Samples
  - Stratified Sampling
  - Cluster Sampling
- Non-probability Samples - Almost all qualitative research methods rely on this type of sampling

- Judgment Sampling
    - The researcher relies on his or her own judgment to select the subjects
- Non-probability Samples
  - Quota Sampling
  - Convenience Sampling
    - The selection is based on availability of the subjects
- Institutional Review Board (IRB)
  - Responsible for reviewing all research projects for approval
  - IRB is the common name, but can be called by whatever name an organization chooses
  - Purpose is to ensure that steps are being taken by the researcher to protect the rights and welfare of anyone participating in the research study
- Institutional Review Board
  - Responsible for reviewing the research procedures before the study is begun
  - May require periodic reviews during the study
  - May approve, revise or deny requests for research
- Institutional Review Board
  - Researcher must obtain permission from the IRB before any research study is started and prepare a plan for the research
    - This includes
      - Preparing the instrument

- Deciding how to select a sample
    - Deciding how to collect date
    - Getting informed consent forms for the subjects to read and sign

- Institutional Review Board
  - Organizations may also have a Monitoring Committee to ensure that there are no averse effects on participants
  - May recommend closure of research that is not meeting safety standards, does not have scientific merit or is not meeting the goals of the research
- Sample Size
  - The larger the sample, the smaller the magnitude of sampling error
  - Survey studies ordinarily have a larger sample size then experimental studies
- Sample Size
  - If you mail questionnaires, your response rate could be a low as 20 to 30 percent so a large initial sample is recommended
  - Subject availability and costs are legitimate considerations in determining a sample size
- Analyze the Data
  - The investigator tries to determine what the data disclose
    - There are a variety of statistics used to analyze data
    - Most investigators use a variety of techniques to describe the data

- Two types of statistical applications are relevant to most studies – Descriptive and Inferential statistics

- Analyze the Data
  - Descriptive statistics
    - These are statistics that describe the data
    - Includes measures of central tendency and measures of variation
  - Inferential statistics
- Analyze the Data
  - Statistics Software Packages
    - Help the researcher produce descriptive statistics, graphs and charts
- Draw Conclusions
  - Any conclusions should be related to the hypothesis, or if using the qualitative approach, then the problem identified
  - Results for each hypothesis should be described
  - Any limitations that the researcher discovered during data analysis should be reported
  - New hypotheses may be proposed if the data do not support the original hypotheses

Researchers usually include tables and graphs in this section of the research report to clarify the data

- How to develop research skills
  - Take at least one course in statistics and research methodologies

- Begin to read research studies in professional journals to see how others have performed their research
  - Agree to work with a skilled health information researcher on a study to gain experience

Learn how to present data effectively both in written form and verbally

- Draw Conclusions
  - Research always raises new questions about what to study
    - These become suggestions for future research
  - The researcher should always include conclusions as to whether or not this problem is better understood or perhaps even resolved from the research.

## 16. THEORETICAL STUDY, APPLIED STUDY AND CREATIVITY

### 16.1. Two types of research - basic and applied

- Basic research
  - The investigator is not concerned with the immediate applicability of his results but rather he or she tries to look for understanding of natural processes
- Applied research
  - The investigator has some kind of application in mind and wishes to solve a problem or in some way contribute to society

### 16.2. Creativity and its development.

We will consider by reviewing of [17-20] the notion of creativity, methods to development of creativity, conceptual map.

### 16.2.1. The notion of creativity.

The notion of creativity is connected with an ability, with an attitude and is represented as a process.

A simple definition is that creativity is the ability to imagine or invent something new. As we will see below, creativity is not the ability to create out of nothing (only God can do that), but the ability to generate new ideas by combining, changing, or reapplying existing ideas. Some creative ideas are astonishing and brilliant, while others are just simple, good, practical ideas that no one seems to have thought of yet.

Believe it or not, everyone has substantial creative ability. Just look at how creative children are. In adults, creativity has too often been suppressed through education, but it is still there and can be reawakened. Often all that's needed to be creative is to make a commitment to creativity and to take the time for it.

Creativity is also an attitude: the ability to accept change and newness, a willingness to play with ideas and possibilities, a flexibility of outlook, the habit of enjoying the good, while looking for ways to improve it. We are socialized into accepting only a small number of permitted or normal things, like chocolate-covered strawberries, for example. The creative person realizes that there are other possibilities, like peanut butter and banana sandwiches, or chocolate-covered prunes.

Creative people work hard and continually to improve ideas and solutions, by making gradual alterations and refinements to their works. Contrary to the mythology surrounding creativity, very, very few works of creative excellence are produced with a single stroke of brilliance or in a frenzy of rapid activity. Much closer to the real truth are the stories of companies who had to take the invention away from the inventor in order to market it because the inventor would have kept on tweaking it and fiddling with it, always trying to make it a little better.

The creative person knows that there is always room for improvement.

### 16.2.2. Creative Methods.

Several methods have been identified for producing creative results. Here are the five classic ones: evolution, synthesis, revolution, reapplication, changing direction.

Evolution is the method of incremental improvement. New ideas stem from other ideas, new solutions from previous ones, the new ones slightly improved over the old ones. Many of the very sophisticated things we enjoy today developed through a long period of constant incrementation. Making something a little better here, a little better there gradually makes it something a lot better--even entirely different from the original.

**Example**. Let us look at the history of the automobile or any product of technological progress. With each new model, improvements are made. Each new model builds upon the collective creativity of previous models, so that over time, improvements in economy, comfort, and durability take place. Here the creativity lies in the refinement, the step-by-step improvement, rather than in something completely new. Another example would be the improvement of the common wood screw by what are now commonly called drywall screws. They have sharper threads which are angled more steeply for faster penetration and better holding. The points are self tapping. The shanks are now threaded all the way up on lengths up to two inches. The screws are so much better that they can often be driven in without pilot holes, using a power drill.

The evolutionary method of creativity also reminds us of that critical principle: Every problem that has been solved can be solved again in a better way. Creative thinkers do not subscribe to the idea that once a problem has been solved, it can be forgotten, or to the notion that "if it ain't broke, don't fix it." A creative thinker's philosophy is that "there is no such thing as an insignificant improvement."

With the method of synthesis, two or more existing ideas are combined into a third, new idea. Combining the ideas of a magazine and an audio tape gives the idea of a magazine you can listen to, one useful for blind people or freeway commuters.

For example, someone noticed that a lot of people on dates went first to dinner and then to the theater. Why not combine these two events into one? Thus, the dinner theater, where people go first to eat and then to see a play or other entertainment.

Revolution is in some cases required (but also in some cases the dangerous step). Sometimes the best new idea is a completely different one, an marked change from the previous ones. While an evolutionary improvement philosophy might cause a professor to ask, "How can I make my lectures better and better?" a revolutionary idea might be, "Why not stop lecturing and have the students teach each other, working as teams or presenting reports?"

For example, the evolutionary technology in fighting termites eating away at houses has been to develop safer and faster pesticides and gasses to kill them. A somewhat revolutionary change has been to abandon gasses altogether in favor of liquid nitrogen, which freezes them to death or microwaves, which bake them. A truly revolutionary creative idea would be to ask, "How can we prevent them from eating houses in the first place?" A new termite bait that is placed in the ground in a perimeter around a house provides one answer to this question.

One operational meaning of reapplication is looking at something old in a new way. Go beyond labels. Unfixate, remove prejudices, expectations and assumptions and discover how something can be reapplied. One creative person might go to the junkyard and see art in an old model T transmission. He paints it up and puts it in his living room. Another creative person might see in the same transmission the necessary gears for a multi-speed hot walker for his horse. He hooks it to some poles and a motor and puts it in his corral. The key is to see beyond the previous or stated applications for some idea, solution, or thing and to see what other application is possible.

Many creative breakthroughs occur when attention is shifted from one angle of a problem to another. This is sometimes called creative insight and may be connected with a changing direction.

A classic example is that of the highway department trying to keep kids from skateboarding in a concrete-lined drainage ditch. The highway department put up a fence to keep the kids out; the kids went around it. The department then put up a longer fence; the kids cut a hole in it. The department then put up a stronger fence; it, too, was cut. The department then put a threatening sign on the fence; it was ignored. Finally, someone decided to change direction, and asked, "What really is the problem here? It's not that the kids keep getting through the barrier, but that they want to skateboard in the ditch. So how can we keep them from skateboarding in the ditch?" The solution was to remove their desire by pouring some concrete in the bottom of the ditch to remove the smooth curve. The sharp angle created by the concrete made skateboarding impossible and the activity stopped. No more skateboarding problems, no more fence problems.

This example reveals a critical truth in problem solving: **the goal is to solve the problem, not to implement a particular solution**. When one solution path is not working, shift to another. There is no commitment to a particular path, only to a particular goal. Path fixation can sometimes be a problem for those who do not understand this; they become overcommitted to a path that does not work and only frustration results.

### 16.2.3. Concept mapping versus topic maps and mind mapping.

A **concept map** is a diagram showing the relationships among concepts. They are graphical tools for organizing and representing knowledge.

Concepts, usually represented as boxes or circles, are connected with labeled arrows in a downward-branching hierarchical structure. The relationship between concepts can be articulated in linking phrases such as "gives rise to", "results in", "is required by," or "contributes to".[1]

The technique for visualizing these relationships among different concepts is called "Concept mapping".

An industry standard that implements formal rules for designing at least a subset of such diagrams is the Unified Modeling Language (UML).

Concept maps are rather similar to topic maps (in that both allow to connect concepts or topics via graphs), while both can be contrasted with the similar idea of mind mapping, which is often restricted to radial hierarchies and tree structures. Among the various schema and techniques for visualizing ideas, processes, organizations, concept mapping, as developed by Joseph Novak is unique in philosophical basis, which "makes concepts, and propositions composed of concepts, the central elements in the structure of knowledge and construction of meaning."[5] Another contrast between Concept mapping and Mind mapping is the speed and spontaneity when a Mind map is created. A Mind map reflects what you think about a single topic, which can focus group brainstorming. A Concept map can be a map, a system view, of a real (abstract) system or set of concepts. Concept maps are more free form, as multiple hubs and clusters can be created, unlike mind maps which fix on a single conceptual center.

General algorithm of the conceptual map application:

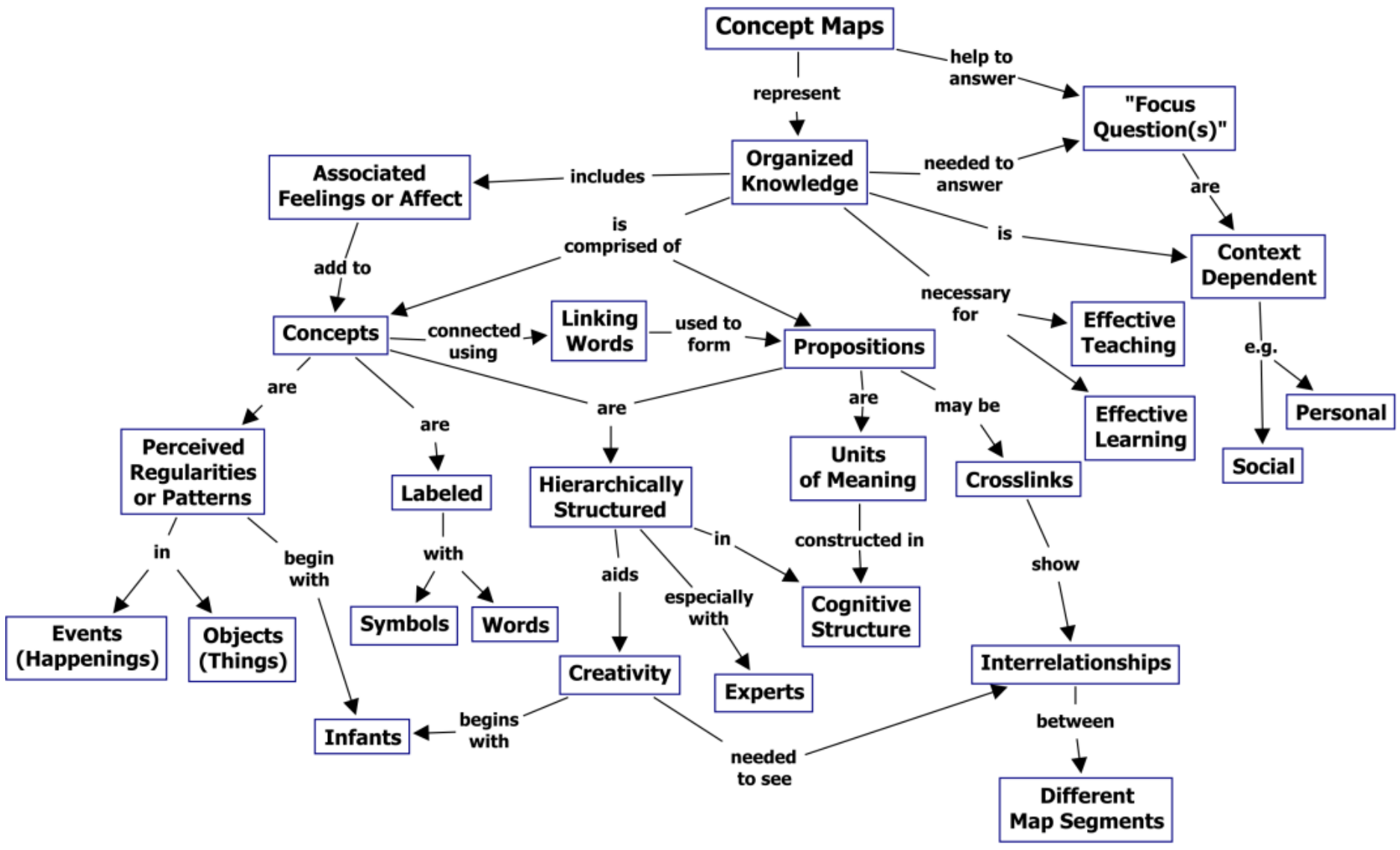

## 16.2.4. Use of Concept Maps

Concept maps are used to stimulate the generation of ideas, and are believed to aid creativity. For example, concept mapping is sometimes used for brain-storming. Although they are often personalized and idiosyncratic, concept maps can be used to communicate complex ideas.

Formalized concept maps are used in software design, where a common usage is Unified Modeling Language diagramming amongst similar conventions and development methodologies.

Concept mapping can also be seen as a first step in ontology-building, and can also be used flexibly to represent formal argument.

Concept maps are widely used in education and business for:

- Note taking and summarizing gleaning key concepts, their relationships and hierarchy from documents and source materials
- New knowledge creation: e.g., transforming tacit knowledge into an organizational resource, mapping team knowledge
- Institutional knowledge preservation (retention), e.g., eliciting and mapping expert knowledge of employees prior to retirement
- Collaborative knowledge modeling and the transfer of expert knowledge
- Facilitating the creation of shared vision and shared understanding within a team or organization
- Instructional design: concept maps used as Ausubelian "advance organizers" which provide an initial conceptual frame for subsequent information and learning.
- Training: concept maps used as Ausubelian "advanced organizers" to represent the training context and its relationship to their jobs, to the organization's strategic objectives, to training goals.
- Increasing meaningful learning
- Communicating complex ideas and arguments

- Examining the symmetry of complex ideas and arguments and associated terminology
- Detailing the entire structure of an idea, train of thought, or line of argument (with the specific goal of exposing faults, errors, or gaps in one's own reasoning) for the scrutiny of others.
- Enhancing metacognition (learning to learn, and thinking about knowledge)
- Improving language ability
- Knowledge Elicitation
- Assessing learner understanding of learning objectives, concepts, and the relationship among those concepts

Example of the conceptual map:

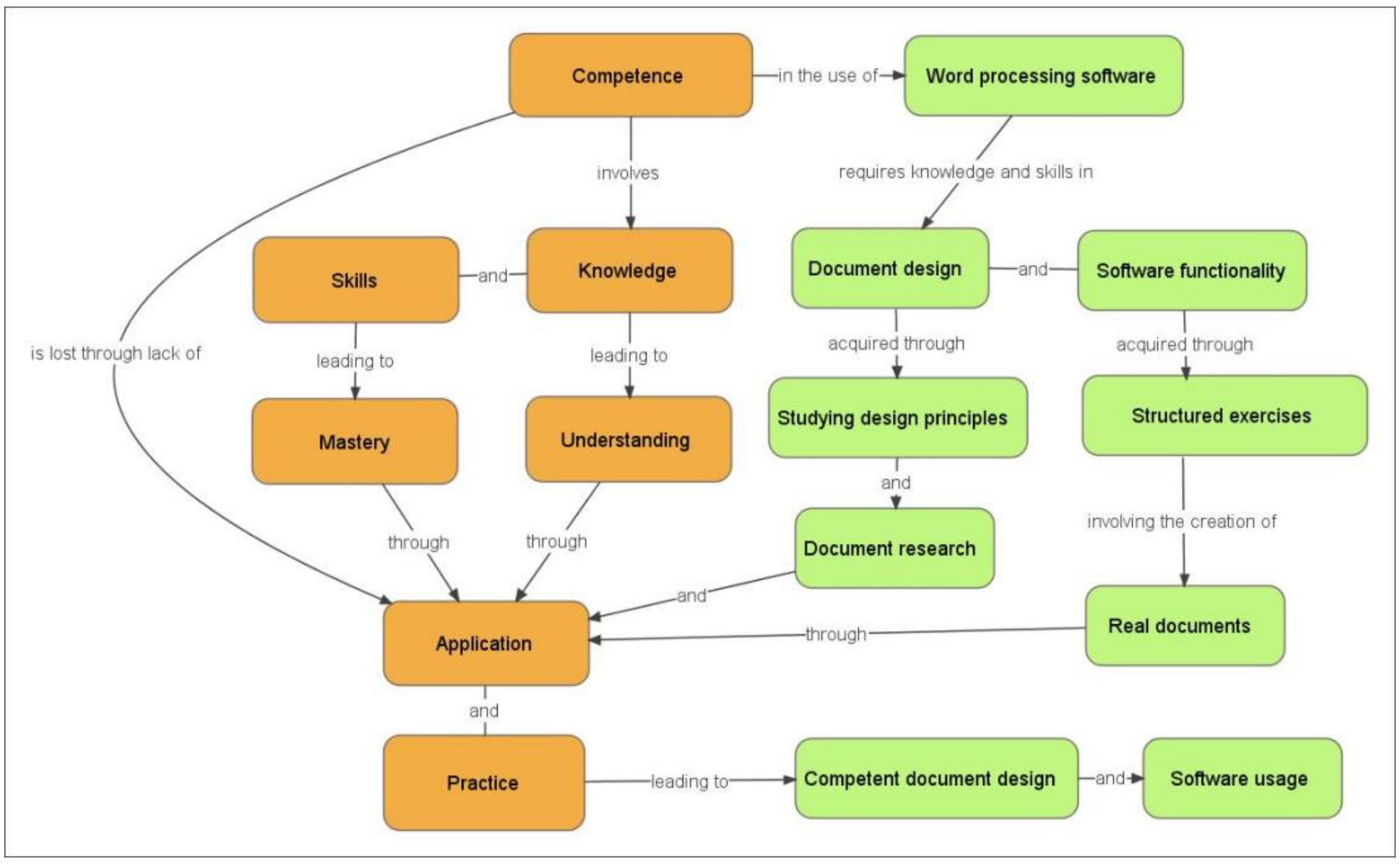

## 17. TYPES OF SCIENTIFIC RESEARCH

Here we will based on the previous considerations and indicate shortly next provisions in the framework of types of scientific research:

classification by organization activity;

classification by nature of cognitive activity.

nature of activity
cognitive activity
question design
personal safety
collective activity

Organization activity: personal, collective activity.
Nature of cognitive activity: experimental, theoretical.

Theoretical research:
Induction and deduction;
Analysis and synthesis;
Logical and historical methods.
hypothetico-deductive method
Logical method includes hypothetico-deductive method and axiomatic method.

### 17.1. Elements of scientific method.

There are different ways of outlining the basic method used for scientific inquiry. The scientific community and philosophers of science generally agree on the following classification of method components. These methodological elements and organization of procedures tend to be more characteristic of natural sciences than social sciences. Nonetheless, the cycle of formulating hypotheses, testing and analyzing the results, and formulating new hypotheses, will resemble the cycle described below.

Four essential elements of a scientific method are iterations, recursions, interleavings, and orderings of the following:

- Characterizations (observations, definitions, and measurements of the subject of inquiry)
- Hypotheses (theoretical, hypothetical explanations of observations and measurements of the subject)
- Predictions (reasoning including logical deduction from the hypothesis or theory)
- Experiments (tests of all of the above)

Each element of a scientific method is subject to peer review for possible mistakes. These activities do not describe all that scientists do (see below) but apply mostly to experimental sciences (e.g., physics, chemistry). The elements above are often taught in the educational system.

Scientific method is not a recipe: it requires intelligence, imagination, and creativity. It is also an ongoing cycle, constantly developing more useful, accurate and comprehensive models and methods. For example, when Einstein developed the Special and General Theories of Relativity, he did not in any way refute or discount Newton's *Principia*. On the contrary, if the astronomically large, the vanishingly small, and the extremely fast are reduced out from Einstein's theories — all phenomena that Newton could not have observed — Newton's equations remain. Einstein's theories are expansions and refinements of Newton's theories and, thus, increase our confidence in Newton's work.

A linearized, pragmatic scheme of the four points above is sometimes offered as a guideline for proceeding:

1. Define the question
2. Gather information and resources (observe)
3. Form hypothesis
4. Perform experiment and collect data
5. Analyze data

6. Interpret data and draw conclusions that serve as a starting point for new hypothesis
7. Publish results
8. Retest (frequently done by other scientists)

The iterative cycle inherent in this step-by-step methodology goes from point 3 to 6 back to 3 again.

While this schema outlines a typical hypothesis/testing method, it should also be noted that a number of philosophers, historians and sociologists of science (perhaps most notably Paul Feyerabend) claim that such descriptions of scientific method have little relation to the ways science is actually practiced.

The "operational" paradigm combines the concepts of operational definition, instrumentalism, and utility:

The essential elements of a scientific method are operations, observations, models, and a utility function for evaluating models.

- Operation - Some action done to the system being investigated
- Observation - What happens when the operation is done to the system
- Model - A fact, hypothesis, theory, or the phenomenon itself at a certain moment
- Utility Function - A measure of the usefulness of the model to explain, predict, and control, and of the cost of use of it. One of the elements of any scientific utility function is the refutability of the model. Another is its simplicity, on the Principle of Parsimony also known as Occam's Razor.

### 17.2. Overview of the Scientific Method

The scientific method is a process for experimentation that is used to explore observations and answer questions. Scientists use the scientific method to search for **cause and effect** relationships in nature. In other words, they design an experiment so that changes to one item cause something else to vary in a predictable way.

Just as it does for a professional scientist, the scientific method will help you to focus your science fair project question, construct a hypothesis, design, execute, and evaluate your experiment.

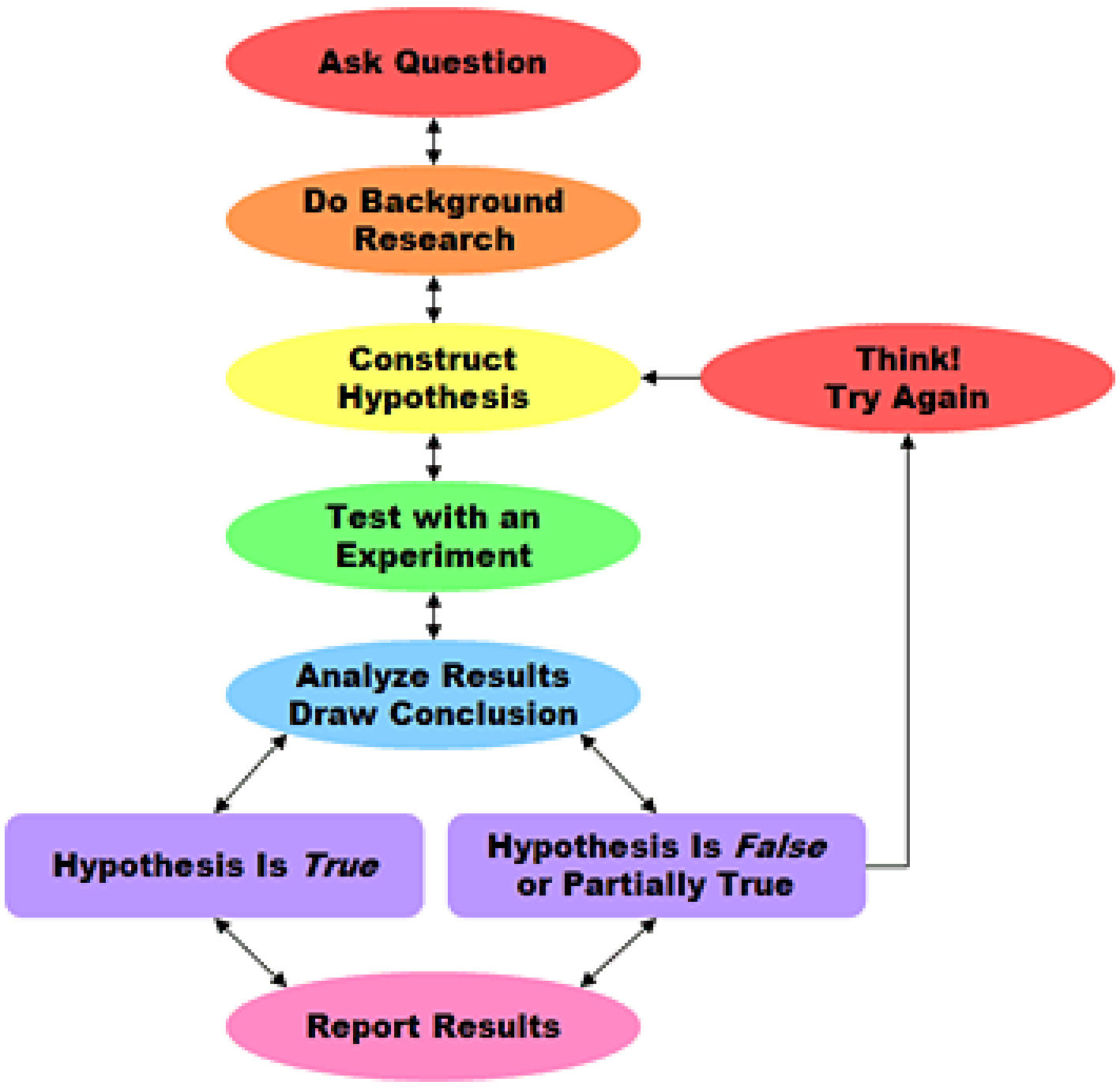

Even though we show the scientific method as a series of steps, keep in mind that new information or thinking might cause a scientist to back up and repeat steps at any point during the process. A process like the scientific method that involves such backing up and repeating is called an **iterative process**.

Throughout the process of doing your science fair project, you should keep a journal containing all of your important ideas and information. This journal is called a laboratory notebook.

### 17.3. What is the purpose of the Scientific Method?

The scientific method is the means by which researchers are able to make conclusive statements about their studies with a minimum of bias. The interpretation of data, for example the result of a new drug study, can be laden with bias. The researcher often has a personal stakes in the results of his work. As any skilled debater knows, just about any opinion can be justified and presented as fact. In order to minimize the influence of personal stakes and biased opinions, a standard method of testing a hypothesis is expected to be used by all members of the scientific community.

## 17.4. How does the Scientific Method Work?

The first step to using the scientific method is to have some basis for conducting your research. This is based on observed phenomena that is either directly or indirectly related to the specific subject matter of your proposed research. These are the observations you make in your first step of using the Scientific Method.

The next step is to form a hypothesis to explain some aspect of your observations.

Now that you have a hypothesis, you are ready to test it. You must now use your hypothesis to predict other phenomena that have not yet been observed.

The final step of the scientific method is to rigorously test your prediction. Remember, you cannot "prove" your hypothesis. You can only fail to disprove it. While this is an example of how the scientific method is used in everyday research and hypothesis testing, it is also the basis of creating theories and laws.

The scientific method requires a hypothesis to be eliminated if experiments repeatedly contradict predictions. No matter how great a hypothesis sounds, it is only as good as it's ability to consistently predict experimental results. It should also be noted that a theory or hypothesis is not meaningful if it is not quantitative and testable. If a theory does not allow for predictions and experimental research to confirm these predictions, than it is not a scientific theory.

## 17.5. What is a Hypothesis?

It is important to distinguish between a hypothesis, and a theory or law. Although in everyday language, people sometimes use these terms interchangeably, they have very distinct connotations in the scientific community.

A hypothesis is a 'small' cause and effect statement about a specific set of circumstances. It represents a belief that a researcher possesses before conducting a satisfactory number of experiments that could potentially disprove that belief.

A theory or law in the world of science is a hypothesis, or many hypotheses, which have undergone rigorous tests and have never been disproved. There is no set number of tests or a set length of time in which a hypothesis can become a theory or a law. A hypothesis becomes a theory or law when it is the general consensus of the scientific community that it should be so. Theories and laws are not as easily discarded as hypotheses.

## 17.6. Misapplications of the Scientific Method

A common error encountered by people who claim to use the scientific method is a lack of testing. A hypothesis brought about by common observations or common sense does not have scientific validity. As stated above, even though a good debater may be quite convincing as he conveys the merits of his theory, logical arguments are not an acceptable replacement for experimental testing.

Although the purpose of the scientific method is to eliminate researcher bias, an investigation of the raw data from an experiment is always a good idea. Researchers sometimes toss out data that does not support their hypothesis. This isn't necessarily done with the intent of deception, it is sometimes done because the researcher so passionately believes in his hypothesis that he assumes unsupportive data must have been obtained in error. Other times, outside forces (such as the corporation sponsoring and conducting the research) may put extreme pressure on the researcher to get specific results.

## 17.7. Problem of optimization of scientific creativity

By the 2010 Wiki [5, wiki/Creativity] creativity is the ability to generate innovative ideas and manifest them from thought into reality. The process involves original thinking and then producing.

The description of the discovering process by the French mathematician Henri Poincaré (1854-1912) was the beginning of creativity research.

Follow to [13,14] and another publications the definition of creativity has three parts:
- Creativity is a complex process, subject to studies by Psychology, Psychometrics, Cognitive Science, Artificial Intelligence, Philosophy, Economics, Business and Management, etc.
- Creativity is an interpretation of past knowledge and experience in a new way.
- Creativity contributes to the enrichment of the existing knowledge base.

George Keller expressed this definition as “Creativity, it has been said, consists largely of re-arranging what we know in order to find out what we do not know”.

Jacques Hadamard, in his book *Psychology of Invention in the Mathematical Field*, uses introspection to describe mathematical thought processes. In contrast to authors who identify language and cognition, he describes his own mathematical thinking as largely wordless, often accompanied by mental images that represent the entire solution to a problem. He surveyed 100 of the leading physicists of his day (ca. 1900), asking them how they did their work. Many of the responses mirrored his own.

Hadamard described the experiences of the mathematicians/theoretical physicists Carl Friedrich Gauss, Hermann von Helmholtz, Henri Poincaré and others as viewing entire solutions with “sudden spontaneity.”

The same has been reported in literature by many others, such as mathematicians G. H. Hardy and B. L. van der Waerden.

To elaborate on one example, Einstein, after years of fruitless calculations, suddenly had the solution to the general theory of relativity revealed in a dream “like a giant die making an indelible impress, a huge map of the universe outlined itself in one clear vision.”

Researchers have considered creativity as a process.

Hadamard described the process as having steps (i) preparation, (ii) incubation, (iv) illumination, and (v) verification of the five-step Graham Wallas creative-process model, leaving out (iii) intimation.

Creative problem solving has been defined as a process of applied creativity to the finding or solving of complex problems and having an actual behavioral creative product or plan as the final result. The process model of creative problem solving may be described by a cycle of four distinct sequential stages called Generation (problem finding), Conceptualization (problem formulation), Optimization (problem solving) and Implementation (solution implementation).

Some researchers of creativity consider the creative process as a five-step one: [13].

- Fact-finding
- Problem-finding
- Idea-finding
- Solution-finding
- Acceptance-finding.

Certain cognitive characteristics contribute to the one’s creative behaviour [10]:

- Fluency
- Flexibility
- Visualisation
- Imagination

- Expressiveness
- Openness.

There are many methods to stimulate your creativity.

We mentioned here, follow to [13] some selected methods to stimulate your creativity:

Fluency is the production of multiple problems, ideas, alternatives or solutions. It has been shown that the more ideas we produce, the more likely we are to find a useful idea or solution. Fluency is a very important ability especially in the creative problem solving process.

Brainstorming is a creative tool, which has been widely used with big success for generating many ideas.

Flexibility is the ability to process ideas or objects in many different ways given the same stimulus. It is the ability to delete old ways of thinking and begin in different directions. It is adaptive when aimed at a solution to a specific problem, challenge or dilemma. Flexibility is especially important when logical methods fail to give satisfactory results.

Verbal checklists is a family of creative tools which has been created to enhance flexibility in the creative process. Usually this is a checklist of questions about an existing product, service, process, or other item to yield new points of view and thereby lead to innovation.

Originality means getting away from the obvious and commonplace or breaking away from routine bound thinking. Original ideas are statistically infrequent. Originality is a creative strength, which is a mental jump from the obvious. Original ideas are usually described as unique, surprising, wild, unusual, unconventional, novel, weird, remarkable or revolutionary.

Picture Stimulation is a very popular technique used to provide ideas beyond those that might be obtained using brainstorming.

Originality can also be enhanced by analogies and metaphors.

Mind Mapping is a visual and verbal tool usually used to structure complex situations in a radial and expanding way during the creative problem solving process. A mind map is by definition a creative pattern of related ideas, thoughts, processes, objects, etc.

## 17.8. Principles of optimization scientific creativity

### Some optimization techniques

In the above we introduced functions of the form $f(\mathbf{a};\mathbf{Y})$ which measure the fit of a model instance with **n** parameters a to some set of data Y. We are interested in the optimal choice of parameters, those which give the best fit to the data. This involves finding the optimum (maximum or minimum) of the function $f(\mathbf{a};\mathbf{Y})$ with respect to a. For notational simplicity we will use $f(\mathbf{a}) = f(\mathbf{a};\mathbf{Y})$. Since any maximum of $f(\mathbf{a})$ is a minimum of $-f(\mathbf{a})$ we will only consider minimization. Formally $\mathbf{a}$ is a minimum point of $f(\mathbf{a})$ if there exists a region about a of radius $\epsilon$ such that

$$f(\mathbf{a}+\mathbf{x}) > f(\mathbf{a}) \quad \forall \quad |\mathbf{x}| < \epsilon$$

The maxima and minima of a function can either be *global* (the highest or lowest value over the whole region of interest) or *local* (the highest or lowest value over some small neighbourhood). We are usually most interested in finding the global optimum (such as the model parameters which give the best match to some image data), but this can be very difficult. Often a problem will have many local optima (perhaps caused by image noise or clutter) which means that locating the single global optima can be tricky.

The most suitable methods to locate minima depend upon the nature of the function we are dealing with. There are two broad classes of algorithms.

- *Local* minimizers that, given a point in a `valley' of the function, locate the lowest point on the valley.
- *Global* minimizers that range over a region of parameter space attempting to find the bottom of the deepest valley.

If a good estimate of the position of the minimum exists we need only use a local minimizer to improve it and find the optimum choice of parameters. If no such estimate exists some global method must be used. The simplest would be to generate a set of possible start points, locally optimize each and choose the best. However, this may not be the most efficient approach.

Often an application will require both local and global methods. For instance, in a tracking problem initializing a model on the first frame may require a global search, but subsequent frames would only require a local search about the current best estimate.

The choice of which local minimization technique to use will depend upon

- Whether a is one or many-dimensional,
- Whether $f(\mathbf{a})$ can be differentiated efficiently,
- How noisy $f(\mathbf{a})$ is.

In the following we will give an overview of some of the methods for locating both global and local minima. For a more comprehensive survey, including algorithmic details, see [6].

## Minimization in One Dimension

The simplest minimization problems are those in which there is only a single variable. Later we will see how minimization in many dimensional space can be broken down into a series of 1-D `line' minimizations.

One major advantage of 1-D over multi-D minimization is that it is easy to define a region in which we are sure a minima must exist.

If we can find three points **a**, **b** and **c** with **a<b<c** and $f(a) > f(b) < f(c)$ then there must exist at least one minimum point in the interval $(a,c)$. The points **a** , **b** and **c** are said to *bracket* the minimum.

## Multi-Dimensional Search

Finding the **n** parameters a which minimize $f(\mathbf{a})$ when **n>1** is a much harder problem. The **n=2** case is like trying to find the point at the bottom of the lowest valley in a range of mountains. Higher dimensional cases are hard to visualize at all. This author usually just pretends they are 2-D problems and tries not to worry too much.

When presented with a new problem it is usually fruitful to get a feel for the shape of the function surface by plotting $f(\mathbf{a})$ against $a_i$ for each $i = 1..n$, in each case keeping $a_j \ (j \neq i)$ fixed.

As mentioned above the choice of local minimizer will depend upon whether it is possible to generate first or second derivatives efficiently, and how noisy the function is. All the local minimizers below assume that the function is locally relatively smooth - any noise will confuse the derivative estimates and may confound the algorithm. For such functions the Simplex Method is recommended.

Most of the local minimizers below rely on approximating the function about the current point, a, using a Taylor series expansion

At the minimum the gradient is zero (if it wasn't we could move further downhill), so $\mathbf{b} = \mathbf{0}$and

$$f(\mathbf{a}_{min} + \mathbf{x}) \approx c + \frac{1}{2}\mathbf{x}^T \mathbf{A}\mathbf{x}$$

(where A is the Hessian evaluated at the minimum point $\mathbf{a}_{min}$).

**Multi-Dimensional Search (With First Derivatives)**

Suppose that we can efficiently calculate the gradient vector $\nabla f(\mathbf{a})$at a point (by `efficiently' we mean analytically, requiring less computation than **n** individual evaluations of the original function). There are cunning `Conjugate Gradient' algorithms available which use this information to build up conjugate sets much faster than Powell's method (**n** line minimizations rather than $n^2$).

**Global Optimization**

Global optimizers are useful when the search space is likely to have many minima, making it hard to locate the true global minimum. In low dimensional or constrained problems it may be enough to apply a local optimizer starting at a set of possible start points, generated either randomly or systematically (for instance at grid locations), and choose the best result. However this approach is less likely to locate the true optimum as the ratio of volume of the search region to number of starting points increases. Correctly applied, the Simulated Annealing and Genetic Algorithm approaches described below can explore the search space better than a grid search for a given number of function evaluations, and are more likely to find the true global minimum. Note that both these approaches involve a stochastic element and so may fail to find the true minimum. In addition, since they are better at global searching than local optimization, it is usually worthwhile to `polish' any final solution using one of the local optimizers above.

The most widely used optimization methods are the following:

- Simplex Algorithm
- Simulated Annealing
- Genetic Algorithms
- Multi-Resolution Methods and Graduated Non-Convexity
- The Hough Transform

Subgradient

$$f(x)-f(x_0)\geq(g_f(x_0),x-x_0)$$

Sub differential

$$\partial f(x_0)=\left\{x^*\in X^*\mid f(x)-f(x_0)\geq<x^*,x-x_0>,\forall x\in X\right\}$$

Subgradient method

$$x_{k+1}=x_k-h_k(x_k)g_f(x_k)$$

Sub gradient method with the space dilation

$$x=\gamma_\varsigma(x)\varsigma+d_\varsigma(x)$$

Where

$$\gamma_\varsigma(x)=(x,\varsigma),\qquad (x,d_\varsigma(x))=0,$$

$$R_\alpha(\varsigma)x=\alpha\gamma_\varsigma(x)\varsigma+d_\varsigma(x).$$

## 18. TYPES OF SCIENTIFIC RESEARCH: FORMS OF PRESENTATION OF MATERIALS OF SCIENTIFIC RESEARCH.

materials of a conference

approbation, approval testing

Stage of approval of results of scientific research.

paper work

preparation

Preparation of publication.

The structure of publication.

## 18. Scientific publications. Types, structure, logic of publications. Rules to righting of student works: papers, abstract, diploma thesis.

### 18.1. Types of scientific publications

We will consider next types of scientific publications: paper, abstract, diploma thesis. We will illustrate structure and logic of scientific publications on the example of scientific paper.

### 18.2. Structure of scientific publication.

1. Title
2. Author
3. Abstract
4. Introduction

- Object
- Subject
- Aim of the research

5. Content (sections)
6. Conclusion

References

**18.3. The structure of a scientific paper.** All scientific papers have the same general format. They are divided into distinct sections and each section contains a specific type of information. The number and the headings of sections may vary among journals, but for the most part a basic structure is maintained. Typically, scientific papers in physics, chemistry, biology and behavioral sciences, astronomy and astrophysics, geophysics, sociology are comprised of the following parts:

- Title
- Abstract
- Introduction
- Methods
- Results
- Discussion
- Acknowledgments
- Literature sited

A scientific paper in the disciplines is a written report describing original research results. The format of a scientific paper has been defined by centuries of developing tradition, editorial practice, scientific ethics and the interplay with printing and publishing services. A scientific paper should have, in proper order, a Title, Abstract, Introduction, Materials and Methods, Results, and Discussion.

**Title.** A title should be the fewest possible words that accurately describe the content of the paper. Omit all waste words such as "A study of ...", "Investigations of ...", "Observations on ...", etc. Indexing and abstracting services depend on the accuracy of the title, extracting from it keywords useful in cross-referencing and computer searching. An improperly titled paper may never reach the audience for which it was intended, so be specific. If the study is of a particular species, name it in the title. If the inferences made in the paper are limited to a particular region, then name the region in the title.

**Keyword List.** The keyword list provides the opportunity to add keywords, used by the indexing and abstracting services, *in addition* to those already present in the title. Judicious use of keywords may increase the ease with which interested parties can locate your article.

**Abstract.** A well prepared abstract should enable the reader to identify the basic content of a document quickly and accurately, to determine its relevance to their interests, and thus to decide whether to read the document in its entirety. The abstract should concisely state the principal objectives and scope of the investigation where these are not obvious from the title. More importantly, it should concisely summarize the results and principal conclusions. Do not include details of the methods employed unless the study is methodological, i.e. primarily concerned with methods.

The abstract must be concise, not exceeding 250 words. If you can convey the essential details of the paper in 100 words, do not use 200. Do not repeat information contained in the title. The abstract, together with the title, must be self-contained as it is published separately from the paper in abstracting services such as Zentralblatt MATH or Mathematical Review. Omit all references to the literature and to tables or figures, and omit obscure abbreviations and acronyms even though they may be defined in main body of the paper.

## Introduction

The Introduction should begin by introducing the reader to the pertinent literature. A common mistake is to introduce authors and their areas of study in general terms without mention of their major findings.

It has to be informative lead-in to the literature, and it will enable the reader to clearly place the current work in the context of what is already known. An important function of the introduction is to establish the significance of the current work: Why was there a need to conduct the study?

Having introduced the pertinent literature and demonstrated the need for the current study, you should state clearly the scope and objectives. Avoid a series of point-wise statements -- use prose. A brief description of the region in which the study was conducted, and of the taxa in question, can be included at this point. The introduction can finish with the statement of objectives or, as some people prefer, with a brief statement of the principal findings. Either way, the reader must have an idea of where the paper is heading in order to follow the development of the evidence.

## Materials and Methods

The main purpose of the Materials and Methods section is to provide enough detail for a competent worker to repeat your study and reproduce the results. The scientific method requires that your results be reproducible, and you must provide a basis for repetition of the study by others.

Often in field-based studies, there is a need to describe the study area in greater detail than is possible in the Introduction. Usually authors will describe the study region in general terms in the Introduction and then describe the study site and climate in detail in the Materials and Methods section. The sub-headings "Study Site", "General Methods" and "Analysis" may be useful, in that order.

Equipment and materials available off the shelf should be described exactly (Supercomputer Earth Simulator (ES), 35.86 TFLOPS, 700 terabytes of disk storage (450 for the system and 250 for the users) and 1.6 petabytes of mass storage in tape drives) and sources of materials should be given if there is variation in quality among supplies. Modifications to equipment or equipment constructed specifically for the study should be carefully described in detail. The method used to prepare reagents, fixatives, and stains should be stated exactly, though often reference to standard recipes in other works will suffice.

The usual order of presentation of methods is chronological, however related methods may need to be described together and strict chronological order cannot always be followed. If your methods are new (unpublished), you must provide all of the detail required to repeat the methods. However, if a method has been previously published in a standard journal, only the name of the method and a literature reference need be given.

Be precise in describing measurements and include errors of measurement. Ordinary statistical methods should be used without comment; advanced or unusual methods may require a literature citation.

Show your materials and methods section to a colleague. Ask if they would have difficulty in repeating your study.

## Results

In the results section you present your findings. Present the data, digested and condensed, with important trends extracted and described. Because the results comprise the new knowledge that you are contributing to the world, it is important that your findings be clearly and simply stated.

The results should be short and sweet, without verbiage. However, do not be too concise. The readers cannot be expected to extract important trends from the data unaided. Few will bother. Combine the use of text, tables and figures to condense data and highlight trends. In doing so be sure to refer to the guidelines for preparing tables and figures below.

## Discussion

In the discussion you should discuss the results. What cybernetical principles have been established or reinforced? What generalizations can be drawn? How do your findings compare to the findings of others or to expectations based on previous work? Are there any theoretical/practical implications of your work? When you address these questions, it is crucial that your discussion rests firmly on the evidence presented in the results section. Continually refer to your results (but do not repeat them). Most importantly, do not extend your conclusions beyond those which are directly supported by your results. Speculation has its place, but should not form the bulk of the discussion. Be sure to address the objectives of the study in the discussion and to discuss the significance of the results. Don't leave the reader thinking "So what?".

**Conclusions**

In this conclution section you have to give a short summary or conclusion regarding the significance of the work.

**Note**. In mathematics, computer science and softwere engineering the structure of scientific paper follow to the structure of scientific publication:

1.Title

2.Author

3.Abstract

4.Introduction

- Object
- Subject
- Aim of the research

5.Content (sections)

## 6.Conclusion

## References

# CONCLUSIONS

In the manual we present various known results and approaches at the field of scientific research.

It consists of 18 sections and some ideas of the manual can be seen from their (and their subsections) titles: 1. General notions about scientific research (SR). 1.1. Scientific method. 1.2.Basic research. 1.3. Information supply of scientific research.

2. Ontologies and upper ontologies. 2.1 Concepts of Foundations of Research Activities. 2.2 Ontology components. 2.3. Ontology for the visualization of a lecture.
3. Ontologies of object domains. 3.1. Elements of the ontology of spaces and symmetries. 3.1.1. Concepts of electrodynamics and classical gauge theory.
4. Examples of Research Activity. 4.1. Scientific activity in arithmetics, informatics and discrete mathematics. 4.2. Algebra of logic and functions of the algebra of logic. 4.3. Function of the algebra of logic.
5. Some Notions of the Theory of Finite and Discrete Sets.
6. Algebraic Operations and Algebraic Structures.
7. Elements of the Theory of Graphs and Nets.
8. Scientific activity on the example “Information and its investigation”**.**
9. Scientific research in Artificial Intelligence.
10. Compilers and compilation.
11. Objective, Concepts and History of Computer security.
12. Methodological and categorical apparatus of scientific research.
13. Methodology and methods of scientific research. 13.1. Methods of theoretical level of research. 13.1.1. Induction. 13.1.2. Deduction. 13.2. Methods of empirical level of research.
14. Scientific idea and significance of scientific research.
15. Forms of scientific knowledge organization and principles of SR. 15.1. Forms of scientific knowledge. 15.2. Basic principles of scientific research.
16. Theoretical study, applied study and creativity. 16.1. Two types of research - basic and applied. 16.2. Creativity and its development. 16.2.1. The notion of creativity. 16.2.2. Creative Methods. 16.2.3. Concept mapping versus topic maps and mind mapping. 16.2.4. Use of Concept Maps.
17. Types of scientific research: theoretical study, applied study. 17.1. Elements of scientific method. 17.2. Overview of the Scientific Method. 17.3. What is the purpose of the Scientific Method? 17.4. How does the Scientific Method Work? 17.5. What is a Hypothesis? 17.6. Misapplications of the Scientific Method. 17.7. Problem of optimization of scientific creativity. 17.8. Principles of optimization scientific creativity’
18. Types of scientific research: forms of representation of material.